%% file: main.tex
\documentclass[12pt,twoside]{mitthesis}
\usepackage{lgrind}
\usepackage{cmap}
\usepackage[T1]{fontenc}
\usepackage{lmodern}
\def\all{all}

\usepackage{amsmath,amsthm,amssymb,amsfonts,chngcntr,setspace,array,graphicx,multirow,booktabs,tensor,enumerate}
\usepackage[dvipsnames]{xcolor}
\usepackage{lscape}
\usepackage{verbatim}
\usepackage{tabularx}
\usepackage{mathtools}
\usepackage{systeme}
\usepackage{float}
\usepackage{numprint}
\usepackage{caption}
\usepackage{subcaption}
\usepackage{multirow}
\usepackage{etoolbox}
\usepackage{blkarray, bigstrut} 
\usepackage{setspace}
\usepackage{url}
\usepackage{seqsplit}
\usepackage{pdflscape}
\usepackage[nobiblatex]{xurl}
\usepackage[numbers,square,sort]{natbib}
\usepackage[utf8]{inputenc}
\usepackage{hyperref}
\usepackage{stackengine,scalerel}
\usepackage[normalem]{ulem}

\usepackage{color,soul}

\usepackage{algorithm}
\usepackage{algpseudocode}

\usepackage{makecell}

\usepackage[shortlabels]{enumitem}

\usepackage[left=1.1in,right=1.1in,top=1.15in,bottom=1.3in,footskip=.7in]{geometry}

\usepackage{mathrsfs}

\DeclareFontFamily{U}{calligra}{}
\DeclareFontShape{U}{calligra}{m}{n}{<->callig15}{}

\usepackage{chngcntr}
 
\counterwithout{figure}{chapter}
\counterwithout{table}{chapter}

\usepackage{caption}

\DeclarePairedDelimiter\floor{\lfloor}{\rfloor}

\newtheorem{proposition}{Proposition}
\newtheorem{definition}{Definition}
\newtheorem{example}{Example}
\newtheorem{property}{Property}

\def\tsc#1{\csdef{#1}{\textsc{\lowercase{#1}}\xspace}}
\tsc{WGM}
\tsc{QE}
\tsc{EP}
\tsc{PMS}
\tsc{BEC}
\tsc{DE}

\newcommand\overstar[1]{\ThisStyle{\ensurestackMath{%
  \setbox0=\hbox{$\SavedStyle#1$}%
  \stackengine{0pt}{\copy0}{\kern.2\ht0\smash{\SavedStyle*}}{O}{c}{F}{T}{S}}}}

\definecolor{RR_color}{RGB}{0, 0, 0}

\newcommand{\addRR}[1]{\textcolor{RR_color}{{#1}}}

\definecolor{rp_color}{rgb}{1, 0, 0} 

\usepackage{setspace}
\begin{document}

\include{cover}
\pagestyle{plain}
\include{contents}

\include{chap0}
\include{chap1}
\include{chap2}
\include{chap3}
\include{chap4}
\include{chap5}
\appendix
\include{appa}
\include{appb}
\include{appc}
\include{biblio}
\end{document}

%% file: cover.tex
%
%
%
%
%
%
%
%
%
%
%

\title{Exact and Heuristic Algorithms for Energy-Efficient Scheduling}

\author{Roberto Ronco}

\department{Department of Computer Science, Bioengineering, Robotics, and Systems Engineering}

\degree{Doctor of Philosophy in Computer Science and Systems Engineering}

\degreemonth{January}
\degreeyear{2022}
\thesisdate{October 31, 2021}


\supervisor{Massimo Paolucci}{Associate Professor}
\supervisor{Raffaele Pesenti}{Full Professor}

\chairman{Chairman name}{Chairman, Department Committee on Doctoral Theses}

\maketitle



\cleardoublepage
\setcounter{savepage}{\thepage}



\cleardoublepage

%% file: contents.tex
\tableofcontents
\newpage
\listoffigures
\newpage
\listoftables

%% file: chap0.tex
\chapter*{Preface}

The combined increase of energy demand and environmental pollution at a global scale is entailing a rethinking of the production models in sustainable terms.
As a consequence, energy suppliers are starting to adopt strategies that flatten demand peaks in power plants by means of pricing policies that stimulate a change in the consumption practices of customers.
A representative example is the \emph{Time-of-Use} (TOU)-based tariffs policy, which encourages electricity usage at off-peak hours by means of low prices, while penalizing peak hours with higher prices.
To avoid a sharp increment of the energy supply costs, manufacturing industry must carefully reschedule the production process, by shifting it towards less expensive periods.
The TOU-based tariffs policy induces an implicit partitioning of the time horizon of the production into a set of time slots, each associated with a non-negative cost that becomes a part of the optimization objective.

This thesis focuses on a representative bi-objective energy-efficient job scheduling problem on parallel identical machines under TOU-based tariffs by delving into the description of its inherent properties, mathematical formulations, and solution approaches.
Specifically, the thesis starts by reviewing the flourishing literature on the subject, and providing a useful framework for theoreticians and practitioners.
Subsequently, it describes the considered problem and investigates its theoretical properties.
In the same chapter, it presents a first mathematical model for the problem, as well as a possible reformulation that exploits the structure of the solution space so as to achieve a considerable increase in compactness.
Afterwards, the thesis introduces a sophisticated heuristic scheme to tackle the inherent hardness of the problem, and an exact algorithm that exploits the mathematical models.
Then, it shows the computational efficiency of the presented solution approaches on a wide test benchmark.
Finally, it presents a perspective on future research directions for the class of energy-efficient scheduling problems under TOU-based tariffs as a whole.

%% file: chap1.tex

\chapter{Introduction}
\label{chap:intro}

This thesis deals with an energy-efficient job scheduling problem on parallel identical machines under TOU-based tariffs that requires to find a schedule so as to simultaneously minimize the maximum completion time of the jobs, and a measure of the energy consumption cost of the schedule.
The purpose of the thesis is to tackle such problem by means of a fast heuristic scheme and an efficient exact algorithm that exploits structural properties of the solutions space.

This chapter is organized as follows.
Section \ref{sec:motivation} stresses the compelling need and hard challenge of properly integrating energy efficiency criteria and objectives in the production processes.
Afterwards, Section \ref{sec:basic-definition-and-notation} introduces a formal framework that proves useful to provide a solid foundation for the basic definitions, properties, and notation used in scheduling under TOU-based tariffs.
Then, Section \ref{sec:literature-classification} presents a thorough literature review on the subject.
Section \ref{sec:thesis-structure} concludes by describing the structure of the remainder of the thesis.

\section{Motivation}
\label{sec:motivation}

The dramatic rise of worldwide energy consumption in the last decades, together with the related environmental pollution, became a compelling matter for global demand supply. The World Energy Outlook 2019 reported 14314 million tonnes of oil-equivalent demand from primary energy consumption, corresponding to 33.2 gigatonnes of CO$_2$ emission in the sole 2018~\citep{IEA_2019}. The sectors related to manufacturing, agriculture, mining, and construction collectively consumed 54\% of the worldwide energy demand in 2012~\citep{IEA_2016}. In 2018, the total energy consumption of the sole manufacturing sector amounted to 19.436 trillion British thermal units \citep{MECS_2018} and, all together, the above sectors were forecast to grow at a rate of 1.2\% per year until 2040~\citep{IEA_2016}. 

Rethinking the production processes under a sustainable lens, and simultaneously fostering environment-aware consumption practices in customers, appear to be a necessary condition to invert this trend on the long term. One of the first actions undertaken by energy suppliers to relieve this trend in the short term consisted of flattening the peaks of demand in power plants by means of strategies aimed at reducing the high economic burdens related to the generation of high energy loads in short periods of time and, consequently, the environmental impact related to energy production. These strategies mostly consist of pricing policies that stimulate a change in the consumption practices of customers~\citep{TOU3_Hu2018ASD}. One example of such policies is provided by the \emph{Time-of-Use} (TOU)-based tariffs, that spur  electricity usage at off-peak hours by means of low prices, while penalizing peak hours with higher prices. The identification of the optimal prices with regard to the inherent stochasticity of the demand can be carried out, e.g., by means of robust optimization approaches, such as those described in~\citep{TOU3_Hu2018ASD}, or stochastic programming models, such as those discussed in \citep{TOU7_Nikzad2020IntegrationOO}. Overall, TOU-based tariff policies proved effective to flatten the peaks of demand so far, by ensuring, at the same time, a good service stability~\citep{TOU1_Chawla2017StabilityOS,TOU3_Hu2018ASD}. How long they will succeed to contain a world-scale increase in energy consumption, however, still remain obscure to predict. 

In the context of manufacturing, an immediate reaction to TOU-based tariffs consisted of carefully rescheduling the production processes during periods characterized by low energy supply costs. 
One of the requirements to accomplish this task is the rescheduling the jobs involved in such processes. 
Job Scheduling under TOU-based tariffs (or \emph{TOU scheduling} for short) is similar to classical job scheduling problems, in that a processing order of a number of jobs must be identified and carried out on one or possibly multiple machines~\citep{Pinedo2016}. The processing of the jobs must comply with time requirements (e.g., a due date or a common deadline)~\citep{Pinedo2016}. However, the presence of TOU-based tariffs induces an implicit partitioning of the time horizon of the production into a set of time slots, each associated with a non-negative cost called \emph{TOU costs} \citep{TimeSch3_chen_scheduling_2019} or \emph{TOU prices} \citep{TimeSch31_ding_parallel_2016}. This partitioning is usually referred to as \emph{TOU pricing scheme} \citep{TimeSch30_che_energy-conscious_2017,TimeSch31_ding_parallel_2016,TimeSch48_soares_deterministic_2021}.
The sum of the TOU costs of slots associated with some job processing in the considered time horizon, also referred to as the \emph{Total Energy Cost} (TEC), becomes part of the classical objectives of a scheduling problem or, in the context of multi-objective optimization problems, adds to the classical objectives in scheduling, e.g., the flow time or the total weighted tardiness~\citep{Pinedo2016}.
In addition, the TEC, being a non-regular performance measure~\citep{Pinedo2016}, often represents a conflicting objective with respect to the usual classical ones.

The literature on sustainable manufacturing systems offers a number of recent surveys on energy-efficient scheduling. For instance, \citet{Survey3_Gahm2016EnergyefficientSI} study the impact of energy-aware practices in sustainable production and propose a classification of energy-efficient job scheduling based on energetic coverage, energy supply and energy demand. \citet{Survey4_supplyChain2020} focus on the technological aspects of manufacturing, by analyzing the service supply chain and its environmental-related concerns. \citet{Survey1_Despeisse2012IndustrialEA} propose a conceptual framework to model single factory units as ecosystems aimed at enabling sustainable performance improvements while limiting environmental pollution and unrestrained exploitation of natural resources. \citet{Survey2_Giret2015SustainabilityIM} focus on energy-efficiency operations scheduling by exploring the conditions for its sustainability. They distinguish between (i) approaches based on the input data (e.g., as machines, jobs, and scheduling horizon), 
(ii) approaches based on the environmental output (e.g., pollution, waste), and 
(iii) approaches based on mixed objectives. Finally, \citet{Survey5_Gao2020ARO} focus on energy-efficient scheduling in intelligent production systems, by proposing a general classification of the most important problems in sustainable manufacturing and of the corresponding solution approaches. None of the above surveys, however, comprehensively discusses the models, methods, and algorithms for job scheduling under TOU-based energy tariffs presented in the literature. 
This chapter closes this gap, by providing researchers and practitioners with (i) a framework that may summarize the most important theoretical results and practical applications on the topic as well as (ii) a guide that may direct new research efforts towards unexplored directions in TOU scheduling for sustainable manufacturing. 


\section{Basic definitions and notation}
\label{sec:basic-definition-and-notation} 
This section formalizes the job scheduling problems in presence of TOU-based energy tariffs, by discussing the properties that time slots, jobs, and machines may have, as well as the different problems that occur in the literature. 

Let us introduce some definitions and notation. 
Given three positive integers $N$, $M$, and $K$, let $\mathcal{J} = \{1, 2, \ldots, N\}$, $\mathcal{H} = \{1, 2, \ldots, M\}$, and $\mathcal{T} = \{1, 2, \ldots, K\}$, be a set of jobs, machines, and time slots, respectively.
Each job in $\mathcal{J}$ has to be processed on \addRR{a machine of the set} $\mathcal{H}$, until its completion, in a subset of the $K$ consecutive time slots in $\mathcal{T}$ that constitute the time horizon of the production.
Furthermore, each machine $h \in \mathcal{H}$ can process at most one job at a time.

Let $p_{j, h}$ denote a positive integer encoding the \emph{processing time} of the job $j \in \mathcal{J}$ on machine $h \in \mathcal{H}$, i.e., the number of time slots required to process $j$ on $h$. 
If the processing time of the jobs does not depend on $h$, such as in the case of parallel identical machines, we omit the second subscript from $p_{j, h}$.
Let also $c_t$ denote a positive real representing the TOU cost associated with the time slot $t \in \mathcal{T}$. 
The sequence $\{c_t, t = 1, 2, \ldots, K\}$ is referred to as $\{c_t\}$. Moreover, $\{c_t\}$ is \emph{pyramidal} \citep{TimeSch2_fang_scheduling_2016} if there is some slot $t'$ such that $c_1 < c_2 < \ldots < c_{t' - 1} < c_{t'} > c_{t' + 1} > \ldots > c_{K - 1} > c_K$. When the costs $c_t$ are non-increasing (non-decreasing) with respect to $t = 1, 2, \ldots, K$, $\{c_t\}$ is simply said to be non-increasing (non-decreasing).

\begin{figure}[t]
    \captionsetup{font=footnotesize}
     \centering
     \begin{subfigure}[b]{0.28\textwidth}
         \centering
         \includegraphics[width=\textwidth]
         {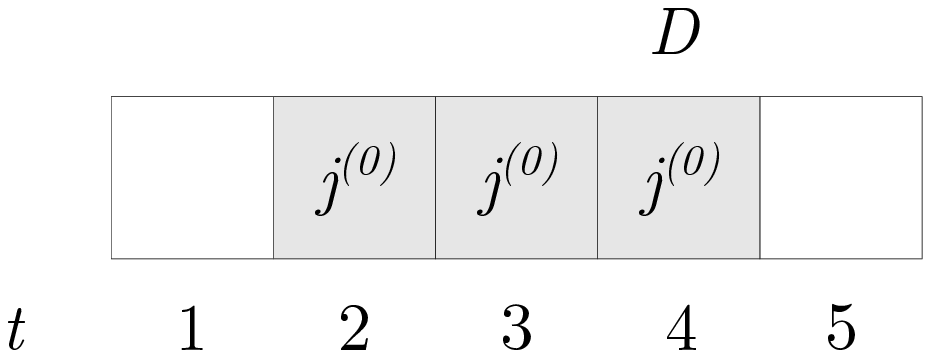}
         \caption{}
         \label{fig:three-schedules-a}
     \end{subfigure}
     \hfill
     \begin{subfigure}[b]{0.28\textwidth}
         \centering
         \includegraphics[width=\textwidth]
         {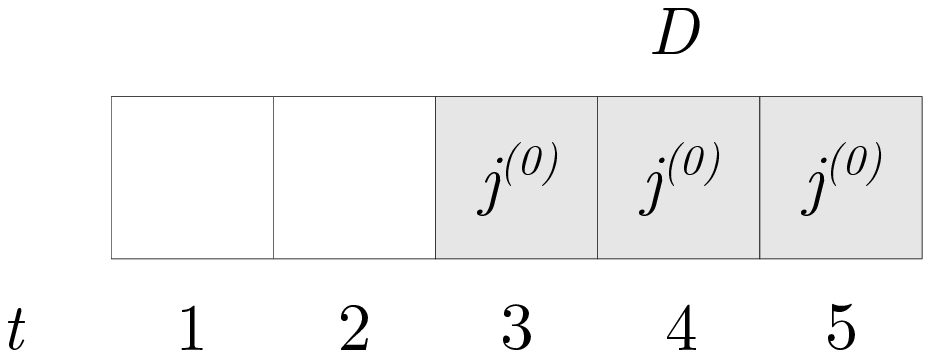}
         \caption{}
         \label{fig:three-schedules-b}
     \end{subfigure}
     \hfill
     \begin{subfigure}[b]{0.28\textwidth}
         \centering
         \includegraphics[width=\textwidth]
         {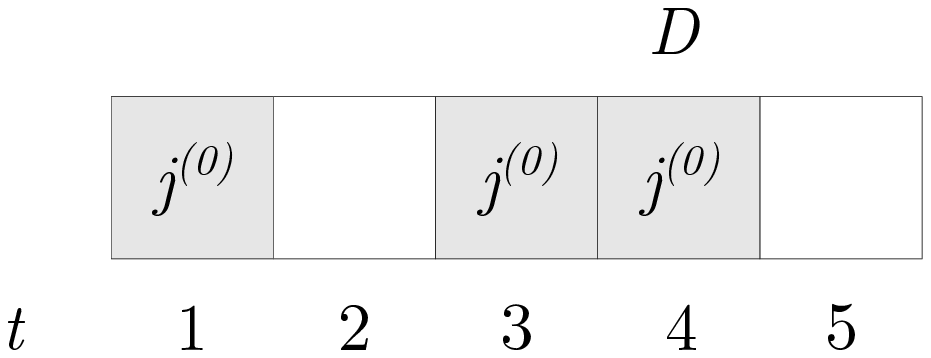}
         \caption{}
         \label{fig:three-schedules-c}
     \end{subfigure}
        \caption{An example of three possible schedules for a dummy instance of the JSTP including a single job $j$ with processing time $3$, a time horizon consisting of 5 time slots, a deadline $D=4$, and TOU costs all equal to one and omitted for the sake of clarity. The schedule shown in Figure~\ref{fig:three-schedules}(a) is non-preemptive, while the one shown in Figure~\ref{fig:three-schedules}(c) is preemptive. The schedule shown in Figure~\ref{fig:three-schedules}(a) is feasible as the processing of the job is completed within the deadline. The other two schedules are \addRR{infeasible}. In particular, the schedule shown in Figure~\ref{fig:three-schedules}(b) violate the deadline and the schedule shown in Figure~\ref{fig:three-schedules}(c) violates non-preemption.}
        \label{fig:three-schedules}
\end{figure}

Let us define a \emph{job-machine-time slots assignment} as a triplet $(j, h_j, \mathcal{T}_j)$ such that $j$ is a job in $\mathcal{J}$, $h_j$ is a machine in $\mathcal{H}$, and $\mathcal{T}_j$ is a non-empty subset of $\mathcal{T}$ encoding the time slots during which $h_j$ processes $j$.
Then, let us define a \emph{schedule} $\mathcal{S}$ as a set of job-machine-time slots assignments satisfying the following two conditions: (i) exactly one pair $(j, \mathcal{T}_j)$ exists for each $j\in\mathcal{J}$ and (ii) for any pair of distinct jobs $j', j'' \in \mathcal{J}$, the intersection $\mathcal{T}_{j'} \cap \mathcal{T}_{j''}$ is empty if $h_j = h_{j'}$, i.e.,
\begin{align}\label{eq:multi-machines-schedule}
  \mathcal{S} = \{(j, h_j, \mathcal{T}_j) : \emptyset \neq \mathcal{T}_j \subseteq \mathcal{T}, \forall \, j \in \mathcal{J}, \mathcal{T}_{j'} \cap \mathcal{T}_{j''}=\emptyset~\forall~j', j'' \in \mathcal{J}, j'\neq j'', h_j = h_{j'}\}.
\end{align}
Then, the \emph{makespan} $C_\text{max}$ of $\mathcal{S}$ is
\begin{align}\label{eq:single-machine-schedule-makespan}
  C_\text{max}(\mathcal{S}) = \displaystyle \max_{t \in \bigcup_{j \in \mathcal{J}} \mathcal{T}_j} t,
\end{align}
i.e., it is the overall amount of time necessary to process the jobs in $\mathcal{J}$ on the given machine. 

A schedule $\mathcal{S}$ is \emph{preemptive} if, for some $j \in \mathcal{J}$, $\mathcal{T}_j$ contains $p_j$ non consecutive slots, and \emph{non-preemptive} otherwise. For example, Figure~\ref{fig:three-schedules} shows three possible schedules for a job $j$ characterized by a processing time $p_j=3$ over a time horizon of 5 slots. The schedule shown in Figure~\ref{fig:three-schedules}(a) is non-preemptive as job $j$ is processed consecutively during 3 time slots. The schedule shown in Figure~\ref{fig:three-schedules}(c) is instead preemptive, as job $j$ is not processed consecutively during a number of time slots equal to $p_j$.

Each machine $h \in \mathcal{H}$ is associated with a positive energy consumption rate $u_h$.
The energy consumption rate reflects into the energy cost of the scheduled jobs.
Formally, if job $j \in \mathcal{J}$ is scheduled in $\mathcal{T}_j$ on machine $h_j \in \mathcal{H}$, the cost associated with the processing of the job $j$ is $u_{h_j} \sum_{k \in \mathcal{T}_j} c_k$.
Then, let us define the \emph{Total Energy Cost} (TEC) of a schedule $\mathcal{S}$ as 
\begin{align}\label{eq:TEC-single-machine}
  E(\mathcal{S}) = \sum_{j \in \mathcal{J}} u_{h_j} \sum_{k \in \mathcal{T}_j} c_k,
\end{align}
i.e., as the sum of the TOU costs associated to the job-machine-time slots assignments in $\mathcal{S}$. 

\bigskip

This thesis deals with the problem of finding a schedule of a set $\mathcal{J}$ of independent jobs, with no release times and no due dates, on a set $\mathcal{H}$ of parallel, identical machines, over the time horizon given as the set of time slots $\mathcal{T}$, so as to simultaneously minimize \eqref{eq:single-machine-schedule-makespan} and \eqref{eq:TEC-single-machine}.
However, with intent of laying a basic foundation to the discussion, let us begin from one of the simplest and most representative TOU scheduling problems, and then progressively expand over it to obtain a comprehensive view of the subject.

\begin{figure}[t]
    \captionsetup{font=footnotesize}
    \centering
    \includegraphics[scale=0.5]{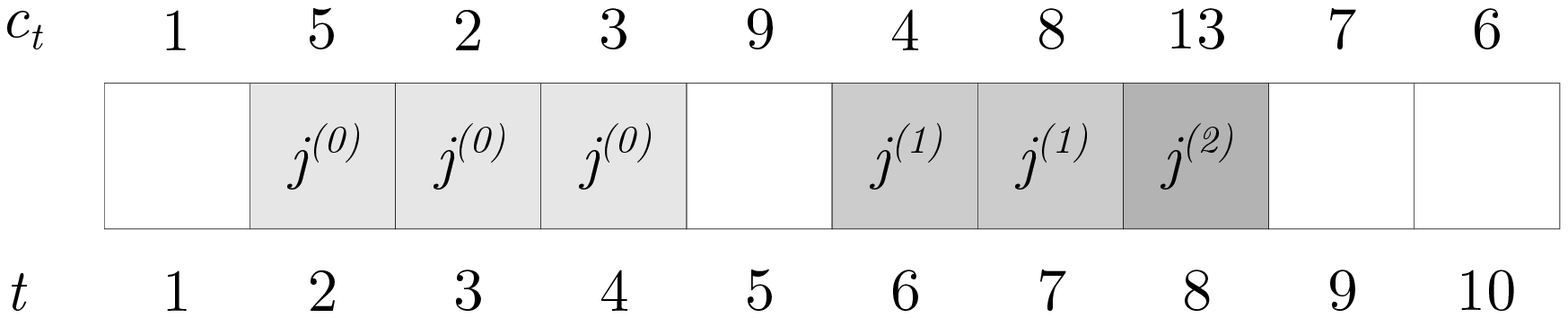}
    \caption{An example of a possible schedule for an instance of the JSTP having three jobs $\{j^{(0)},j^{(1)},j^{(2)}\}$, processing times $\{3, 2, 1\}$, a time horizon of 10 time slots, and TOU costs $c_t=\{1,5,2,3,9,4,8,13,7,6\}$. 
    \addRR{The energy consumption rate of the machine is $1$, and t}he deadline $D$ is assumed to be equal to 10.}
    \label{fig:schedule-example}
\end{figure}

Given a set of jobs $\mathcal{J}$, a set of integer processing times associated with the jobs in $\mathcal{J}$, a time horizon $\mathcal{T}$, a set $\{c_k\}$ of TOU costs associated with the time slots in $\mathcal{T}$, and a positive integer $D$, called \emph{deadline}, the Job Scheduling with Time-of-Use Cost Problem (JSTP) is to find a non-preemptive schedule $\mathcal{S}$ on a single machine that minimizes the total energy cost~\eqref{eq:TEC-single-machine}, and whose makespan~\eqref{eq:single-machine-schedule-makespan} is smaller than or equal to the given deadline, i.e., 
\begin{align}\label{eq:TECProb}
    \min_{\mathcal{S}} &\quad E(\mathcal{S}) \\
    s.t. &\quad C_{max}(\mathcal{S}) \leq D. \label{eq:C1}
\end{align}
\noindent As an example, by referring again to the three alternative schedules shown in Figure \ref{fig:three-schedules}, the schedule shown in Figure \ref{fig:three-schedules}(a) is a feasible solution for the toy instance of the JSTP including the single job $j$ with processing time 3, a time horizon of 5 time slots, a deadline $D=4$, and TOU costs all equal to 1 (and omitted for the sake of clarity), as its makespan satisfies~\eqref{eq:C1}. 
Instead, the schedule in Figure \ref{fig:three-schedules}(b) is infeasible, as part of job $j$ is processed after the deadline.
Similarly, the schedule in Figure \ref{fig:three-schedules}(c) satisfies the deadline $D$, but it is not feasible as well because it is preemptive. 
Figure \ref{fig:schedule-example} shows instead an example of a possible schedule for an instance of the JSTP.
This schedule has makespan and TEC equal to 8 and 35, respectively. Observe that if job $j$ was instead assigned to slot $1$, the resulting schedule would achieve a makespan equal to $7$ and a TEC equal to $23$, which is provably minimum. 

The literature on TOU scheduling describes several \emph{versions} of the JSTP. As a significant example, \citet{TimeSch3_chen_scheduling_2019} considered a JSTP version 
called \emph{Scheduling with Time-of-Use Costs Problem} (STOUCP). The difference between the JSTP and the STOUCP lies in the definition of the time horizon, and in the consequent relaxation of the integrality of the start times of the jobs. 
Specifically, in the STOUCP, for a given positive integer $K$, the time horizon is given as a continuous interval $[0, K] \subseteq \mathbb{R}$. As a consequence, a job $j \in \mathcal{J}$ can start at any $s_j \in [0, K]$.
Moreover, the set of time slots $\mathcal{T}$ is partitioned into the set $\pi$ of \emph{time periods}, or \emph{time intervals}. 
Each time interval in $\pi$ is a set of time slots that have the same TOU cost.
The time horizon is often presented as a set of time periods in TOU scheduling problems \citep{TimeSch2_fang_scheduling_2016,TimeSch4_chen_optimal_2021,TimeSch33_wang_bi-objective_2018}.

In some JSTP versions, the machines in $\mathcal{H}$ can be regulated by an ``on/off'' switching mechanism. In such case, each machine can be in one of these three states: (a) ``processing'', (b) ``idle'', or (c) ``shutdown''. A machine can process a job, i.e., be in the state (a), only while ``on''. If a machine is still ``on'' but it is not processing any job, then it is in state (b). In such state, a machine requires energy despite not processing any job, but it is ready to start processing a job at any time.
In state (c), the machine is shut down, i.e., it is ``off''. Its energy consumption is zero, but it cannot process any job without first transitioning into state (a). However, in order to do so, a machine requires time and energy.
As an example, \citet{TimeSch32_shrouf_optimizing_2014} tackle a JSTP version that is identical to the JSTP, except for the ``on/off'' switching mechanism that adds to the machine functionality. The objective function takes into account both the TEC of the schedule and the energy consumed by the machines due to the state transitions.

For each $j \in \mathcal{J}$, a power demand $q_j > 0$ may be associated with job $j$, such as in the one of the problems studied by \citet{TimeSch2_fang_scheduling_2016}. From a manufacturing standpoint, the introduction of power demands for jobs reflects the need to model energy-intensive and non-energy-intensive jobs.
Formally, if $j \in \mathcal{J}$ is processed in $\mathcal{T}_j \in \mathcal{T}$ on machine $h$, the cost of processing $j$ is $q_j \sum_{t \in \mathcal{T}_j} c_t$. 
More generally, if the power demand of $j \in \mathcal{J}$ also depends on the machine $h \in \mathcal{H}$ that processes $j$, then the power demand of $j$ is expressed as $q_{j, h} > 0$.

In some JSTP versions, the jobs in $\mathcal{J}$ are constrained to follow a specific processing sequence on machines. 
Such processing sequence induces a total ordering on the start times $s_j$, $j \in \mathcal{J}$.
It is interesting to observe that some JSTP versions with such a total ordering can be solved to optimality in polynomial \citep{TimeSch102_aghelinejad_complexity_2019,TimeSch103_aghelinejad_single-machine_2019} or pseudo-polynomial \citep{TimeSch101_aghelinejad_energy_2018} time.



\begin{table}[!t]
\captionsetup{font=footnotesize}
\centering
\scalebox{0.825}{
\begin{tabular}{lccc} 
\toprule
\textbf{Property}                                           & \textbf{Notation ($\beta$ field)} \\

Jobs power demands                                           & $q_j$ \\

\midrule

Fixed jobs processing sequence (on a single machine)         & $seq$ \\

\midrule

Continuous start time of the jobs                                  & $s_j \in \mathbb{R}_{0^+}$ \\

\midrule

Machines regulation via ``on/off'' switching mechanism       & $\text{on/off}$ \\

\bottomrule
\end{tabular}
}
\caption{Three-field notation useful for TOU scheduling.}\label{table:TOU-abg-notation}
\end{table}

In order to describe the TOU scheduling problems, the remainder of the paper makes extensive use of the 
$\alpha | \beta | \gamma$ notation introduced by \citet{Graham1977OptimizationAA}, also referred to as the \emph{three-field} notation.
The $\alpha | \beta | \gamma$ notation is a classical formal tool of the literature of job scheduling that allows \addRR{one }to concisely state different scheduling problems.
Specifically, the $\alpha$ field characterizes the machines environment. For instance, the case of parallel identical machines is represented with $Pm$ in the $\alpha$ field.
Instead, the $\beta$ field specifies the constraints and the processing restrictions, such as jobs preemption, represented with $prmp$.
Finally, the $\gamma$ field contains the optimization objectives, such as the minimization of the makespan.
For more detailed and complete information on the naming conventions used in the three-field notation for classical scheduling, see Appendix \ref{sec:notation} or the seminal book of \citet{Pinedo2016}.

Table \ref{table:TOU-abg-notation} summarizes the three-field notation useful to TOU scheduling.
As a conclusive remark for this section, the presence of TOU costs is usually denoted by $slotcost$ in the $\beta$ field \citep{TimeSch1_wan_scheduling_2010} in the literature. However, since this thesis only consider TOU scheduling problems in this article, it will be disregarded from now on.

\section{A literature review}
\label{sec:literature-classification}

In this section, we propose a systematic classification of the literature of TOU scheduling by dividing the considered problems into two different classes. 

The first class consists of the problems that are easily representable with the three-field notation.
Among the problems in the first class, we further distinguish between (a) problems that are solvable to optimality with a polynomial or a pseudo-polynomial time algorithm (Subsection \ref{sec:poly-pseudo-poly}), and 
(b) problems that have been proven to be NP-hard, or that do not have a polynomial or pseudo-polynomial exact solution algorithm yet (Subsection \ref{sec:np-hard}).

The second class is instead constituted by 
(a) problems with particular characteristics that would require non-standard notation, such as technological features or unusual objective functions (Subsection \ref{sec:special-features}), and 
(b) applicative problems inspired by a practical case study (Subsection \ref{sec:practical-case-studies}).

The problem studied in this thesis is $\mathcal{NP}$-hard and belongs to the first class, as it can be stated as $Pm||C^\text{max}$, $TEC$.
It was first studied in the seminal work of \citet{TimeSch33_wang_bi-objective_2018}, and then thoroughly investigated in our recent article \citep{TimeSch11_Anghinolfi2021ABH}.

The purpose of this section is to provide a thorough reference that should serve as a general overview for scheduling researchers and industry decision makers.

\subsection{Problems with an exact algorithm running in polynomial or pseudo-polynomial time}
\label{sec:poly-pseudo-poly}
We summarize the problems that have been proven to admit a polynomial or a pseudo-polynomial algorithm in Table \ref{table:polynomially}.
The seminal work of \citet{TimeSch1_wan_scheduling_2010} laid the foundations for single-machine scheduling under TOU costs. Specifically, \citet{TimeSch1_wan_scheduling_2010} considered five different JSTP versions, corresponding to five different objective functions obtained by summing the TEC with five regular functions often used in non-TOU scheduling: the total flow time, the maximum lateness, the maximum tardiness, the weighted number of tardy jobs, and the total tardiness. The authors proved the strong NP-hardness of all the five problems with a reduction from the 3-partition problem.
In addition, they proposed polynomial algorithms for the problems under specific assumptions.
\citet{TimeSch3_chen_scheduling_2019} focused on $1 | s_j \in \mathbb{R}_{0^+} | TEC$, proved its strong NP-hardness again with a reduction from the 3-partition problem.
Similarly, the authors provided different polynomial and pseudo-polynomial algorithms based on several assumption on the time slot costs. Such work is essential in highlighting the relations between the structure of $\{c_k\}$ and the existence of efficient solution algorithms.
\citet{TimeSch2_fang_scheduling_2016} and \citet{TimeSch4_chen_optimal_2021} provided several results for preemptive TOU scheduling. The former authors proved the existence of polynomial algorithms for TOU scheduling with jobs weights and power demands under specific assumptions.
Instead, the latter authors specifically provided optimal and approximation polynomial algorithms for different JSTP versions, considering fixed jobs sequence, jobs weights and release dates.
To the best of our knowledge, \citet{TimeSch5_penn_complexity_2021} are the only authors to provide a polynomial-time algorithm for a preemptive TOU scheduling problem that considers due dates.
Instead, \citeauthor{TimeSch99_aghelinejad_sforza_preemptive_2017} extensively studied the class of single-machine TOU scheduling problems with on/off switching with the objective of minimizing the TEC.
\citeauthor{TimeSch99_aghelinejad_sforza_preemptive_2017} first relaxed the problem by allowing preemptive schedules \citep{TimeSch99_aghelinejad_sforza_preemptive_2017}. In a subsequent work, \citeauthor{TimeSch99_aghelinejad_sforza_preemptive_2017} introduced the assumption of the fixed jobs sequence \citep{TimeSch102_aghelinejad_complexity_2019,TimeSch103_aghelinejad_single-machine_2019}. In this setting, they later considered the jobs' power consumption and different machine speeds  \citep{TimeSch101_aghelinejad_energy_2018}.
Finally, \citet{TimeSch7_wang_scheduling_2018} considered a two-machines permutation flow shop scheduling problem, and they provided a polynomial-time algorithm when the jobs sequence is fixed.

\begin{table}[!ht]
\captionsetup{font=footnotesize}
\centering
\scalebox{0.6}{
\begin{tabular}{lccc} 
\toprule
\textbf{Article} & \textbf{Problem} 
& \begin{tabular}{c}Computational complexity \\of the solution algorithm\end{tabular}
& \textbf{Assumptions}  \\
\midrule
\multirow{5}{*}{\citet{TimeSch1_wan_scheduling_2010}}  & $1 | | \sum_{j \in J} C_j + TEC$              & $O(K^2)$                                        & $\{c_k\}$ non-increasing \\
                                                       & $1 | | L_{max} + TEC$                         & $O(K^2)$                                        & $\{c_k\}$ non-increasing \\
                                                & $1 | | T_{max} + TEC$                         & $O(K^2)$                                        & $\{c_k\}$ non-increasing \\
                                                & $1 | | \sum_{j \in J} w_j U_j + TEC$          & (a) $O(K^2)$ or (b) $O(NK^3)$                   & tardy jobs as (a) lost sales or (b) backlogging \\
                                                & $1 | | \sum_{j \in J} T_j + TEC$              & $O(N^4 K^3)$                                    & $\{c_k\}$ monotone non-increasing \\
\midrule
\multirow{3}{*}{\citet{TimeSch2_fang_scheduling_2016}}           & $1 | q_j, prmp | TEC$                         & No explicit expression.                         &      \\
                                                & $1 | q_j | TEC$                               & No explicit expression.                         & $\{c_k\}$ pyramidal \\
                                                & $1 | w_j, q_j, prmp | TEC$                    & No explicit expression.                         & $c_k > 0, k \in \mathcal{T}$ \\
\midrule
\multirow{5}{*}{\citet{TimeSch3_chen_scheduling_2019}}           & $1 | | TEC$                                   & $O(|\pi|)$                                      & $\pi$ has 1 valley \\
                                                &                                               & $O((\sum_{j\in\mathcal{J}}p_j)^v(N +|\pi|^v))$  & $\pi$ has $v\ge2$ valley \\
                                                &                                               & $O(|\pi|^2 N^{2r - 1})$                         & $\pi$ has 2 valleys, $|\mathcal{P}_\mathcal{J}| = r$ \\
                                                &                                               & $O(|\pi|(|\pi| + N)N^3)$                        & Bounded lateness, $\pi$ has 1 valley \\
                                                &                                               & $O(|\pi|^{\pi + 2} N^{3|\pi| - 1})$             & $\sum_{j \in J} p_j$ bounded; $\pi$ has 1 valley \\
\midrule
\multirow{2}{*}{\citet{TimeSch4_chen_optimal_2021}}              & $1 | prmp | \sum_{j \in J} C_j + TEC$                 & No explicit expression.                         & \\
                                                & $1 | prmp, seq, w_j | \sum_{j \in J} w_j C_j + TEC$   & $O(NK)$                                         & \\
\midrule
\citet{TimeSch5_penn_complexity_2021}           & $1 | r_j, d_j, prmp | TEC$                            & $O(log_2 K (K + N^2 log_2 K))$                  & \\
\midrule
\citet{TimeSch99_aghelinejad_sforza_preemptive_2017}                                                    & $1 | \text{on/off}, prmp | TEC$                       & $O(K^3)$                        & \\
\midrule
\citet{TimeSch102_aghelinejad_complexity_2019,TimeSch103_aghelinejad_single-machine_2019}               & $1 | \text{on/off}, seq | TEC$                        & $O(K^3)$                        & \\ 
\midrule
\citet{TimeSch101_aghelinejad_energy_2018}                                                              & $1 | \text{on/off}, seq, q_j | TEC$                   & $O(K^3)$                        & \\ 
                                                                                                        & $1 | \text{on/off}, seq, q_j, \text{speed} | TEC$     & $O(T^2 V (T + V))$              & \\
\midrule
\citet{TimeSch7_wang_scheduling_2018}           & $F2 | prmu, seq | TEC$                                & $O(NK^4)$                                       & \\
\bottomrule
\end{tabular}
}
\caption{Optimal polynomial and pseudo-polynomial time algorithms for different JSTP versions.}\label{table:polynomially}
\end{table}

\subsection{Problems with approximation or heuristic algorithms}
\label{sec:np-hard}

In this section, we deal with TOU scheduling problems that have not been proven to be solvable in polynomial or pseudo-polynomial time.
Table \ref{table:np-hard-problems} reports a classification of such problems according to the type of the processing machines and the number of the objective functions.
Let us first deal with single-objective single-machine TOU scheduling problems.
\citet{TimeSch3_chen_scheduling_2019} provided a FPTAS for the STOUCP by exploiting the notion of \emph{valley} for a set of time periods $\pi$, which is a time interval with a TOU cost lower than its neighbors.
Specifically, the authors found that the STOUCP admits a FPTAS when the set of time periods $\pi$ has at most $2$ valleys.
\citet{TimeSch4_chen_optimal_2021} and \citet{TimeSch40_Kulkarni2012AlgorithmsFC} are the only authors, at the best of our knowledge, to provide a PTAS for a preemptive scheduling problem.
\citet{TimeSch2_fang_scheduling_2016} provided a $\sum_{k = 1}^K (c_k \, / \, \min_{t \in \mathcal{T}}{c_t})^{\alpha \, / \, / (\alpha - 1)} $-approximation algorithm for $1 | w_j, q_j | TEC$ (the authors also studied the preemptive version of such problem, see Section \ref{sec:poly-pseudo-poly}).
Instead, \citet{TimeSch79_che_efficient_2016} considered the problem $1 | s_j \in \mathbb{R} | TEC$, proven to be strongly NP-hard by \citet{TimeSch2_fang_scheduling_2016}.
\citeauthor{TimeSch79_che_efficient_2016} proposed a simple MILP formulation with two different types of assignment variables to handle the condition $s_j \in \mathbb{R}$, and they presented a greedy insertion heuristic by building upon the work of \citet{TimeSch2_fang_scheduling_2016}.
\citeauthor{TimeSch32_shrouf_optimizing_2014} were the first authors to study $1 | \text{on/off} | TEC$ \citep{TimeSch32_shrouf_optimizing_2014}, as well as to provide a MILP formulation and to develop a GA for the problem.
\citeauthor{TimeSch98_aghelinejad_machine_2016} also studied $1 | \text{on/off} | TEC$: by building upon \citeauthor{TimeSch32_shrouf_optimizing_2014}'s effort, they provided an improved MILP formulation \citep{TimeSch98_aghelinejad_machine_2016} that they later strenghtened with further lower bounds \citep{TimeSch102_aghelinejad_complexity_2019}. In \citep{TimeSch100_aghelinejad_production_2018}, \citeauthor{TimeSch98_aghelinejad_machine_2016} also developed a heuristic and a GA as well.
Finally, \citet{TimeSch10_cheng_single-machine_2016} presented and compared two different MILP formulations for $1 | batch(b) | TEC$.

In the literature, the only problems that have been proven to admit a PTAS only involve a single machine and a single objective.
One of the first efforts in single-machine TOU scheduling with multiple objectives is due to \citet{TimeSch86_cheng_bi-objective_2014}, who developed a bi-objective MILP formulation for the NP-hard problem $1 | batch(b) | C_{max}, TEC$. 
The authors developed an algorithm based on the $\epsilon$-constraint method for multi-objective optimization that is based on solving a sequence of single-objective formulations of the original problems. In \citep{TimeSch78_cheng_bi-objective_2016}, \citeauthor{TimeSch78_cheng_bi-objective_2016} simplified their original MILP formulation \citep{TimeSch86_cheng_bi-objective_2014}, and they built a faster optimal $\epsilon$-constraint algorithm upon their results.
The general idea at the core of such algorithm was further used, developed and refined in subsequent works involving bi-objective scheduling \citep{TimeSch27_wu_large-scale_2021,TimeSch33_wang_bi-objective_2018,TimeSch11_Anghinolfi2021ABH}. Specifically, for the same batch scheduling problem tackled by \citeauthor{TimeSch86_cheng_bi-objective_2014}, \citet{TimeSch27_wu_large-scale_2021} proposed two new fast $\epsilon$-constraint-based constructive heuristics that improve \citeauthor{TimeSch86_cheng_bi-objective_2014}'s algorithm from a computational standpoint. 
\citet{TimeSch54_zhou_energy-efficient_2020} also considered a batch scheduling problem with release dates, that requires the simultaneous minimization of the makespan and TEC. 
\citet{TimeSch63_rubaiee_energy-aware_2019} considered the same objectives in a single-machine problem with preemption.

\citet{TimeSch33_wang_bi-objective_2018} considered one of the foundational problems in multi-objective parallel identical machines TOU scheduling, $Pm | | C_{max}, TEC$, providing a constructive heuristic and a MILP formulation. \citet{TimeSch11_Anghinolfi2021ABH} built upon this work and provided a heuristic constituted by an improved constructive heuristic and a local search algorithm; furthermore, they presented an improved MILP formulation.
The authors that considered TOU scheduling problems with parallel batch processing machines provided a MILP formulation and one or a few metaheuristics \citep{TimeSch52_qian_multi-objective_2020,TimeSch53_rocholl_bi-criteria_2020,TimeSch65_zhou_multi-objective_2018,TimeSch57_tan_genetic_2019}. At the best of our knowledge, \citet{TimeSch57_tan_genetic_2019} are the only authors to deal with non-identical parallel machines.

In the literature, there are also several research efforts in unrelated parallel machines TOU scheduling.
The foundational problem $Rm | | TEC$ has first been studied by \citet{TimeSch31_ding_parallel_2016}, who showed the NP-hardness of the problem and proposed a MILP formulation, as well as a Dantzig–Wolfe decomposition based reformulation of the problem along with a column generation heuristic. 
\citet{TimeSch9_cheng_improved_2018} improved the MILP formulation proposed by \citet{TimeSch31_ding_parallel_2016} by using less decision variables, 
and \citet{TimeSch38_saberi-aliabad_energy-efficient_2020} further improved the formulation by introducing several strengthening inequalities.
\citet{TimeSch30_che_energy-conscious_2017} instead considered a version of the problem characterized by continuous start times.
The problem $Rm | | K C_{max} + TEC$, for a positive real $K$, received particular attention during the last decade, with the presentation of multiple MILP formulations and metaheuristics \citep{TimeSch22_moon_optimization_2013,TimeSch93_kurniawan_genetic_2017,TimeSch58_cheng_mixed-integer_2019,TimeSch37_kurniawan_triplechromosome_2020}.
\citet{TimeSch16_pei_approximation_2021} instead tackle $Rm | | C_{max}, TEC$, the only unrelated parallel machines problem with multiple objectives. The authors first linearly combined the two objectives and provided a MINLP formulation for the resulting single-objective problem; subsequently, they exploited a relaxation of the problem to develop an approximate algorithm.

As regards flow shop scheduling, \citet{TimeSch7_wang_scheduling_2018} studied a two-machines permutation flow shop scheduling problem with the usual objective of minimizing the TEC. The authors proposed a MILP formulation, as well as a Johnson's rule and a dynamic programming based heuristic. \citet{TimeSch6_ho_electricity_2021} further extended \citeauthor{TimeSch7_wang_scheduling_2018}'s work by proposing a novel formulation and heuristic for the problem. 
\citeauthor{TimeSch105_aghelinejad_energy-cost-aware_2020} instead performed a mathematic modeling effort for the two-machines flow shop scheduling problem with on/off switching mechanisms \citep{TimeSch104_aghelinejad_multi-state_2020,TimeSch105_aghelinejad_energy-cost-aware_2020}.

\citet{TimeSch87_luo_hybrid_2013} are among the first authors to consider a single-objective flow shop TOU scheduling problems. In more detail, \citeauthor{TimeSch87_luo_hybrid_2013} studied a flexible (or hybrid) flow shop scheduling problem requiring the minimization of the TEC, and proposed a ACO-based multi-objective metaheuristics.
\citet{TimeSch76_zhang_new_2017} extended \citeauthor{TimeSch87_luo_hybrid_2013}'s work by considering different machine speeds, and they proposed a MILP formulation for the problem as well as a modified "biogeography-based" algorithm combined with VNS.
\citet{TimeSch39_zhang_optimization_2019} also extended \citeauthor{TimeSch87_luo_hybrid_2013}'s work, but opposite to \citet{TimeSch76_zhang_new_2017}, they considered an on/off switching mechanism for the considered machines. 
\citet{TimeSch88_peng_integer_2021} and \citet{TimeSch15_cheng_permutation_2016} proposed several metaheuristics for two distinct flow shop scheduling problems that require the minimization of the TEC. The first is a general multiple machines flow shop, while the second is a no-wait permutation flow shop problem.

Finally, we deal with the few multi-objective job shop scheduling efforts in the literature.
Among these, \citet{TimeSch36_kurniawan_distributed-elite_2020} consider a bi-objective job shop scheduling problem requiring to minimize the TEC and the total weighted tardiness. The authors decompose the problem into the subproblems of a) sequencing the different operations on the machines and b) determining their start time. They propose a ``distributed-elite'' local search based upon a genetic algorithm that encodes the operations sequence and their start time into two different gene representations.
\citet{TimeSch73_wang_batch_2018} instead study a batch scheduling setting with the minimization of the makespan and TEC. In their concise work, they present a mathematical formulation along with some experimental results.
Lastly, \citet{TimeSch18_jiang_multi-objective_2020} consider a flexible job shop with the minimization of the makespan and TEC, and present both a MILP model and a hybrid MOEA/D. Such metaheuristic employs specific operators to generate new solutions by exploiting information in their neighborhoods, and it also embeds two different intensification operators.

\begin{landscape}
\begin{table}[!ht]
\captionsetup{font=footnotesize}
\centering
\scalebox{0.5}{
\begin{tabular}{cccc}
\toprule
\textbf{Class} & \textbf{Problem} & \textbf{Article} & \textbf{\makecell{Solution\\Approach}} \\

\midrule

\multirow{12}{15em}{\textbf{\makecell{Single-objective\\single-machine\\scheduling}}} 
& $1 | prmp, w_j | \sum_{j \in J} w_j C_j + TEC$                  & \citet{TimeSch4_chen_optimal_2021}                & \makecell{PTAS}                                              \\
& $1 | prmp, w_j, r_j | \sum_{j \in J} w_j (C_j - r_j) + TEC$     & \citet{TimeSch40_Kulkarni2012AlgorithmsFC}        & \makecell{PTAS (at most two distinct values in $\{c_k\}$)}  \\
& $1 | | TEC$                                             & \citet{TimeSch3_chen_scheduling_2019}                     & \makecell{FPTAS ($\pi$ has at most $2$ valleys)} \\ 
& $1 | w_j | TEC$                                         & \citet{TimeSch29_zhang_new_2018}                          & \makecell{Greedy insertion heuristic algorithm} \\
& $1 | s_j \in \mathbb{R}_{0^+} | TEC$                          & \citet{TimeSch79_che_efficient_2016}                      & \makecell{MILP formulation, Greedy insertion heuristic} \\
& $1 | w_j, q_j | TEC$                                    & \citet{TimeSch2_fang_scheduling_2016}                     & \makecell{$\sum_{k = 1}^K (c_k \, / \, \min_{t \in \mathcal{T}}{c_t})^{\alpha \, / \, / (\alpha - 1)}$-approximation algorithm} \\ 
& $1 | \text{on/off} | TEC$                               & \citet{TimeSch32_shrouf_optimizing_2014}                  & \makecell{MILP, GA} \\                     
& $1 | \text{on/off} | TEC$                               & \citet{TimeSch98_aghelinejad_machine_2016}                & \makecell{MILP} \\
& $1 | \text{on/off} | TEC$                               & \citet{TimeSch100_aghelinejad_production_2018}            & \makecell{MILP, Heuristic, GA} \\
& $1 | \text{on/off} | TEC$                               & \citet{TimeSch102_aghelinejad_complexity_2019}            & \makecell{New lower bounds for MILP formulation} \\
& $1 | batch(b) | TEC$                                    & \citet{TimeSch10_cheng_single-machine_2016}               & \makecell{MILP} \\

\midrule

\multirow{6}{15em}{\textbf{\makecell{Multi-objective\\single-machine\\scheduling}}}
& $1 | prmp | C_{max}, TEC$                               & \citet{TimeSch63_rubaiee_energy-aware_2019}               & \makecell{MILP, ACO-based algorithms} \\
& $1 | | \sum_{j \in \mathcal{J}} w_j T_j$, TEC           & \citet{TimeSch14_kurniawan_mix_2018}                      & \makecell{MILP, Hybrid GA with random insertion} \\
& $1 | batch(b) | C_{max}, TEC$                           & \citet{TimeSch86_cheng_bi-objective_2014}                 & \makecell{MILP} \\
&                                                         & \citet{TimeSch78_cheng_bi-objective_2016}                 & \makecell{MILP, exact $\epsilon$-constraint algorithm} \\
&                                                         & \citet{TimeSch26_cheng_bi-criteria_2017}                  & \makecell{Heuristic-based $\epsilon$-constraint heuristics}\\
&                                                         & \citet{TimeSch64_zhang_improved_2018}                     & \makecell{MILP, Heuristics} \\
&                                                         & \citet{TimeSch27_wu_large-scale_2021}                     & \makecell{MILP, Heuristics} \\ 
& $1 | batch(b), r_j | C_{max}, TEC$                      & \citet{TimeSch54_zhou_energy-efficient_2020}              & \makecell{MILP, Hybrid multi-objective meta-heuristic algorithm} \\

\midrule

\multirow{6}{15em}{\textbf{\makecell{Multi-objective\\parallel\\identical machines}}}
& $Pm | | C_{max}, TEC$                                   & \citet{TimeSch33_wang_bi-objective_2018}                  & \makecell{MILP, Constructive heuristic, GA}  \\ 
&                                                         & \citet{TimeSch11_Anghinolfi2021ABH}                       & \makecell{MILP, "Split-greedy" heuristic, "Exchange-search" local search} \\
& $Pm | batch(b) | C_{max}, TEC$                          & \citet{TimeSch52_qian_multi-objective_2020}               & \makecell{MILP, Multi-objective EA based on adaptive clustering} \\
& $Pm | batch(b) | \sum_{j \in \mathcal{J}} w_j T_j, TEC$ & \citet{TimeSch53_rocholl_bi-criteria_2020}                & \makecell{MILP, NSGA-II with embedded heuristics} \\
& $Pm | batch(b), r_j | C_{max}, TEC$                     & \citet{TimeSch65_zhou_multi-objective_2018}               & \makecell{MILP, Multi-objective differential EA} \\

\midrule

\textbf{\makecell{Non-identical\\parallel machines}} 
& $Qm | batch(b) | TEC$                                   & \citet{TimeSch57_tan_genetic_2019}                        & \makecell{MILP, SPGA, MPGA} \\

\midrule

\multirow{9}{15em}{\textbf{\makecell{Single-objective unrelated\\parallel machines}}}
& $Rm | | TEC$                                            & \citet{TimeSch31_ding_parallel_2016}                      & \makecell{MILP} \\
&                                                         & \citet{TimeSch9_cheng_improved_2018}                      & \makecell{MILP} \\
&                                                         & \citet{TimeSch38_saberi-aliabad_energy-efficient_2020}    & \makecell{MILP} \\
& $Rm | s_j \in \mathbb{R} | TEC$                         & \citet{TimeSch30_che_energy-conscious_2017}               & \makecell{MILP, Two-stage heuristic} \\
& $Rm | prmp | C_{max} + TEC$                             & \citet{TimeSch4_chen_optimal_2021}                        & \makecell{$O(K^2)$ + time for solving $Rm | prmp | C_{max}$ by IP \citep{Sch1_Lawler1978OnPS}} \\
& $Rm | | K C_{max} + TEC$, $K \in \mathbb{R}_{>0}$       & \citet{TimeSch22_moon_optimization_2013}                  & \makecell{MILP, Hybrid GA} \\
&                                                         & \citet{TimeSch93_kurniawan_genetic_2017}                  & \makecell{MILP, GA} \\
&                                                         & \citet{TimeSch58_cheng_mixed-integer_2019}                & \makecell{MILP} \\
&                                                         & \citet{TimeSch37_kurniawan_triplechromosome_2020}         & \makecell{``Triple-chromosome'' GA}\\

\midrule

\textbf{\makecell{Multi-objective\\unrelated\\parallel machines}}  
& $Rm | | C_{max}, TEC$                                   & \citet{TimeSch16_pei_approximation_2021}                  & \makecell{MILP, Approximation algorithm} \\

\midrule

\multirow{10}{15em}{\textbf{\makecell{Single-objective\\flow shops}}}
& $F2 | prmu | TEC$                                       & \citet{TimeSch7_wang_scheduling_2018}                     & \makecell{MILP, ILS} \\
&                                                         & \citet{TimeSch6_ho_electricity_2021}                      & \makecell{Heuristics} \\
& $F2 | \text{on/off} | TEC$                              & \citet{TimeSch104_aghelinejad_multi-state_2020}           & \makecell{Two MIP formulations} \\
&                                                         & \citet{TimeSch105_aghelinejad_energy-cost-aware_2020}     & \makecell{LP formulation} \\
& $Fm | | TEC$                                            & \citet{TimeSch88_peng_integer_2021}                       & \makecell{ILP, PSO} \\
& $Fm | nwt, prmu | TEC$                                  & \citet{TimeSch15_cheng_permutation_2016}                  & \makecell{MILP, GA} \\
& $FF2 (Qm, Rm') | | TEC, m > m'$                         & \citet{TimeSch8_zhang_two-stage_2021}                     & \makecell{MILP, TS - Greedy insertion algorithm} \\
& $FFc | | TEC$                                           & \citet{TimeSch87_luo_hybrid_2013}                         & \makecell{ACO-based metaheuristic} \\
& $FFc | \text{speed} | TEC$                              & \citet{TimeSch76_zhang_new_2017}                          & \makecell{MILP formulation, Modified "biogeography-based" algorithm combined with VNS} \\
& $FFC | \text{on/off} | TEC$                             & \citet{TimeSch39_zhang_optimization_2019}                 & \makecell{SPEA2} \\

\midrule

\multirow{5}{15em}{\textbf{\makecell{Multi-objective\\flow shops}}}
& $Fm | | TEC, \sum_{j \in \mathcal{J}} E_j + T_j$        & \citet{TimeSch21_badri_flow_2021}                         & \makecell{Single-objective LP after multi-objective fuzzy programming} \\
& $Fm | prmu | \sum_{j \in \mathcal{J}} w_j T_j, TEC$     & \citet{TimeSch89_kurniawan_optimization_2020}             & \makecell{SPEA2-based metaheuristic} \\
& $Fm | prmu | C_{max}, TEC$                              & \citet{TimeSch55_wang_discrete_2020}                      & \makecell{MILP} \\
& $FFc (Qm) | d_j | \sum_{j \in \mathcal{J}} T_j, TEC$    & \citet{TimeSch28_ding_energy_2021}                        & \makecell{MILP, Hybrid PSO} \\
& $FFc (Pm, batch(b)) | \text{on/off} | C_{max}, TEC$     & \citet{TimeSch56_wang_energy-efficient_2020}              & \makecell{MILP, Constructive heuristic with local search, TS, ACO} \\

\midrule

\multirow{5}{15em}{\textbf{\makecell{Multi-objective\\job shops}}} \\
& $Jm | | \sum_{j \in \mathcal{J}} w_j T_j, TEC$          & \citet{TimeSch36_kurniawan_distributed-elite_2020}        & \makecell{Distributed-elite LS based on GA} \\
& $Jm | batch(b) | C_{max}, TEC$                          & \citet{TimeSch73_wang_batch_2018}                         & \makecell{MILP} \\
& $FJc | | C_{max}, TEC$                                  & \citet{TimeSch18_jiang_multi-objective_2020}              & \makecell{MILP, Hybrid MOEA/D} \\
\bottomrule
\end{tabular}
}
\caption{Integer programming algorithms, approximation algorithms, heuristics and metaheuristics for different JSTP problem versions.}\label{table:np-hard-problems}
\end{table}
\end{landscape}

\subsection{Problem with particular features or objectives}
\label{sec:special-features}

In this subsection, we consider JSTP versions that do not fit into our classification due to peculiar features of the considered manufacturing system or particular optimization objectives.

\addRR{Let us first specifically consider those problems that require the minimization of an unusual or particular objective function.}
As an example, \citet{TimeSch5_penn_complexity_2021} consider a single-machine problem that consists in non-preemptively assigning jobs to time slots in order to maximize a measure of profit associated to the resulting schedule.
Specifically, profit is obtained as the difference between the total revenue of the jobs, which is given as input data to the problem, and the energy consumption cost associated with the jobs themselves, depending on their assignment in the schedule.
While the general case for such problem is NP-hard, the authors present a pseudo-polynomial time algorithm for the problem when preemption is allowed.
\citet{TimeSch34_zeng_bi-objective_2018} study a uniform parallel machines environment where the total number of machines, a very unusual objective in the literature, has to be minimized along with the TEC.
\citet{TimeSch60_gong_energy-_2019} instead propose a flexible job scheduling problem with job recirculation and operation sequence-dependent setups. Alongside the makespan and the TEC, the authors take into account three other different objectives that may arise in a practical manufacturing setting: the total labor cost, the maximal workload, and the total workload.
\citet{TimeSch25_batista_abikarram_energy_2019} expand the model of \citet{TimeSch32_shrouf_optimizing_2014} by taking into account demand charges as additional energy consumption in a parallel machines environment. \citet{TimeSch66_tan_economic_2018} instead aggregate energy consumption with additional costs due to production load shifting (such as employee overtime costs and gas emissions penalties) in a complex batch scheduling setting.
Finally, \citet{TimeSch92_chen_order_2019} and \citet{TimeSch35_zhang_energy-conscious_2014} also address environmental concerns by integrating carbon and CO$_2$ emissions within the optimization objectives, respectively.
\citet{TimeSch23_lee_dynamic_2017} instead consider the minimization of the sum of the TEC and the just-in-time (JIT) cost of a schedule for a single-machine problem. The JIT cost is given as the sum of the mean squared earliness and the tardiness of the jobs, which may be a relevant performance measures for manufacturing productions that require a certain degree of timeliness. The authors experimentally test their dynamic control method on a real case involving a milling process performed by a HAAS machine.
\citet{TimeSch81_gong_generic_2016} consider manufacturing unit processes by studying the problem of minimizing the TEC of the schedule of a set of jobs with due dates on a single machine. More specifically, the authors use finite state machines to describe the energetic transitions of the machines, and they also implemented a genetic algorithm for the problem. Finally, they validated their approach on a real case study based on the measurements of a griding machines and the energy prices.

Researchers also devoted efforts to model particular situations within the manufacturing chain.
For instance, \citet{TimeSch43_geng_bi-objective_2020} propose a problem based on a type of flexible-flow shop called ``re-entrant'' \citep{GRAVES1983197}, and tackle it with an improved multi-objective ``ant lion'' optimization algorithm.
Differently from \citet{TimeSch43_geng_bi-objective_2020}, \citet{TimeSch12_wang_calibration_2020} consider a single-machine setting with a particular technological feature, which is the calibration phase, required to process jobs. Specifically, if the calibration starts at time $t$, then the machine can process any job in the time slots $t, t + 1, \ldots, t + C$ for some fixed $C \in \mathbb{N}$. The objective is to minimize the TEC, while calibrating the machine at most a fixed amount of times.
\citet{TimeSch44_kong_energy-efficient_2021} instead consider the dynamic disruptions that can affect manufacturing systems. The authors model such incovenients as ``arrival'' jobs, and they formulate a rescheduling problem that requires to schedule such jobs alongside a set of the ``original'' jobs intended for processing in order to minimize the TEC.
\citet{TimeSch90_tan_production_2020} again consider the minimization of the TEC, but they introduce a constraint on the maximum power at the peak usage.

Several research efforts successfully integrated the TOU framework within project scheduling \cite{Brucker1999} problems.
As an example, \citet{TimeSch13_najafzad_multi-skill_2019} consider a bi-objective multi-skill project scheduling problem that consists in scheduling activities in order to minimize the total completion time and project cost. The decision-maker has to perform a trade-off between low-energy processing in off-peak hours and the resulting increase in wages due to shift differentials.
Similarly to \citet{TimeSch13_najafzad_multi-skill_2019}, \citet{TimeSch50_javanmard_bi-objective_2021} consider a project scheduling problem with a multi-skilled workforce, with the aim of the reducing the TEC while respecting the projects priorities as much as possible.
While \citet{TimeSch45_maghsoudlou_framework_2021} present again a multi‑skilled project scheduling problem under TOU pricing with the objective of generating a preemptive schedule to minimize the TEC, \citet{TimeSch59_wu_energy-efficiency-oriented_2019} instead aim at the optimizing the consumption of the multiple resources involved.
\citet{TimeSch80_DU2021114754} propose a bi-objective model for a resource-constrained (project scheduling) problem with activity splitting and recombining, to optimize the project delay and TEC.

Some authors also explored the possibility of optimizing production while taking into account preventive \citep{TimeSch17_sin_bi-objective_2020,TimeSch62_assia_management_2019} and planned \citep{TimeSch19_cui_integrating_2020,TimeSch20_cui_energy-aware_2021} maintenance.
\citet{TimeSch24_zhang_energy-efficient_2015} instead study multiple factories in an electrical grid: facilities can exchange information within the grid, while aiming at the reduction of energy consumption.
Finally, \citet{TimeSch83_sharma_econological_2015} provide an ``econological'' model of a manufacturing enterprise constituted by multiple speed-scaling machines by integrating both economic and ecological objectives.

\subsection{Practical case studies}
\label{sec:practical-case-studies}

The literature also offers several efforts towards the efficient exploitation of TOU pricing schemes as a means of cutting expenses in industry production.
\citet{TimeSch82_hadera_optimization_2015} and \citet{TimeSch84_zeng_short-term_2015} are among the first authors to integrate TOU prices in iron and steel production scheduling.
While the former authors focus on the melt shop section of a stainless steel plant, the latter specifically focus on steam power systems such as boilers and steam turbines, taking into account surplus byproduct gas flows.
\citet{TimeSch94_zhao_optimal_2017} specifically address the ecological concern of reducing the byproduct gases in steel production.
\citet{TimeSch49_cao_efficient_2021} describe the problem of simultaneously minimizing the makespan and TEC for the production planning of an iron-steel plant. However, in this setting, the TEC also depends on both the self-generation electricity costs and by the on-grid electrovalence. The authors present a mathematical model, alongside a version of the SPEA2 metaheuristic tailored for the problem.
\citet{TimeSch68_zhao_integrated_2018} face a multi-stage production problem that specifically models the rolling process of steel within an electric grid. The authors first formulate the problem as a MINLP model with generalized disjunctive programming constraints, and then they reformulate it as a MILP model.
\citet{TimeSch70_yang_robust_2018} integrate uncertainty by considering stochastic metal elements concentration in scrap steel charge within the production scheduling of a scrap steel melt shop. The proposed robust optimization approach is experimentally validated by comparison with two deterministic models and two multi-stage optimization approaches.
\citet{TimeSch74_guirong_solving_2017} also consider uncertain data in scheduling, but as a part of the steelmaking-continuous casting production, and tackle the problem with the Monte Carlo based ``cascade'' cross-entropy algorithm.
\citet{TimeSch91_pan_electrical_2019} consider a full steelmaking-refining-continuous casting framework, where TOU-based tariffs model the different time prices for electrical load tracking scheduling, and propose both a MINLP model and an improved SPEA2 metaheuristic for the problem.
\addRR{As a conclusion to the discussion of the applications in metallurgy,} \citet{TimeSch95_tan_optimizing_2017} manage to combine the Hot Rolling Batch Scheduling Problem with the job-shop scheduling problem.
\addRR{\citet{TimeSch77_wang_bi-objective_2016} face a real-world glass manufacturing problem, that falls within single-machine batch scheduling, with two heuristics based on decomposition concepts to deal with large-scale instances.
\citet{TimeSch69_zeng_energy_2021} instead specifically optimize the production scheduling of tissue article mills, which they tackle by means of a multi-objective evolutionary algorithm based on decomposition and ``teaching-learning'' optimization combined with VNS.}

\addRR{Finally, r}esearchers have also extensively considered energy supply scheduling.
\citet{TimeSch71_Liu2021} study the Virtual Power Plant (VPP), an efficient tool for smarter energy supply in an electric grid under TOU costs with trasversal demand-side resources management.
\citet{TimeSch97_sichilalu_optimal_2014} also focus on the electric grid and present a model for the problem of minimizing the energy cost of photovoltaic systems connected by a grid that supply power to heat pump water heaters.
\citet{TimeSch51_wang_unit_2021} instead investigate the impact of TOU prices on the power system operation within a unit scheduling problem for temperature control appliances;
similarly, \citet{TimeSch85_heydarian-forushani_flexible_2015} aim at determining the optimal TOU prices to ensure demand-side flexibility, while also optimizing the supply-side operations for a more secure and sustainable power grid.

\section{Structure of the thesis}
\label{sec:thesis-structure}

The remainder of this thesis is organized as follows.
Chapter \ref{chap:problem} introduces the problem and provides two mathematical models. The first one is based on a time-indexed formulation of the problem.
The second model builds upon the first one, and exploits inherent symmetries of the solutions space so as to achieve a significant improvement in formulation compactness.
Chapter \ref{chap:algorithms} introduces an effective heuristic scheme, as well as an exact algorithm that heavily relies on the mathematical models, to tackle the hardness of the problem.
Chapter \ref{chap:exp-tests} shows the numerical results obtained with the solution approaches presented in the thesis on a large test benchmark.
Chapter \ref{chap:conclusions} concludes the thesis by showing future research directions for a broad class of TOU scheduling problems.
Appendix \ref{sec:notation} reports notation for classical scheduling that is useful for TOU scheduling as well. Moreover, it reports the acronyms of the discrete optimization algorithms used in the literature of TOU scheduling, and presented in this chapter. 
Appendix \ref{chap:symbol-table} presents a compendium of the main mathematical symbols introduced in the definitions throughout the thesis.
Finally, Appendix \ref{chap:reference} reports a brief summary of the main notions, definitions, and algorithms introduced in thesis, so as to provide a convenient reference.

%% file: chap2.tex

\chapter{The scheduling problem}
\label{chap:problem}

The chapter begins with the introduction of the problem at the core of the thesis in Section 
\ref{sec:problem:problem-statement}. 
Subsequently, it provides two different mathematical models in Section \ref{sec:problem:mathematical-models}. 
Specifically, after the presentation of the first mathematical model, it describes a characterizing property of the solutions space that \addRR{enables} a compact reformulation of the model.
The chapter concludes with an in-depth comparison of the two models aimed at laying the foundations of the exact approach.

\section{Problem statement}
\label{sec:problem:problem-statement}

Let $\mathcal{J}=\{1,\ldots,N\}$ be the set of jobs, $\mathcal{H}=\{1,\ldots,M\}$ the set of identical machines, and $\mathcal{T}=\{1,\ldots,K\}$ the set of available time slots. 
Jobs are non-preemptible, and are characterized by an integer \emph{processing time}
$p_j \le K$,
$j \in \mathcal{J}$, corresponding to an integer number of distinct time slots.
Machines are endowed with an \emph{energy consumption rate} $u_h \ge 0$, $h \in \mathcal{H}$.
Moreover, a non-negative cost $c_t \ge 0$, $t\in\mathcal{T}$, is associated with each time slot.
An \emph{assignment} of a job $j \in \mathcal{J}$ to a subset $\mathcal{T}_j \subseteq \mathcal{T}$ of $p_j$ time slots on machine $h \in \mathcal{H}$ entails the processing of $j$ during the time slots in $\mathcal{T}_j$ by machine $h$.
In such case, job $j$ is \emph{scheduled} in the time slots in $\mathcal{T}_j$ on machine $h$.
Then, we recall that a schedule
\begin{align}\label{eq:schedule}
    \mathcal{S} = \{(j, h_j, \mathcal{T}_j) : h_j \in \mathcal{H}, \mathcal{T}_j \subseteq \mathcal{T}, \forall j \in \mathcal{J}\}
 \end{align}
is a set of the assignments of the jobs in $\mathcal{J}$ such that each $j \in \mathcal{J}$ is scheduled on one and only one machine $h_j \in \mathcal{H}$, and at most a single job in $\mathcal{J}$ is assigned to each time slot in $\mathcal{T}$ on each machine in $\mathcal{H}$.
If $\mathcal{T}_j$ is a set of $p_j$ consecutive time slots, then the schedule $\mathcal{S}$ is \emph{feasible}.
In this section, whenever $\mathcal{S}$ is referred to simply as a schedule, it is implied that $\mathcal{S}$ is feasible.

The \emph{completion time} of a job $j \in \mathcal{J}$ is the largest time slot in $\mathcal{T}_j$, i.e., $C_j(\mathcal{S}) = \max_{t \in \mathcal{T}_j}t$, $j \in \mathcal{J}$.
Then, the makespan $C^\text{max}$ of a schedule $\mathcal{S}$ is the largest among the completion times of the jobs in $\mathcal{J}$, i.e., 
\begin{align}
    C^\text{max}(\mathcal{S}) = \max\{C_j(\mathcal{S}), j \in \mathcal{J}\}.
    \label{eq:Cmax}
\end{align}
\addRR{The energy cost associated with the processing of job $j$ on machine $h_j$ in $\mathcal{S}$ is $u_{h_j} \sum_{t \in \mathcal{T}_j} c_t$.
As a consequence, the TEC of $\mathcal{S}$ is given by
\begin{align}\label{eq:TEC}
    E(\mathcal{S}) = \sum_{j \in \mathcal{J}} u_{h_j} \sum_{t \in \mathcal{T}_j} c_t.
\end{align}
\noindent
Then, the \emph{Bi-objective Identical Parallel Machine Scheduling with Time-of-Use Costs Problem} (BPMSTP) consists in finding a schedule $\mathcal{S}$ that simultaneously minimizes \eqref{eq:Cmax} and \eqref{eq:TEC}.}
The BPMSTP can be stated as $Pm | | C^{max}, TEC$ by means of the $\alpha | \beta | \gamma$ notation.
Observe that since both $Pm | | C^{max}$ \citep{Garey1978S} and $1 | | TEC$ \citep{TimeSch3_chen_scheduling_2019} are strongly $\mathcal{NP}$-hard, the BPMSTP is strongly $\mathcal{NP}$-hard as well.

The ordered tuple $(\mathcal{J}, \{p_j, j \in \mathcal{J}\}, \mathcal{H}, \{u_h, h \in \mathcal{H}\}, \mathcal{T}, \{c_t, t \in \mathcal{T}\})$ will be referred to as an \emph{instance} $\mathcal{I}$ of the BPMSTP.
Since a schedule $\mathcal{S}$ is a feasible solution to $\mathcal{I}$, the expressions ``schedule'' and ``feasible solution'' will be used interchangeably in the rest of the chapter.
Moreover, to avoid burdening the notation, the dependence of $C_j$, $C^\text{max}$ and $E$ on $\mathcal{S}$ is omitted from now on.
As a final remark, this thesis will also consider the BPMSTP from a practical standpoint, so as to comply with the needs of the decision makers in manufacturing industry as well.
Specifically, finding all the Pareto-optimal solutions to $\mathcal{I}$ will be a further purpose of the thesis, and the following chapters will be developed accordingly.

\section{Mathematical models}
\label{sec:problem:mathematical-models}

Let us describe the first mixed-integer programming model for the BPMSTP \citep{TimeSch11_Anghinolfi2021ABH}, referred to as ``Formulation 1''.
To this end, let us denote by $x_{j, h, t} \in \{0, 1\}$, $j \in \mathcal{J}$, $h \in \mathcal{H}$, $t \in \mathcal{T}$, a decision variable that is equal to $1$ if $t$ is the \addRR{first time slot }of job $j$ on machine $h$, and $0$ otherwise. 
Moreover, let us express the makespan in $\eqref{eq:Cmax}$ and the TEC in $\eqref{eq:TEC}$ with the decision variables $C^{max} \in \mathbb{N}$ and $E \ge 0$, respectively.

\bigskip

\hypertarget{F1}{\noindent\textbf{Formulation 1.}}
\begin{align}
    &\min C^\text{max} \label{eq:model1:CMAX}, \\
    &\min E \label{eq:model1:TEC},
\end{align}
subject to
\begin{align}
    & E = \sum_{h \in \mathcal{H}} u_h \sum_{j \in \mathcal{J}}\sum_{t = 1}^{K - p_j + 1} x_{j, h, t}\left(\sum_{i = t}^{t + p_j - 1} c_i \right), \label{eq:model1:E} \\
    & \sum_{h \in \mathcal{H}}\sum_{t = 1}^{K - p_j + 1} x_{j, h, t} = 1, \quad j \in \mathcal{J}, \label{eq:model1:sj} \\
    & \sum_{j \in \mathcal{J}}\sum_{i = \max\{1, t - p_j + 1\}}^{t} x_{j, h, i} \le 1, \quad h \in \mathcal{H}, t \in \mathcal{T}, \label{eq:model1:NoOverl} \\
    & \sum_{h \in \mathcal{H}}\sum_{t = 1}^{K - p_j + 1} (t + p_j - 1) x_{j, h, t} \le C^\text{max}, \quad j \in \mathcal{J}, \label{eq:model1:Cj}\\
    & C^\text{max} \le K, \label{eq:model1:Cmax-ub}\\
    & C^\text{max}\ge 0, \quad 
    E \ge 0, \quad 
    x_{j, h, t} \in \{0, 1\}, j \in \mathcal{J}, h \in \mathcal{H}, t \in \mathcal{T}.\label{eq:model1:belong}
\end{align}

The objectives \eqref{eq:model1:CMAX} and \eqref{eq:model1:TEC} account for the minimization of the makespan and the TEC, respectively, according to the definition of the makespan introduced in \eqref{eq:Cmax} and of the TEC in \eqref{eq:model1:E}. Constraints \eqref{eq:model1:sj} impose that each job $j \in \mathcal{J}$ starts in a single slot on a single machine. Constraints \eqref{eq:model1:NoOverl} avoid that more than one job is processed in the same time slot on the same machine. The left-hand side of \eqref{eq:model1:Cj} defines the completion time of each job in $\mathcal{J}$, which must not exceed the makespan $C^\text{max}$. 
In turn, the makespan cannot be greater than the number of time slots $K$ owing to \eqref{eq:model1:Cmax-ub}. 
Lastly, \eqref{eq:model1:belong} defines the decision variables.

\hyperlink{F1}{Formulation 1} employs $NMK + 2$ decision variables and $2N + MK + 2$ constraints. 
The former number is due to the $NMK$ variables $x_{j, h, t}, j \in \mathcal{J}, h \in \mathcal{H}, t \in \mathcal{T}$, together with $C^\text{max}$ and $E$, while the latter one is due to constraints \eqref{eq:model1:E}--\eqref{eq:model1:Cmax-ub}.
The main drawback of \hyperlink{F1}{Formulation 1} lies in the number of decision variables that may become large as the size of the BPMSTP instances increases and, above all, presents many equivalent solutions.

Next, we introduce a novel model that overcomes the above limitations. Toward this end, we first formally define the concept of equivalence.
Let
\begin{align}\label{eq:PJ}
    \mathcal{P}_{\mathcal{J}'} \coloneqq \{ d : \exists\, j \in \mathcal{J}', p_j = d \}
\end{align}
be the set of distinct processing times of the jobs in $\mathcal{J}' \subseteq \mathcal{J}$.
Moreover, let
\begin{align}\label{eq:Jd}
    \mathcal{J}_d \coloneqq \{j: j \in \mathcal{J}, p_j = d\}, \quad d \in \mathcal{P} 
\end{align}
be the subset of jobs with processing time equal to $d$.
Two feasible solutions $\mathcal{S}$ and $\mathcal{S}'$ to the BPMSTP are \emph{equivalent} if they have the same value for $C^\text{max}$ and $E$.
Then, we show that the BPMSTP may present exponentially many equivalent solutions.
\begin{property}
    For each feasible solution $\mathcal{S}$ to the BPMSTP, there are at least $\prod_{d \in \mathcal{P}_\mathcal{J}} |\mathcal{J}_d|\,! - 1$ other different, equivalent feasible solutions.
\end{property}
\begin{proof}
    Let the schedule $\mathcal{S}$ be given as in \eqref{eq:schedule}.
    Let also $\mathcal{Z} = \{\{h_j, \mathcal{T}_j\}, j \in \mathcal{J}\}$ be the set of all distinct unordered pairs $\{h_j, \mathcal{T}_j\}$ such that there is a job $j \in \mathcal{J}$ scheduled in the time slots in $\mathcal{T}_j$ on machine $h_j$ in the schedule $\mathcal{S}$.
    Observe that $\mathcal{Z}$ can be rewritten as $\bigcup_{d \in \mathcal{P}_\mathcal{J}} \mathcal{Z}_d$, where $\mathcal{Z}_d = \{\{h_j, \mathcal{T}_j\}, j \in \mathcal{J}_d\}$.
    Since all the jobs in $\mathcal{J}_d$ require the same number of time slots, all the assignments of the jobs in $\mathcal{J}_d$ to the elements of $\mathcal{Z}_d$, for each $d \in \mathcal{P}_\mathcal{J}$, generate schedules that are equivalent to $\mathcal{S}$.
    As the number of distinct assignments of the jobs in $\mathcal{J}_d$ to $\mathcal{Z}_d$ corresponds to the number of permutations of the jobs in $\mathcal{J}_d$, i.e., $|\mathcal{J}_d|!$, then the distinct number of assignments of the jobs in $\mathcal{J}$ to $\mathcal{Z}$ is the product of $|\mathcal{J}_d|!$ for each $d \in \mathcal{P}_\mathcal{J}$.
    Observing that the schedule $\mathcal{S}$ is one of such assignments concludes the proof.
\end{proof}

Finally, let $b_{d, t} = \sum_{k = t}^{t + d - 1} c_k$, $d \in \mathcal{P}_\mathcal{J}$, $t = 1, \ldots, K - d + 1$, be the cumulative cost associated with the time slots $t, t + 1, \ldots, t + d - 1$. 
As a consequence, any job $j$ with processing time $p_j = d$ assigned to machine $h$ starting at time slot $t$ is characterized by an energy cost equal to $u_h \, b_{d, t}$.
Let also $y_{d, h, t} \in \{0, 1\}$, $d \in \mathcal{P}_\mathcal{J}$, $h \in \mathcal{H}$, $t \in \mathcal{T}$, be a binary decision variable that is equal to $1$ if $t$ is the start time of a job with processing time equal to $d$ on machine $h$, and $0$ otherwise.\\

\hypertarget{F2}{\noindent\textbf{Formulation 2.}}
\begin{align}
    &\min C^\text{max}, \label{eq:model2:CMAX} \\
    &\min E, \label{eq:model2:TEC}
\end{align}
subject to
\begin{align}
    & E = \sum_{h \in \mathcal{H}} u_h \sum_{d \in \mathcal{P}_\mathcal{J}}\sum_{t = 1}^{K - d + 1} b_{d, t} \: y_{d, h, t}, \label{eq:model2:E} \\
    & \sum_{h \in \mathcal{H}}\sum_{t = 1}^{K - d + 1} y_{d, h, t} = |\mathcal{J}_d|, \quad d \in \mathcal{P}_\mathcal{J},  \label{eq:model2:TotProc} \\
    & \sum_{d \in \mathcal{P}_\mathcal{J}}\sum_{i = \max\{1, t - d + 1\}}^{t} y_{d, h, i} \le 1, \quad h \in \mathcal{H}, t \in \mathcal{T}, \label{eq:model2:NoOverl} \\
    & (t + d - 1) \, y_{d, h, t} \le C^\text{max}, \quad d \in \mathcal{P}_\mathcal{J}, h \in \mathcal{H}, t = 1, \ldots, K - d + 1, \label{eq:model2:Cj}\\
    & C^\text{max} \le K, \label{eq:model2:Cmax-ub} \\
    & C^\text{max}\ge 0, \quad 
    E\ge 0, \quad
    y_{d, h, t} \in \{0, 1\}, d \in \mathcal{P}_\mathcal{J}, h \in \mathcal{H}, t \in \mathcal{T}. \label{eq:model2:belong}
\end{align}

The objectives \eqref{eq:model2:CMAX} and \eqref{eq:model2:TEC} account for the minimization of the makespan and the TEC, respectively, with the TEC here given by \eqref{eq:model2:E}. 
Constraints \eqref{eq:model2:TotProc} impose that, for each distinct processing time $d \in \mathcal{J}_d$, exactly $|\mathcal{J}_d|$ jobs with processing time $d$ are assigned to some subsets of slots on the machines. 
Equation \eqref{eq:model2:NoOverl} guarantees that, on each machine, at most a single job is processed in a time slot.
The left-hand side of \eqref{eq:model2:Cj} defines the completion time of jobs, which must be less than or equal to the makespan $C^\text{max}$. 
Similarly to \hyperlink{F1}{Formulation 1}, $C^\text{max}$ must not exceed the scheduling horizon $K$, owing to \eqref{eq:model2:Cmax-ub}. 
Lastly, \eqref{eq:model2:belong} defines the domain of the decision variables.

Each feasible solution of \hyperlink{F2}{Formulation 2} defines a class of equivalent schedules. Indeed,
\hyperlink{F2}{Formulation 2} guarantees that, for each $y_{d, h, t} = 1$, a job $j \in \mathcal{J}$ with processing time $d$ is non-preemptively scheduled in the slots $t, t + 1, \ldots, t + d - 1$ on machine $h$.
\hyperlink{F2}{Formulation 2} also ensures that each job $j \in \mathcal{J}$ is scheduled once and only once.

Algorithm \ref{alg:fromYtoS} generates a possible schedule out of the class of schedules defined by a solution of \hyperlink{F2}{Formulation 2}.
In more detail, Algorithm \ref{alg:fromYtoS} first initializes the schedule $\mathcal{S}$ to the empty set at line \ref{alg:fromYtoS:S-init} and the sets needed for subsequent computations at lines \ref{alg:fromYtoS:Jd-init-begin}--\ref{alg:fromYtoS:Jd-init-end}.
Then, for each $d$, $h$, and $t$ such that $y_{d, h, t} = 1$, a job in $\mathcal{J}_d'$ is assigned to $d$ consecutive slots on machine $h$ starting from slot $t$ (see lines \ref{alg:fromYtoS:main-for-begin}--\ref{alg:fromYtoS:main-for-end}).
Finally, the computed schedule $\mathcal{S}$ is returned at line \ref{alg:fromYtoS:return}.
At the end of the algorithm, $\mathcal{J}'_d = \emptyset$ for each $d \in \mathcal{P}_\mathcal{J}$, all the jobs in $\mathcal{J}$ are assigned, and there are not slots on the same machine assigned to more than one job.

\begin{algorithm}[tb]
\begin{tabularx}{\textwidth}{lX}
\textbf{Input:} & The assignment variables $y_{d, h, t}$, $d \in \mathcal{P}_\mathcal{J}, h \in \mathcal{H}, t \in \mathcal{T}$ \\
\textbf{Output:} & A feasible schedule $\mathcal{S}$ \\
\hline
\end{tabularx}
\begin{algorithmic}[1]
    \State Let $\mathcal{S} \leftarrow \emptyset$
    \label{alg:fromYtoS:S-init}
    \For{$d \in \mathcal{P}_\mathcal{J}$}
    \label{alg:fromYtoS:Jd-init-begin}
        \State Let $\mathcal{J}_d' \leftarrow \mathcal{J}_d$
    \EndFor
    \label{alg:fromYtoS:Jd-init-end}
    \For{$(\hat d, \hat h, \hat t) \in \{(d, h, t) : y_{d, h, t} = 1, d \in \mathcal{P}_\mathcal{J}, h \in \mathcal{H}, t \in \mathcal{T}\}$}
    \label{alg:fromYtoS:main-for-begin}
        \State Let $j \in \mathcal{J}'_{\hat d}$
        \State $\mathcal{S} \leftarrow \mathcal{S} \cup (j, \hat h, \{\hat t, \hat t + 1, \ldots, \hat t + \hat d - 1\})$
        \State $\mathcal{J}'_{\hat d} \leftarrow \mathcal{J}'_{\hat d} \, \setminus \, \{j\}$
    \EndFor
    \label{alg:fromYtoS:main-for-end}
    \Return{$\mathcal{S}$} \label{alg:fromYtoS:return}
\caption{Generate-Schedule}
\label{alg:fromYtoS}
\end{algorithmic}
\end{algorithm}

\hyperlink{F2}{Formulation 2} employs $|\mathcal{P}_\mathcal{J}|MK + 2$ decision variables and $|\mathcal{P}_\mathcal{J}|$ + $MK$ + $|\mathcal{P}_\mathcal{J}|M$ $\sum_{d \in \mathcal{P}_\mathcal{J}}(K - d + 1)$ + $2$ constraints. 
The former number is due to the $|\mathcal{P}_\mathcal{J}|MK$ variables $y_{d, h, t}, j \in \mathcal{J}, h \in \mathcal{H}, t \in \mathcal{T}$, together with $C^\text{max}$ and $E$, while the latter one is due to constraints \eqref{eq:model2:E}--\eqref{eq:model2:Cmax-ub}.
Let us finally compare the number of variables needed by \hyperlink{F1}{Formulation~1} and \hyperlink{F2}{Formulation~2}.
The worst case for \hyperlink{F2}{Formulation~2} occurs when $|\mathcal{P}_\mathcal{J}| = N$, i.e., when all the processing times in $\mathcal{J}$ are distinct. 
In this case, \hyperlink{F2}{Formulation 2} has the same number $NMK + 2$ of decision variables as \hyperlink{F1}{Formulation 1}.
On the contrary, the most convenient situation for \hyperlink{F2}{Formulation 2} occurs when the processing times of all the jobs in $\mathcal{J}$ are equal, i.e., when $|\mathcal{P}_\mathcal{J}| = 1$. 
In such case, \hyperlink{F1}{Formulation 1} is still characterized by $NMK + 2$ decision variables, while \hyperlink{F2}{Formulation 2} has only $MK + 2$ variables. 
Thus \hyperlink{F2}{Formulation 2} uses less decision variables than \hyperlink{F1}{Formulation 1}, except for the case $|\mathcal{P}_\mathcal{J}| = N$ when the two models are equivalent.

Let us now better characterize the worst case for the number of decision variables of \hyperlink{F2}{Formulation 2}. 
To this end, observe that a necessary condition for an instance of the BPMSTP to admit at least a feasible solution is that the sum of all the time slots required by the jobs in $\mathcal{J}$ does not exceed the overall number $MK$ of slots available for the scheduling, i.e.,
\begin{align}
    N \le \sum_{j \in \mathcal{J}}p_j \le MK,
    \label{eq:sched-nec-cond}
\end{align}
where the equality $\sum_{j \in \mathcal{J}}p_j = N$ holds when $p_j = 1$ for each $j \in \mathcal{J}$.
Let us then formulate the following stronger necessary condition for feasibility by building upon \eqref{eq:sched-nec-cond}.

\begin{proposition}[Necessary condition for the existence of a solution]
\label{prop:nec-cond-worst-case-comp}
For a BPMSTP instance that admits at least a feasible solution, the following inequality holds:
\begin{align}\label{eq:prop-nec-cond-2-thesis}
    |\mathcal{P}| \le \left\lfloor\frac{-1 + \sqrt{1 + 8MK}}{2}\right\rfloor.
\end{align}
\end{proposition}
\begin{proof}
First,
\begin{align}
    \sum_{j \in \mathcal{J}} p_j = \sum_{d \in \mathcal{P}} |\mathcal{J}_d| d \ge \sum_{i = 1}^{|\mathcal{P}|} i = \frac{|\mathcal{P}| (|\mathcal{P}| + 1)}{2}
    \label{eq:distinct-pt-sum-lb}
\end{align}
since $|\mathcal{J}_d| \ge 1$ and the elements in $\mathcal{P}$ are pairwise distinct positive integers.
By combining \eqref{eq:sched-nec-cond} with \eqref{eq:distinct-pt-sum-lb}, we obtain
\begin{align*}
    \frac{|\mathcal{P}| (|\mathcal{P}| + 1)}{2} \le MK,
\end{align*}
which entails $|\mathcal{P}|^2 + |\mathcal{P}| - 2 MK \le 0$, and therefore
\begin{align*}
    0 \le |\mathcal{P}| \le \frac{-1 + \sqrt{1 + 8MK}}{2}.
\end{align*}
\end{proof}

We observe that since \eqref{eq:sched-nec-cond} and \eqref{eq:prop-nec-cond-2-thesis} only depend on the parameters of the BPMSTP, they are valid for both \hyperlink{F1}{Formulation~1} and \hyperlink{F2}{Formulation~2}. In more detail, Proposition \ref{prop:nec-cond-worst-case-comp} is useful to identify a larger class of unfeasible solutions with respect to \eqref{eq:sched-nec-cond}, and therefore it enables avoiding solving several instances for \hyperlink{F2}{Formulation~2} by simply checking the validity of \eqref{eq:prop-nec-cond-2-thesis} beforehand.

In order to illustrate how Proposition \ref{prop:nec-cond-worst-case-comp} provides a better description of the worst case of \hyperlink{F2}{Formulation~2} as regards the number of decision variables, we consider an instance with $K = 200$ and $M = 10$ as a simple example.
The greatest value of $N$ for the existence of at least a feasible solution corresponds to the case $p_j = 1$ for all $j \in \mathcal{J}$, and it is equal to $MK = 2 \cdot 10^3$, owing to \eqref{eq:sched-nec-cond}. In this case, the number of decision variables of \hyperlink{F1}{Formulation~1} is $4 \cdot 10^6 + 2$, whereas it is equal to $2 \cdot 10^3 + 2$ for \hyperlink{F2}{Formulation~2} since $|\mathcal{P}| = 1$. 
Observe that, for such an instance, condition \eqref{eq:prop-nec-cond-2-thesis} also holds.
Instead, if $|\mathcal{P}| = N$, the number of decision variables for \hyperlink{F1}{Formulation~1} and \hyperlink{F2}{Formulation~2} is the same.
In particular, according to Proposition \ref{prop:nec-cond-worst-case-comp}, a necessary condition for feasibility is $N \le \lfloor{(-1 + \sqrt{16001}) \, / \, 2}\rfloor = 62$.
Hence, in order for the considered instance to be possibly feasible, the number of the variables has to be no greater than $1.24 \cdot 10^5 + 2$.
The necessary condition \eqref{eq:sched-nec-cond} would instead provide the higher upper bound $M^2K^2 + 2 = 4 \cdot 10^6 + 2$.

To complete the comparison of \hyperlink{F1}{Formulation~1} and \hyperlink{F2}{Formulation~2}, we also have to take into account the number and nature of the sets of constraints.
However, as it will be discussed in Chapter~\ref{chap:algorithms},
these constraints can be neglected in the framework of the developed exact algorithm.
Moreover, Chapter \ref{chap:exp-tests} reports the significantly lower computational effort required to solve Formulation 2 with respect to \hyperlink{F1}{Formulation 1} in all the considered experimental tests.

%% file: chap3.tex

\chapter{Algorithms} 
\label{chap:algorithms}

This chapter describes the different original solution approaches for the BPMSTP.
First, it introduces Split-Greedy Heuristic (SGH) in Section \ref{sec:SGH}. 
SGH enhances the heuristic algorithm proposed by \citet{TimeSch33_wang_bi-objective_2018} by considering a larger space of greedy decisions.
SGH is used as a sub-routine by the Split-Greedy Scheduler (SGS) algorithm, described in Section \ref{sec:SGS}, so as to find all the Pareto-optimal solutions to the problem.
Next, Section \ref{sec:ES} introduces Exchange Search (ES) \citep{TimeSch11_Anghinolfi2021ABH}, which is a local search algorithm that uses improving moves enabled by inherent characteristics of the BPMSTP.
Lastly, this chapter concludes by describing the complete heuristic scheme used to solve the BPMSTP, called Split-Greedy Scheduler (SGS-ES), which is a natural extension of SGS obtained by combining SGH and ES.

The main contributions of this section are the introduction of SGH and ES.
Specifically, SGH enables to increase the quality and the number of the feasible heuristic solutions with respect to the heuristic algorithm of \citeauthor{TimeSch33_wang_bi-objective_2018}.
The idea at the core of SGH is to greedily assign jobs in order to minimize the TEC, while possibly allowing particular violations to the non-preemption constraint.
It is however shown that such an \addRR{infeasible} solution can always be converted into a feasible and equivalent one.
ES searches for specific improving moves so as to improve the TEC of such a feasible schedule.
Despite its computational burden, it proves very useful in improving the outcomes of the greedy choices performed by SGH.

\section{Split-Greedy Heuristic}
\label{sec:SGH}

Let us first introduce some useful definitions.
Two slots $t$ and $t + k$, $1 \le t \le K - 1$, on a machine $h \in \mathcal{H}$ are \emph{adjacent} if $k = 1$.
Moreover, for a given schedule $\mathcal{S}$, a slot $t \in \mathcal{T}$ on machine $h \in \mathcal{H}$ is \emph{free}, or \emph{idle}, if no job is assigned to $t$ on $h$.

\begin{definition}\label{def:free-consecutive-slots}
\emph{(Free-consecutive slots)}\,
For a given schedule $\mathcal{S}$, the slots $t$ and $t + k$, $1 \le t \le t + 2 \le t + k \le K$, are \textbf{\emph{free-consecutive}} on machine $h \in \mathcal{H}$ if $t$ and $t + k$ are free, and slots $t + 1, t + 2, \ldots, t + k - 1$ are not free in $\mathcal{S}$.
\end{definition}

\begin{figure}[b]
    \captionsetup{font=footnotesize}
    \centering
    \includegraphics[scale=0.5]{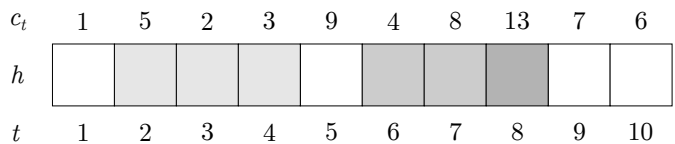}
    \caption{An example of three distinct locations on machine $h$, colored with different shades of grey.
    The location highlighted with the lightest shade is $(h, \{2, 3, 4\})$, the one with the intermediate shade is $(h, \{6, 7\})$, while the one with the darkest shade is $(h, \{8\})$.}
    \label{img:locations}
\end{figure}

Let us refer to a pair $l = (h, \mathcal{F})$, with $h \in \mathcal{H}, \mathcal{F} \subseteq \mathcal{T}$, as a \emph{location}.
Furthermore, let $u_h \sum_{t \in F} c_t$ be the \emph{energy cost} of location $l$.
Let us consider the example in Figure \ref{img:locations}. The sets of slots with different shades of grey correspond to the locations $(h, \{2, 3, 4\})$, $(h, \{6, 7\})$, and $(h, \{8\})$, from the lightest to the darkest shade, with energy costs $10$, $12$, and $13$, respectively.

\begin{definition}\label{def:free-location}
\emph{(Free location)}\,
For a given schedule $\mathcal{S}$, a \textbf{\emph{free location}} for a job $j$ is a location $(h, \mathcal{F})$ such that $\mathcal{F}$ contains only free adjacent and/or free-consecutive slots on machine $h$ in $\mathcal{S}$, and $|\mathcal{F}| = p_j$.
\end{definition}

\begin{figure}[b]
    \captionsetup{font=footnotesize}
    \centering
    \includegraphics[scale=0.525]{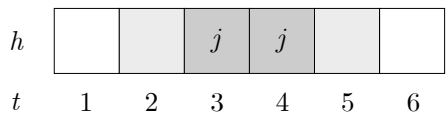}
    \caption{
    An example of free-consecutive and adjacent slots.
    Slots 2 and 5 are free-consecutive, while slots 3 and 4 are the adjacent assigned slots of job $j$.}
    \label{img:free-consecutive}
\end{figure}

Figure \ref{img:free-consecutive} reports an example of $K = 6$ slots on a machine $h$ where a single job $j$ with processing time $p_j= 2$ is scheduled.
Slots $2$ and $5$ are free-consecutive slots on $h$. 
Furthermore, the location $(h, \mathcal{A})$, $\mathcal{A} = \{2, 5\}$, is a free location for a job $j'$ with processing time $p_{j'} = 2$ on $h$. 
It is also worth to observe that, if $j'$ was assigned to $\mathcal{A}$, the resulting schedule would be \addRR{infeasible} due to the preemption of $j'$.

\begin{definition}\label{def:assigned-location}
\emph{(Assigned location)}\,
For a given schedule $\mathcal{S}$, if a job $j$ is assigned to the set $\mathcal{S}_j$ of $p_j$ adjacent slots on machine $h_j$ in $\mathcal{S}$, then the location $(h_j, \mathcal{S}_j)$ is the \textbf{\emph{assigned location}} of $j$.
\end{definition}

If $l_j = (h_j, \mathcal{S}_j)$ is the assigned location of $j$, we also simply say that $j$ is assigned to $l_j$.
In such case, the start time of $j$ is $s_j \coloneqq min_{t \in \mathcal{S}_j}\{t\}$.
In Figure \ref{img:free-consecutive}, $(h,\{3, 4\})$ is the assigned location of job $j$.

Let us now describe the intuition underlying SGH and its novelty for the BPMSTP.
Toward this end, let us first outline the heuristic proposed by \citet{TimeSch33_wang_bi-objective_2018}.
The core of such heuristic consists in the sequential application of a constructive heuristic and a local search algorithm, which is iterated in order to find all the feasible optimal solutions to the problem.
We refer to this heuristic as CH because \citet{TimeSch33_wang_bi-objective_2018} adopt this naming convention while presenting their computational results.

CH is based on the $\epsilon$-constraint paradigm \citep{Haimes_epsilon} for multi-objective optimization.
The basic idea of such paradigm is to minimize (or maximize) one of the objectives while the other ones are constrained to be lower (or greater) than fixed values.
Specifically, given an instance $\mathcal{I}$ of the BPMSTP, CH first (i) sets an upper bound on the makespan, and then (ii) minimizes the TEC.
CH iterates over steps (1) and (2) by progressively reducing the upper bound, until a lower bound for the makespan is reached.
In this way, CH finds a set of heuristic solutions to the problem that approximate the set of non-dominated points $\mathcal{I}$. Specifically, CH consists in the following steps:
\begin{enumerate}
    \item the jobs in $\mathcal{J}$ are sorted according to the Longest Processing Time (LPT) rule, i.e., in non-increasing order of processing times;
    \item the maximum makespan $K^\text{max}$ for the current iteration is initialized to $K$, and its lower bound is set to $K^{min} = \sum_{j \in J}p_j / M$.
    Moreover, the set of computed solutions $\mathcal{N}_s$ is initialized to $\emptyset$;
    \item if $K^\text{max} < K^{min}$, then CH terminates since no feasible solution with makespan less than or equal to \addRR{$K^\text{min}$} exists.
    Otherwise, for each job $j \in \mathcal{J}$, CH determines the location $l^{min}$ with the smallest cost among the set of free locations for $j$ including only free adjacent slots not exceeding $K^\text{max}$.
    Then, $j$ is assigned to $l^{min}$.
    Possible ties are broken by choosing a location so that the start time of $j$ is the smallest possible;
    \item a local search tries to improve the TEC of the solution computed at the previous step by shifting ``blocks'' of contiguous jobs on each machine in $\mathcal{H}$ without worsening the makespan.
    Specifically, for each $h \in \mathcal{H}$, such a local search explores a neighborhood of moves that can modify the schedule on $h$ in an effort of improving the TEC, while preserving the processing sequence of the jobs on $h$;
    \item the makespan $C^\text{max}$ and the TEC of the solution improved at the previous step are computed.
    Then, the solution is added to $\mathcal{N}_s$.
    Furthermore, the maximum makespan is updated as $K^\text{max} \leftarrow C^\text{max} - 1$, and the algorithm's control flow returns to step 3.
    Observe that, after the update, the condition $K^\text{max} \ge K^{min}$ at step $3$ does not hold if there are no other feasible solutions with makespan less than $C^\text{max}$;
    \item the set of the non-dominated solutions in $\mathcal{N}_s$ is returned.
\end{enumerate}
It is assumed that the ties at step 1 are broken randomly, as there is no specific indication in this regard by the authors.

Observe that, in general, there may not be a feasible solution for \addRR{some} $K^\text{max} \ge K^{min}$, as $K^{min}$ is not a tight lower bound for the makespan.
Hence, the solution computed at step (3) might be \addRR{infeasible}.
However, CH does not verify its feasibility. As a consequence, CH may return a solution set that contains \addRR{infeasible} schedules.
Moreover, according to the original description in \citep{TimeSch33_wang_bi-objective_2018}, CH may not be able to build feasible schedules, even though a feasible schedule exists, when no location including only free adjacent slots is available for a job.
Specifically, CH may return an \addRR{infeasible} schedule as a solution to the problem without reporting its \addRR{infeasibility}.
Example \ref{example:CH-unfeasible} presents an instance that elicits such behavior.

\begin{figure}[t]
    \captionsetup{font=footnotesize}
    \centering
    \includegraphics[scale=0.525]{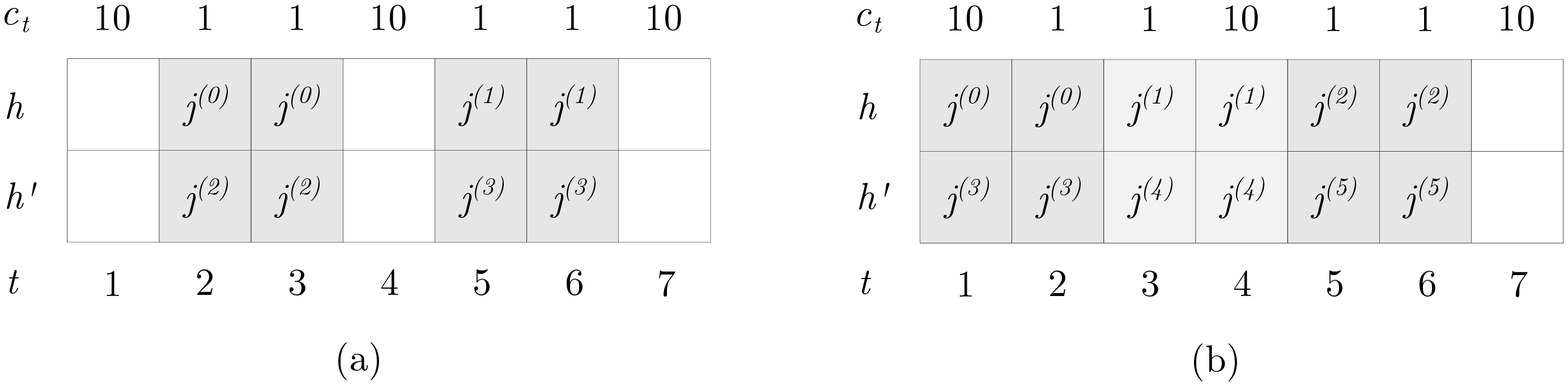}
    \caption{An example of the shortcomings of CH while solving specific instances.  
    Figure \ref{figure:example-CH-unfeasible}(a) shows the partial schedule generated by CH for Example \ref{example:CH-unfeasible}.
    Instead, Figure \ref{figure:example-CH-unfeasible}(b) shows a feasible schedule for the same example.}
    \label{figure:example-CH-unfeasible}
\end{figure}

\begin{example}\label{example:CH-unfeasible}
\normalfont \addRR{Let us consider two machines $h, h' \in \mathcal{H}$ with energy consumption rate $u_h = 1$ and $u_{h'} = 2$, and a set $\hat{\mathcal{J}} = \{j^{(i)}, i = 0, 1, \ldots, 5\}$ of jobs, with processing time $p_j = 2$ for each $j \in \hat{\mathcal{J}}$.
The number $K$ of time slots is $7$, and for $t = 1, \ldots, 7$, the energy costs $c_t$ are $10, 1, 1, 10, 1, 1, 10$, respectively. 
Since the jobs in $\hat{\mathcal{J}}$ have the same processing time, CH considers the jobs in $\hat{\mathcal{J}}$ in random order.
Figure \ref{figure:example-CH-unfeasible}(a) shows one among the possible smallest cost alternative assignments of jobs $j^{(0)}$, $j^{(1)}$, $j^{(2)}$ and $j^{(3)}$ performed by CH. 
However, after such assignment, there is no valid location left for $j^{(4)}$ and $j^{(5)}$.
Hence, CH stops iterating over $\hat{\mathcal{J}}$, and considers such partial (i.e., \addRR{infeasible}) schedule as a valid return value (i.e., a feasible solution).
Furthermore, as shown in Figure \ref{figure:example-CH-unfeasible}(b), a feasible solution however actually exists for the problem.}
\end{example}

In order to overcome this issue, SGH is able to temporarily assign jobs to free locations that, according to Definition \ref{def:free-location}, may include free-consecutive slots. 
Let us formalize this intuitive notion.

\begin{definition}\label{def:assigned-consecutive-slots}
\emph{(Assigned-consecutive slots)}\,
For a given schedule $\mathcal{S}$, the slots $t$ and $t + k$, $1 \le t \le t + 2 \le t + k \le K$, are \textbf{\emph{assigned-consecutive}} on machine $h \in \mathcal{H}$ if $t$ and $t + k$ are assigned slots of some job $j \in \mathcal{J}$, and $t + 1, t + 2, \ldots, t + k - 1$ are assigned slots of a subset of jobs of $\mathcal{J} \setminus \{j\}$ in $\mathcal{S}$.
\end{definition}

\begin{definition}\label{def:split-location}
\emph{(Split-location)}\,
For a given schedule $\mathcal{S}$, a \textbf{\emph{split-location}} is a free location with at least two free-consecutive slots, or an assigned location with at least two assigned-consecutive slots, in $\mathcal{S}$.
A split-location is \textbf{\emph{free}} for a job $j$ if it is also a free location for $j$. 
A split-location $(h_j, \mathcal{T}_j)$ is \textbf{\emph{assigned}} if there is a job $j \in \mathcal{J}$ such that $j$ is assigned to all and only the slots in $\mathcal{T}_j$.
\end{definition}

\begin{figure}[b]
    \captionsetup{font=footnotesize}
    \centering
    \includegraphics[scale=0.525]{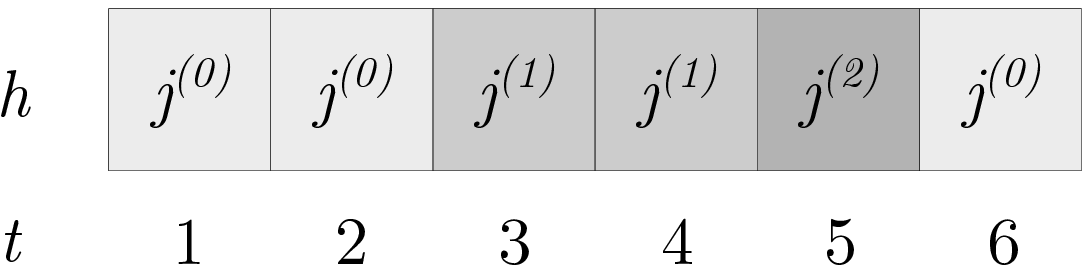}
    \caption{An example of a split-location.
    The location $(h, \{1, 2, 6\})$ is the assigned split-location of job $j$.}
    \label{figure:split-location}
\end{figure}

A job assigned to a split-location $l$ is \emph{split-scheduled} in $l$.
In Figure \ref{figure:split-location}, \addRR{$j^{(0)}$} is assigned to the location $(h, \{1, 2, 6\})$, which is a split-location since slots $1$ and $2$ are adjacent, and slots $2$ and $6$ are free-consecutive.

\begin{definition}\label{definition:split-schedule}
\emph{(Split-schedule)}\,
A \textbf{\emph{split-schedule}} $\mathcal{S}$ is a schedule as in \eqref{eq:schedule} such that, for some $j \in \mathcal{J}$, $(h_j, \mathcal{T}_j)$ is a split-location.
\end{definition}

\begin{figure}[t]
    \captionsetup{font=footnotesize}
    \centering
    \includegraphics[scale=0.525]{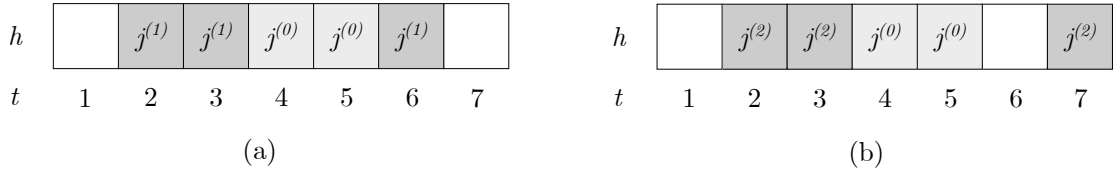}
    \caption{
    An example of a split-schedule (a), and a preemptive schedule which is not a split-schedule (b).
    In particular, the schedule in (a) is a split-schedule since $j^{(0)}$ is assigned to adjacent slots, while $j^{(1)}$ is assigned \addRR{both to adjacent slots and} free-consecutive slots.
    In the schedule in (b), $j^{(0)}$ is assigned to the same adjacent slots, but two of the assigned slots of $j^{(1)}$, i.e., slots $3$ and $7$, are neither adjacent, nor free-consecutive.}
    \label{figure:split-vs-preemptive-schedule}
\end{figure}

Observe that the schedule in Figure \ref{figure:split-location} is a split-schedule.
In fact, as opposed to feasible schedules, in a split-schedule at least one job is assigned to a split-location, according to Definition \ref{definition:split-schedule}.
Hence, split-schedules are not feasible and they are a subset of preemptive schedules.
Figure \ref{figure:split-vs-preemptive-schedule} highlights the difference between a split-schedule and a preemptive schedule which is not a split-schedule.
In more detail, Figure \ref{figure:split-vs-preemptive-schedule}(a) shows a split-schedule with a split-scheduled job $j^{(1)}$, whereas Figure \ref{figure:split-vs-preemptive-schedule}(b) shows a schedule with a preempted job $j^{(2)}$. Such \addRR{a }job is not split-scheduled since slots \addRR{3 and 7 are not assigned-consecutive as slot 6 is free}.

\begin{definition}\label{definition:split-schedule-block}
\emph{(Schedule block)}\,
For a given schedule $\mathcal{S}$, a \textbf{\emph{schedule block}} $\mathcal{B}$ on a machine $h \in \mathcal{H}$ of $\mathcal{S}$ is a schedule which involves only subsets of non-free slots that are delimited by two free slots.
Formally, $\mathcal{B}$ is a set $\{\{j, h, \mathcal{G}_j\}$, $j \in \mathcal{J}' \subseteq \mathcal{J}\}$, such that $\bigcup_{j \in \mathcal{J}'} \mathcal{G}_j$ is equal to a subset $\{i, i + 1, \dots, i + b - 1\} \subseteq \mathcal{T}$ of $b \ge 0$ assigned adjacent slots on $h$ and slots $i - 1$ and $i + b$, if they are in $\mathcal{T}$, are free.
If $\mathcal{B}$ contains at least a split-location, the block is called a split-schedule block; otherwise, it is called a feasible schedule block.
\end{definition}

\begin{figure}[b]
    \captionsetup{font=footnotesize}
    \centering
    \includegraphics[scale=0.476]{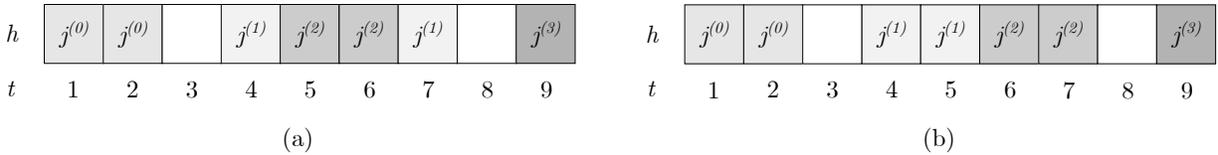}
    \caption{
    An example of split-schedule block and feasible schedule block.
    In Figure \ref{figure:schedule-blocks}(a), slots $3$, $4$, $5$, and $6$ are in a split-schedule block involving jobs $j^{(1)}$ and $j^{(2)}$.
    Instead, in Figure \ref{figure:schedule-blocks}(b), such slots are in a feasible schedule block involving the same two jobs.}
    \label{figure:schedule-blocks}
\end{figure}

A split-schedule block identifies a set of consecutive time slots that are assigned to split-scheduled jobs in the block. 
An example of a split-schedule block and a feasible schedule block is given in Figure \ref{figure:schedule-blocks}. 
The following proposition guarantees the correctness of SGH.

\pagebreak

\begin{proposition}\label{proposition:split-schedule-conversion}
Let $\mathcal{I}$ be an instance of the BPMSTP.
Then, for each split-schedule $\mathcal{S}$ of $\mathcal{I}$, there is a feasible schedule $\mathcal{S}'$ equivalent to $\mathcal{S}$.
\begin{proof}
Let us consider the schedule $\mathcal{S}_h \subseteq \mathcal{S}$ on each machine $h \in \mathcal{H}$.
A split-schedule $\mathcal{S}_h$ on a machine can be partitioned into $b \ge 0$ split-schedule blocks $\mathcal{B}_{h, i}$, $i = 1, \ldots, b$.
For each $h \in \mathcal{H}$, neither the makespan or the TEC of $\mathcal{S}_h$ is affected by different processing sequences of the jobs within each block $\mathcal{B}_{h, i}$, since there are no free slots in any of them.
Indeed, let us observe that each split-schedule block is a solution of an instance $\tilde{\mathcal{I}}$ of $1 \, | \, prmp \, | \, C^\text{max}, TEC$.
Moreover, let us recall the fact that for each such preemptive solution, say, $\mathcal{Z}$ of 
$\tilde{\mathcal{I}}$, 
there is a non-preemptive solution $\mathcal{Z}'$ of $\tilde{\mathcal{I}}$ with the same makespan of $\mathcal{Z}$.
Then, for each split-schedule block, there is an equivalent feasible schedule block with a different jobs processing sequence.
Observing that 
$C^\text{max}(\mathcal{S}) = \max_{h \in \mathcal{H}} C^\text{max}(\mathcal{S}_h)$ and 
$E(\mathcal{S}) = \sum_{h \in \mathcal{H}} E(\mathcal{S}_h)$ concludes the proof.
\end{proof}
\end{proposition}

Proposition \ref{proposition:split-schedule-conversion} ensures the possibility of converting any split-schedule into an equivalent feasible schedule. 
Algorithm \ref{alg:convert-schedule} performs this task. 
Specifically, among the alternative equivalent feasible schedules that can be obtained from a split-schedule $\mathcal{S}$, Algorithm \ref{alg:convert-schedule} generates the feasible schedule $\mathcal{S}$ such that, for each $j$ and $j'$ in $\mathcal{J}$ scheduled on the same machine, the inequality relation between the start times $s_j$ and $s_j'$ of $j$ and $j'$ in $\mathcal{S}$ holds in $\mathcal{S}'$ as well.
The efficiency of Algorithm \ref{alg:convert-schedule} lies in the fact that it considers each job, and possibly changes its assignment, at most once.

Towards the end of describing Algorithm \ref{alg:convert-schedule}, as well as the algorithms in the rest of the chapter, let
\begin{align}\label{eq:instance-hat}
    \mathcal{D}(\hat{K}) = \{(\mathcal{J}, \{p_j, j \in \mathcal{J}\}, \mathcal{H}, \{u_h, h \in \mathcal{H}\}, \hat{\mathcal{T}} = \{1, 2, \ldots, \hat{K}\} \subseteq \mathcal{T}, \{c_t, t \in \hat{\mathcal{T}}\})\}
\end{align}
be an instance of the BPMSTP for a given positive integer $\hat{K} \le K$.
Intuitively, instance \eqref{eq:instance-hat} disregards all the time slots $t$ such that $\hat{K} < t \le K$.
Algorithm \ref{alg:convert-schedule} takes a solution $\mathcal{S}$ for an instance $\mathcal{D}(\hat{K})$ as in \eqref{eq:instance-hat} as an input and returns a feasible schedule $\mathcal{S}'$ equivalent to $\mathcal{S}$.
For each machine $h \in \mathcal{H}$, Algorithm \ref{alg:convert-schedule} iterates over the time slots in non-decreasing order to determine the start time $s_j$ of each job $j$ scheduled on machine $h$ (lines $4$--$9$) in the split-schedule $\mathcal{S}$, and
as soon as it identifies the starting time of a job $j$, it assigns $j$ to $p_j$ consecutive time slots, starting from the first free slot greater than or equal to $s_j$ (lines $5$--$7$).

For the sake of computational efficiency, Left-Leaning Red-Black (LLRB) trees \citep{sedgewick2011Algorithms4E} are used to represent schedules in the implementation of Algorithm \ref{alg:convert-schedule}, as well as in the implementations of the other algorithms presented in this section.
The computational complexity of such algorithms is analysed accordingly.
LLRBs are based on Red-Black trees \citep{cormen2009}, which are a type of self-balancing Binary Search Tree (BST).
BSTs are typically used to represent ordered sets so as to enable logarithmic-time worst-case computational complexity for retrieval, insertion, and deletion operations. 
With such data structures, a schedule is stored as a set of $M$ LLRBs, each of them associated with a distinct machine in $\mathcal{H}$. In more detail, each LLRB stores the start times of the jobs as the keys of the tree, augmenting each node with the location of the related job as additional information. 
For each $h \in \mathcal{H}$, lines $4 - 9$ are executed $O(\Omega)$ times, where
\begin{align*}
    \Omega = \min\{N, \hat{K}\},
\end{align*}
since the maximum number of jobs scheduled on a machine is $N$, and it is upper bounded by the number of time slots $\hat{K}$ in $\mathcal{D}(\hat{K})$.
Lines $3$ and $5$ both take $O(1)$. The extraction of the entry with the lowest key among the ones with key greater than a given one from a LLRB (line $4$) consisting of $n$ elements is $O(log_2\,n)$. The insertion operation on a LLRB (line $7$) has the same computational complexity. Therefore, lines $4$ and $7$ have a $O(log_2\,\Omega)$ complexity as considered in an iteration of lines $4$--$9$.
\begin{algorithm}[!t]
\begin{tabularx}{\textwidth}{ c X }
    \textbf{Input} & A (possibly \addRR{infeasible}) schedule $\mathcal{S}$ for a BPMSTP instance $\mathcal{D}(\hat{K})$ as in \eqref{eq:instance-hat}. \\
    \textbf{Output} & A feasible schedule $\mathcal{S}'$ equivalent to $\mathcal{S}$. \\
\hline
\end{tabularx}
\begin{algorithmic}[1]
    \State {$\mathcal{S}' \leftarrow \{\mathcal{S}'_h$, $h \in \mathcal{H}\}$, a collection of empty sets $\mathcal{S}'_h = \emptyset, h \in \mathcal{H}$
    }
    \For{\textbf{each} $h \in H$}
        \State{\addRR{$s \leftarrow 0$,\, $k \leftarrow 1$}}
        \While{\addRR{it exists a job $j$ such that $\hat{s}_j=\min\{s_i : i \in \mathcal{J}, (i, h, \mathcal{T}_i) \in \mathcal{S}_h, s_{i} > s$\}}}
            \State{$s'_j$ $ \leftarrow \addRR{\max\{k, \hat{s}_j\}}$ \quad //\textit{The actual start time of $j$ in the converted schedule}}
            \State{$l' \leftarrow \{h, \{s'_j, s'_j + 1,..., s'_j + p_j - 1\}\}$}
            \State{Reassign $j$ to $l'$ by adding $(j, h, \{s'_j, s'_j + 1,..., s'_j + p_j - 1\})$ to $\mathcal{S}_h'$}
            \State\addRR{$s \leftarrow \hat{s}_j$,\, $k \leftarrow s_j' + p_j$}
        \EndWhile
    \EndFor
    \State \textbf{return} $\mathcal{S}'$
\caption{Convert-Schedule}
\label{alg:convert-schedule}
\end{algorithmic}
\end{algorithm}
In the pseudo-code, line $6$ is stated consistently with the notation of locations introduced earlier in this section. However, since the slots $s'_j, s'_j + 1,..., s'_j + p_j - 1$ are consecutive, line $6$ can be easily implemented to store only the starting time \addRR{$s_j'$} and the end time \addRR{$s_j' + p_j - 1$} of each job $j \in \mathcal{J}$. Therefore, the complexity of line $6$ is $O(1)$.
Since lines $2$--$10$ are executed $M$ times, the worst-case computational complexity of Algorithm \ref{alg:convert-schedule} is $O(M \Omega \log_2{\Omega})$.

Finally, let us proceed to describe SGH in detail.
SGH constructively computes a heuristic schedule for a BPMSTP instance by minimizing the TEC while disregarding the makespan, similarly to step 3) of CH.
However, as opposed to it, SGH:
\begin{enumerate}[(a)]
    \item returns no feasible solution if $K^\text{max}$ is too tight \addRR{for SGH }to schedule all the jobs in $\mathcal{J}$;
    \item it temporarily allows the generation of split-schedule blocks while scheduling the jobs in $\mathcal{J}$;
    \item it finally converts any split-schedule block obtained with previous computations into an equivalent feasible one by means of Algorithm \ref{alg:convert-schedule}.
\end{enumerate}

The pseudo-code of SGH is reported in Algorithm \ref{alg:SGH}.
Let us first describe the local variables used by Algorithm \ref{alg:SGH} for its computations.
The set $\mathcal{S}$ is a schedule that is partitioned into $M$ schedules $\mathcal{S}_h$, $h \in \mathcal{H}$, which are empty at line $1$, and they are progressively filled as jobs are assigned to locations.
Instead, $P_\mathcal{J}$ is a list that, after its initialization at line $2$, contains the distinct processing times of the jobs in $\mathcal{J}$, i.e., the elements of $\mathcal{P}_\mathcal{J}$, in decreasing order.
The set $\mathcal{J}_d$, $d \in \mathcal{P}_\mathcal{J}$, initialized at line $6$, is a subset of $\mathcal{J}$ that contains jobs with processing time equal to $d$, according to \eqref{eq:PJ}.
Finally, $L_{d, h}$, $d \in \mathcal{P}_\mathcal{J}$, $h \in \mathcal{H}$, initialized at line $6$, is a list of the free locations for a job with processing time $d$ on machine $h$, which is progressively updated as jobs are assigned to the locations.

Algorithm \ref{alg:SGH} initializes $\mathcal{S}$ as a set of empty sets at line $1$.
Then, at line $2$, it computes a list $P_\mathcal{J}$ out of $\mathcal{P}_\mathcal{J}$, and then it sorts $P_\mathcal{J}$ in non-decreasing order.
At line $3$, it initializes $\mathcal{J}_d$ by using $\mathcal{J}$ for each $d \in \mathcal{P}_\mathcal{J}$ according to \eqref{eq:Jd}.
At lines $4$--$15$, Algorithm \ref{alg:SGH} computes a (possibly split-) schedule by assigning the jobs in $\mathcal{J}_d$ for each $d \in \mathcal{P}_\mathcal{J}$.
Specifically, for each $d \in \mathcal{P}_\mathcal{J}$, it first builds, for each $h \in \mathcal{H}$, the list $L_{d, h}$ of all the available, free (possibly split-) locations with the smallest cost for a job with processing time $d$ on $h$ (lines $5$--$7$).
Then, it assigns the jobs in $\mathcal{J}_d$ at lines $8$--$14$ as follows.
For each $j \in \mathcal{J}_d$, it checks if there is at least a free location left for $j$ in $\bigcup_{h \in H}\{l : l \in L_{d, h}\}$ at line $9$. If such a location does not exist, it returns an empty schedule.
Otherwise, it performs a random selection among the possible locations with the smallest cost for $j$. 
The motivation behind this specific tie-breaking strategy lies in experimental observations.
In fact, such random selection enabled to achieve better solutions with respect to the deterministic choice that instead favors the earliest starting location (implemented in CH).
At line $11$, Algorithm \ref{alg:SGH} updates $\mathcal{S}_{\hat{h}}$ by adding $(j, \hat{h}, \hat{\mathcal{A}})$ as a result of assigning $j$ to the location $(\hat{h}, \hat{\mathcal{A}})$.
As a consequence, previously free locations are affected, and new ones may be generated.
Specifically, the locations that contain at least a slot $t \in \hat{\mathcal{A}}$ should be removed from the list, since they are not free after the assignment of $j$.
Intuitively, such assignment may also result in new free split-locations containing the slots that delimit $(\hat{h}, \hat{\mathcal{A}})$.
At lines $12$--$13$, Algorithm \ref{alg:SGH} updates $L_{d, \hat{h}}$ accordingly.
Finally, if the resulting schedule $\mathcal{S}$ is a split-schedule, Algorithm \ref{alg:SGH} converts it into a feasible one in line $17$ with Algorithm \ref{alg:convert-schedule}.

Let us now determine the computational complexity of Algorithm \ref{alg:SGH}.
The computational complexity of lines $1$--$3$ is $O(M + |\mathcal{P}_{\mathcal{J}}| \, log_2{|\mathcal{P}_{\mathcal{J}}|} + N)$, since line $1$ initializes $M$ empty schedules as LLRBs, line $2$ requires to sort the list $P_\mathcal{J}$, and line $3$ iterates over the jobs in $\mathcal{J}$ in order to build $\mathcal{F}_d$ for each $d \in \mathcal{P}_\mathcal{J}$.
Then, let us consider the computational complexity of an iteration of lines $5$--$14$.
Let us focus on lines $5$--$7$.
For each $h \in \mathcal{H}$, in order to identify all the free locations for each $j \in \mathcal{J}_d$ on $h$, line $6$ iterates over the time slots in $\mathcal{T}$ by advancing a pointer until $\hat{K}$ is reached. 
A queue with maximum capacity $d$ is employed to store the free slots reached by the pointer. 
As soon as a free slot $t$ is found, $t$ is enqueued until the maximum capacity is reached. When this happens, a new free (possibly split-) location is identified. The start and the end time of such location are respectively equal to the oldest and the most recent elements in the queue. Then, whenever a new free slot $t$ is found, the oldest element in the queue is dequeued and $t$ is enqueued, progressively identifying new locations. 
Querying the job scheduled in each slot, which consists in retrieving an element in a $\mathcal{S}_h$, $h \in \mathcal{H}$, requires $O(\log_2{\Omega})$ as the maximum number of jobs scheduled on each machine is $\Omega$. 
Hence, lines $5$--$7$ take $O(M \hat{K} \log_2{\Omega})$.
Such complexity can be improved to $O(M \Omega \log_2{\Omega})$ by observing that, for each slot $t \in \mathcal{T}$, the smallest key larger than $t$ in $\mathcal{S}_h$, $h \in \mathcal{H}$, i.e., the smallest start time of the jobs in $\mathcal{S}_h$ among the ones greater than $t$, can be retrieved in $O(\log_2{\Omega})$. 
In this way, $\mathcal{S}_h$ has to be queried at most $\Omega$ times, since the job with the smallest start time greater than $t$ can be determined in logarithmic time.
Consider the loop at lines $8$--$14$. 
Line 9 takes $O(1)$. 
Line $10$ performs the random removal of a location from a set of locations with the same cost on $O(M)$ machines. Hence, line $10$ takes $O(M)$.
Line $11$ updates $\mathcal{S}_{\hat{h}}$ with the result of the new assignment, and since the number of elements in $\mathcal{S}_{\hat{h}}$ is $O(\Omega)$, line $11$ takes $O(\log_2{\Omega})$. 
The update of $L_{d, \hat{h}}$ at lines $12$--$13$ takes $O(\Omega \log_2{\Omega})$ (see the analysis of lines $5$--$7$).
Hence, all the iterations performed by lines $4$--$15$ take
\begin{align*}
    O(|\mathcal{P}_\mathcal{J}|M \Omega \log_2{\Omega} + N(M + \Omega \log_2{\Omega}))
\end{align*}
Finally, lines $16$--$18$ take $O(M \, \Omega \log_2{\,\Omega})$, that is the complexity of Algorithm \ref{alg:convert-schedule}.
\begin{algorithm}[!t]
\begin{tabularx}{\textwidth}{ c X }
    \textbf{Input} & A BPMSTP instance $\mathcal{D}(\hat{K})$ as in \eqref{eq:instance-hat}.\\
    \textbf{Output} & A feasible schedule for $\mathcal{D}(\hat{K})$, if it exists; otherwise, an empty schedule.\\
\hline
\end{tabularx}
\begin{algorithmic}[1]
    \State Let $\mathcal{S} \leftarrow \{\mathcal{S}_h$, $h \in \mathcal{H}\}$ be a collection of the empty sets $\mathcal{S}_h = \emptyset, h \in \mathcal{H}$
    \State Let $P_\mathcal{J}$ be a list of the elements in $\mathcal{P}_\mathcal{J}$ in decreasing order
    \State Let $\mathcal{J}_d \leftarrow \{j \in \mathcal{J} \colon p_{j} = d\}$, $d \in \mathcal{P}_\mathcal{J}$
    \For {$d \in \addRR{\mathcal{P}}_{\mathcal{J}}$}
        \For {$h \in \mathcal{H}$}
            \State Let $L_{d, h}$ be a list of the free locations with \addRR{smallest} cost on $h$ for any $j$, $p_j = d$
        \EndFor
        \For{$j \in \mathcal{J}_d$}
            \State{\textbf{if} $L_{d, h} = \emptyset, \forall h \in \mathcal{H}$ \textbf{then} \textbf{return} $\emptyset$ \quad // No free location for $j$}
            \State {Remove a smallest cost
            location $\hat{l} = (\hat{h}, \hat{\mathcal{A}})$ from $\bigcup_{h \in \mathcal{H}}\{l : l \in L_{d, h}\}$
            }
            \State Assign job $j$ to $\hat{l}$ by adding $(j, \hat{h}, \hat{\mathcal{A}})$ to $\mathcal{S}_{\hat{h}}$
            \State Remove any location or split-location $(\hat{h}, \mathcal{A}) : \mathcal{A} \cap \hat{\mathcal{A}} \neq \emptyset$ from $L_{d, \hat{h}}$
            \State Add any split-location $(\hat{h}, \mathcal{A}')$ with $|\mathcal{A}'| = d$ to $L_{d, \hat{h}}$
        \EndFor
    \EndFor
    \If{$\mathcal{S}$ is a split-schedule}
        \State $\mathcal{S} \leftarrow \text{Convert-Schedule}(\mathcal{S}) \quad \text{// Convert }\mathcal{S}\text{ with Algorithm \ref{alg:convert-schedule}}$ 
    \EndIf
    \State \textbf{return} $\mathcal{S}$
\caption{Split-Greedy Heuristic (SGH)}
\label{alg:SGH}
\end{algorithmic}
\end{algorithm}
Therefore, the complexity of Algorithm \ref{alg:SGH} is
\begin{align}\label{eq:SGH-complexity}
\begin{split}
    & O(M + |\mathcal{P}_\mathcal{J}| \log_2{|\mathcal{P}_\mathcal{J}|} + N) + O(|\mathcal{P}_\mathcal{J}| M \Omega \log_2{\Omega}) + 
    O(N M) \\
    & 
    + O(N \Omega \log_2{\Omega}) 
    + O(M \Omega \log_2{\Omega})
\end{split}
\end{align}
The first and the last term of the sum are the computational complexity of lines $1$--$3$ and $16$--$18$, respectively.
Instead, the second, the third, and the fourth term account for all the iterations of lines $5$--$7$, $9$--$10$, $11$, $12$--$14$, respectively.
Finally, \eqref{eq:SGH-complexity} can be more compactly expressed as
\begin{align}\label{eq:SGH-complexity-compact}
    O(
        \Omega (M |\mathcal{P}_\mathcal{J}| + N) \log_2{\Omega} + 
        NM +
        |\mathcal{P}_\mathcal{J}| \log_2{\mathcal{P}_\mathcal{J}}
    ).
\end{align}

In order to show that \eqref{eq:SGH-complexity-compact} is the most compact expression for the complexity of SGH, let us rewrite it as
\begin{align}\label{eq:SGH-complexity-decompacted}
    O(
        \Omega M |\mathcal{P}_\mathcal{J}| \log_2{\Omega} +
        \Omega N \log_2{\Omega} + 
        NM +
        |\mathcal{P}_\mathcal{J}| \log_2{\mathcal{P}_\mathcal{J}}
    ).
\end{align}
\noindent
Let us consider the first and the second term of the sum in \eqref{eq:SGH-complexity-decompacted}, i.e., $\Omega M |\mathcal{P}_\mathcal{J}| \log_2{\Omega}$ and $\Omega N \log_2{\Omega}$, respectively.
Generally, $|P_\mathcal{J}| \le N$, but $M \gtrless N$. Hence, the first and the second term are asymptotically equivalent, i.e., 
$\Theta(\Omega M |\mathcal{P}_\mathcal{J}| \log_2{\Omega}) = \Theta(\Omega N \log_2{\Omega})$.
Let us now consider the second and the third term of the sum, i.e.,
$\Omega N \log_2{\Omega}$ and $NM$, respectively.
Since $\Omega \log_2{\Omega} \gtrless M$, 
then indeed $\Theta(\Omega N \log_2{\Omega}) = \Theta(NM)$.
Finally, as regards the third and the fourth term of the sum,
since $NM \gtrless |\mathcal{P}_\mathcal{J}| \log_2{|\mathcal{P}_\mathcal{J}|}$,
then $\Theta(NM) = \Theta(|\mathcal{P}_\mathcal{J}| \log_2{|\mathcal{P}_\mathcal{J}|})$ as well.
Then, the four terms in the sum are asymptotically equivalent due to the transitive property of the Big Theta notation.

\section{Exchange Search}\label{sec:ES}
This section introduces ES.
First, it gives the intuition underlying ES.
Then, it states the definitions needed to formally describe ES.
Finally, the section presents ES by providing its pseudo-code and an expression of its computational complexity.

The purpose of ES is to improve the TEC of a feasible schedule $\mathcal{S}$. 
It strives to achieve so by changing the assignment of subsets of scheduled jobs to the machines in order to improve TEC while preserving the feasibility, and without worsening the makespan. 
In some cases, ES may however improve the makespan as a byproduct of TEC minimization.

\begin{definition}\label{def:free-consecutive-slots0}
\emph{(Exchangeable Period Sequence)}\,
Let $\mathcal{S}$ be a schedule as in \eqref{eq:schedule} for a BPMSTP instance $\mathcal{I}$.
An \textbf{\emph{Exchangeable Period Sequence}} (EPS) is a subset $\mathcal{E} \subseteq \mathcal{T}$ of adjacent time slots on a machine $h \in \mathcal{H}$ in $\mathcal{S}$
such that, if $h$ processes some job $j \in \mathcal{J}$ during a time slot $t \in \mathcal{E}$ in $\mathcal{S}$, then $\mathcal{T}_j \subseteq \mathcal{E}$.
\end{definition}

\begin{figure}[!t]
    \captionsetup{font=footnotesize}
    \centering
    \includegraphics[scale=0.525]{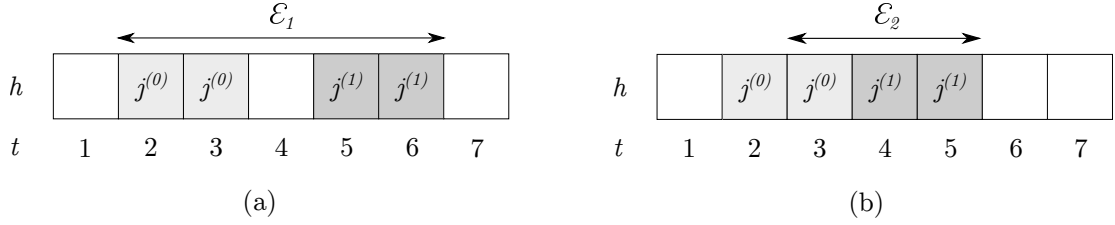}
    \caption{
    An example of two subsets of slots \addRR{$\mathcal{E}_1 = \{2, 3, 4, 5, 6\}$ and $\mathcal{E}_2 = \{3, 4, 5\}$}, of which only $\mathcal{E}_1$ is an EPS.
    In Figure \ref{fig:EPS-example}(a), $\mathcal{E}_1$ is an EPS since $j^{(0)}$ and $j^{(1)}$ are entirely scheduled in $\mathcal{E}_1$.
    On the contrary, in Figure \ref{fig:EPS-example}(b), $\mathcal{E}_2$ is not an EPS since $j^{(0)}$ is not entirely scheduled in $\mathcal{E}_2$.}
    \label{fig:EPS-example}
\end{figure}

Figure \ref{fig:EPS-example} provides an example of EPS. 
The subset of time slots $\mathcal{E}_1$ in Figure \ref{fig:EPS-example}(a) is an EPS with cardinality equal to $5$, since it contains two jobs with processing time equal to $2$ entirely scheduled in $\mathcal{E}_1$, and an idle time slot. 
Differently, Figure \ref{fig:EPS-example}(b) shows a subset of slots $\mathcal{E}_2$ on $h$ that is not an EPS, since job $j^{(0)}$ is not completely scheduled within $\mathcal{E}_2$.

\begin{definition}
\emph{(EPS subschedule)}\,
Let $\mathcal{S}$ be a schedule as in \eqref{eq:schedule} for a BPMSTP instance $\mathcal{I}$.
For a given EPS $\mathcal{E}$ on machine $h \in \mathcal{H}$ in $\mathcal{S}$, an \textbf{\emph{EPS subschedule}} $\mathcal{S}_{\mathcal{E}, h} \subseteq \mathcal{S}$ is a single-machine schedule 
$\{(j, h, \mathcal{T}_j), \forall \, j \in \mathcal{J} : \mathcal{T}_j \cap \mathcal{E} \neq \emptyset\}$.
\end{definition}
For simplicity, we drop the second index from $\mathcal{S}_{\mathcal{E}, h}$ when clear from the context.
For a given EPS $\mathcal{E}$, there always exists a (possibly empty) subschedule $\mathcal{S}_\mathcal{E}$.
In Figure \ref{fig:EPS-example}(a), $\mathcal{S}_{\mathcal{E}_1}$ is the subschedule 
$\{ (j^{(0)}, h, \{2, 3\}), 
(j^{(1)}, h, \{5, 6\}) \}$.
As a remark, observe that for a given free or assigned location $l = (h, \mathcal{A})$ of a job $j \in \mathcal{J}$, where $h \in \mathcal{H}$, $\mathcal{A} \subseteq \mathcal{T}$ is a subset of adjacent slots, and $p_j = |\mathcal{A}|$, $\mathcal{A}$ is an EPS by definition.
For instance, in Figure \ref{fig:EPS-example}(b), $(h, \{4, 5\})$ is the assigned location of $j^{(1)}$, and $(h, \{6, 7\})$ is a free location for a job with processing time equal to $p_{j^{(1)}} = 2$. As a matter of fact, the two sets $\{4, 5\}$ and $\{6, 7\}$ are EPS's.

For a given schedule $\mathcal{S}$, let $\mathcal{J}(\mathcal{S}_\mathcal{E}) \subseteq \mathcal{J}$ be the set of the jobs scheduled in $\mathcal{S}_\mathcal{E}$.

\begin{definition}
    \emph{(EPS swap)}\,
    Let $\mathcal{S}$ be a feasible schedule for an instance $\mathcal{I}$ of the BPMSTP. 
    Moreover, let $\mathcal{E} \subseteq \mathcal{T}$ and $\mathcal{E}' \subseteq \mathcal{T}$ be two EPS’s on $h \in \mathcal{H}$ and $h' \in \mathcal{H}$, respectively, such that $|\mathcal{E}| = |\mathcal{E}'|$, and $\mathcal{E} \cap \mathcal{E}' = \emptyset$ if $h = h'$.
    Then, an \textbf{\emph{EPS swap}} of $\mathcal{E}$ and $\mathcal{E}'$ on $h$ in $\mathcal{S}$ is an algorithm that schedules each $j \in \mathcal{J}(\mathcal{S}_\mathcal{E})$ in $\mathcal{E}'$ on $h'$,
    and each $j' \in \mathcal{J}(\mathcal{S}_{\mathcal{E}'})$ in $\mathcal{E}$ on $h$,
    by generating a new schedule $\mathcal{S}'$ without changing the relative assignments of the jobs in $\mathcal{E}$ and $\mathcal{E}'$, i.e., such that
    \begin{align*}
        C_j(\mathcal{S}') &= C_j(\mathcal{S}) - \addRR{\max_{t \in \mathcal{E}}{t}} + \max_{t \in \mathcal{E}'}{t}, 
        \quad j \in \mathcal{J}(\mathcal{S}_\mathcal{E}),\\
        C_{j'}(\mathcal{S}') &= C_{j'}(\mathcal{S}) - \addRR{\max_{t \in \mathcal{E}'}{t}} + \max_{t \in \mathcal{E}}{t}, 
        \quad j' \in \mathcal{J}(\mathcal{S}_{\mathcal{E}'}).
    \end{align*}
\end{definition}

\begin{figure}[!t]
    \captionsetup{font=footnotesize}
    \centering
    \includegraphics[scale=0.55]{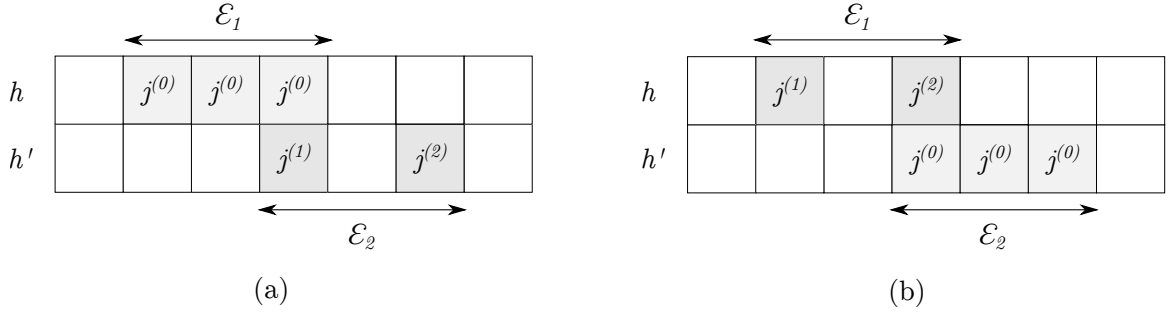}
    \caption{
    An example of an EPS swap of the EPS-J $\mathcal{E}_1$ and the EPS-I $\mathcal{E}_2$ on the machines $h$ and $h'$, respectively.}
    \label{fig:eps-swap}
\end{figure}

Figure \ref{fig:eps-swap} shows an example of an EPS swap of the two EPS's $\mathcal{E}_1$ and $\mathcal{E}_2$.
In Figure \ref{fig:eps-swap}(a), the
EPS $\mathcal{E}_1$ has an associated subschedule $\mathcal{S}_{\mathcal{E}_1}$ that involves job $j^{(0)}$ assigned to the slots in $\mathcal{E}_1$ on machine $h$, while $\mathcal{E}_2$ has an associated subschedule $\mathcal{S}_{\mathcal{E}_2}$ that involves jobs $j^{(1)}$ and $j^{(2)}$ assigned to a subset of the slots in $\mathcal{E}_2$ on machine $h'$.
Figure \ref{fig:eps-swap}(b) shows the result of an EPS swap of the two EPS's.
Observe that an EPS swap does not affect the feasibility of the involved subschedules, as it shifts the start times of the jobs in each subschedule by the same integer.

\begin{definition}\label{def:EPS-rearrangement}
\emph{(EPS rearrangement)}\,
Let $\mathcal{S}$ be a feasible schedule for a BPMSTP instance $\mathcal{I}$, and $\mathcal{E} \subseteq \mathcal{T}$ be an EPS on $h \in \mathcal{H}$ in $\mathcal{S}$.
An \textbf{\emph{EPS rearrangement}} of $\mathcal{E}$ on $h$ in $\mathcal{S}$ is a procedure that reassigns each $j \in \mathcal{J}(\mathcal{S}_\mathcal{E})$ to locations of the form $(h, \mathcal{A})$, where $\mathcal{A} \subseteq \mathcal{E}$ is a subset of 
adjacent slots.
\end{definition}

\begin{figure}[!b]
    \captionsetup{font=footnotesize}
    \centering
    \includegraphics[scale=0.525]{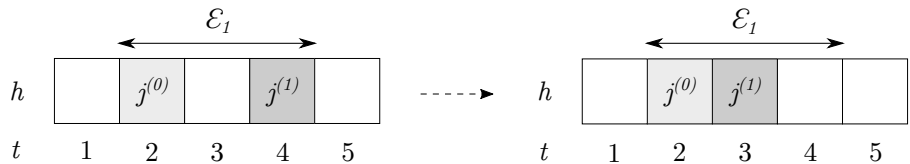}
    \caption{
    An example of an EPS rearrangement of the EPS-I $\mathcal{E}_1$.}
    \label{fig:EPS-rearrangement}
\end{figure}

Figure \ref{fig:EPS-rearrangement} shows an example of an EPS rearrangement applied to the EPS $\mathcal{E}_1$ in the schedule at the left of the arrow, which results in the EPS $\mathcal{E}_2$ in the schedule at the right.

Let us now discuss the idea at the core of ES.
Towards this end, let us consider a schedule $\mathcal{S}$ for a BPMSTP instance $\mathcal{I}$.
First, let us denote an EPS that contains no idle slots, and with a related subschedule with only a single job, as an \emph{EPS-J}. 
Let us also refer to an EPS that contains at least an idle slot as an \emph{EPS-I}.
Then, let $\mathcal{E}$ and $\mathcal{E}'$ be any two EPS's in $\mathcal{S}$ on machines $h$ and $h'$, respectively, such that if $h = h'$, then $\mathcal{E} \cap \mathcal{E}' = \emptyset$.
Hereinafter, we refer to an EPS swap of $\mathcal{E}$ and $\mathcal{E}'$, followed by a rearrangement of $\mathcal{E}$ and $\mathcal{E}'$, as an \emph{EPS move} involving $\mathcal{E}$ and $\mathcal{E}'$.
Suppose that such an EPS move results in a schedule $\mathcal{S}'$.
Generally, $E(\mathcal{S}) \gtrless E(\mathcal{S}')$.
ES searches for each move in the subset of EPS moves involving an EPS-J and an EPS-I, and that results in a schedule $\mathcal{S}'$ with TEC better than $\mathcal{S}$, i.e., $E(\mathcal{S}') < E(\mathcal{S})$.
Each EPS move in such subset involves two EPS's with cardinality no greater than the maximum processing time of the jobs in $\mathcal{J}$, i.e.,
\begin{align*}
    p_\text{max}=\max_{p \in \mathcal{P}_\mathcal{J}} p,
\end{align*}
by construction.
Starting from a feasible schedule $\mathcal{S}$, ES explores such neighborhood of EPS moves, and it modifies $\mathcal{S}$ as soon as it finds an improving EPS move.
Observe that such an EPS move cannot worsen that makespan by definition.
Towards the description of the pseudo-code of ES, let us first define the cumulative energy price 
\begin{align*}
    \mu_{t, h} \coloneqq u_h \sum_{i = 1}^t{c_i}, \quad h \in \mathcal{H}, t = 1, \ldots, K,
\end{align*}
for a given instance $\mathcal{I}$, and $\mu_{0, h} \coloneqq 0, h \in \mathcal{H}$.
Hence, the energy consumption cost of any location
\begin{align*}
    l = (h, \{t, t + 1, \ldots, t + p - 1\}), \quad h \in \mathcal{H}, p \in \mathcal{P}, t = 1, 2, \ldots, K - p + 1,
\end{align*}
can be expressed as $\mu_{t + p - 1, h} - \mu_{t - 1, h}$.
As a computational remark, the use of cumulative energy prices allows \addRR{us }to compute any location cost in $O(1)$, at the expense of precomputing $\mu_{t, h}, t = 1, \ldots, K, h \in \mathcal{H}$ in $O(M K)$.

Secondly, for a given feasible schedule $\mathcal{S}$, let us define $\mathcal{L}^\text{J}_p(\mathcal{S})$ and $\mathcal{L}^\text{I}_p(\mathcal{S})$ as the sets of subschedules of EPS-J's and EPS-I's in $\mathcal{S}$ of cardinality $p \in \mathcal{P}_\mathcal{J}$, respectively. 
Let us also introduce two sets of functions that associate a time slot $t$ on a machine $h$ with a subschedule $\mathcal{S}_\mathcal{E}$ associated with the EPS $\mathcal{E}$ on the same machine, and such that the smallest slot of $\mathcal{E}$ is $t$.
Such functions are very useful from a computational standpoint, as they allow, if implemented as (direct addressing) tables or hash-maps, to verify the existence of an EPS on a machine and retrieve it in $O(1)$. 
With intent of recalling the data structures that enables this computational benefit, we hereinafter refer to such functions as ``maps''.
Formally, let us denote as 
\begin{align*}
    \mathcal{M}^\text{J}_p(\mathcal{S}) : \mathcal{H} \times \mathcal{T} \rightarrow \mathcal{L}^\text{J}_p(\mathcal{S}) \cup \emptyset, \quad p \in \mathcal{P}_\mathcal{J},
\end{align*}
the maps that, for each EPS-J $\{t, t + 1, \ldots, t + p - 1\} = \mathcal{E} : S_\mathcal{E} \in \mathcal{L}^\text{J}_p(\mathcal{S})$, on a machine $h \in \mathcal{H}$, with $t \in \mathcal{T}$ and $p \in \mathcal{P}_\mathcal{J}$, associates $(h, t)$ with $\mathcal{E}$.
If $\mathcal{E}$ is not an EPS-J on $h$ in schedule $\mathcal{S}$, then $\mathcal{M}^\text{J}_p(\mathcal{S})$ associates $(h, t)$ with the special value $\emptyset$.
We denote the analogous maps for the EPS-I's as $\mathcal{M}^\text{I}_p(\mathcal{S}) : \mathcal{H} \times \mathcal{T} \rightarrow \mathcal{L}^\text{I}_p(\mathcal{S})\,\cup\, \emptyset$.
Finally, with a slight abuse of notation, we refer to the two collections of sets $\{\mathcal{L}^\text{J}_p(\mathcal{S}), p \in \mathcal{P}_\mathcal{J}\}$ and $\{\mathcal{L}^\text{I}_p(\mathcal{S}), p \in \mathcal{P}_\mathcal{J}\}$ 
as $\mathcal{L}^\text{J}(\mathcal{S})$ and $\mathcal{L}^\text{I}(\mathcal{S})$, respectively.
Similarly, we denote $\{\mathcal{M}_p^\text{J}(\mathcal{S}), p \in \mathcal{P}_\mathcal{J}\}$ and $\{\mathcal{M}_p^\text{I}(\mathcal{S}), p \in \mathcal{P}_\mathcal{J}\}$ 
as $\mathcal{M}^\text{J}(\mathcal{S})$ and $\mathcal{M}^\text{I}(\mathcal{S})$, respectively. From now on, the dependence on $\mathcal{S}$ is omitted whenever clear from the context.

Algorithm \ref{algorithm:ES} reports the pseudo-code of ES. 
Algorithm \ref{algorithm:ES} takes a feasible schedule $\mathcal{S}$ for a BPMSTP instance $\mathcal{D}(\hat{K})$ as in \eqref{eq:instance-hat} as input, and it returns a feasible schedule $\mathcal{S}'$ obtained by performing all the improving EPS moves for $\mathcal{S}$.
Algorithm \ref{algorithm:ES} accomplishes this task by using the subroutines \textit{FindEPS}, \textit{EvaluateEPSMove}, and \textit{UpdateEPS}, which are not reported in pseudo-code, and they are instead only described for the sake of compactness. 
In this way, the description omits the implementation details that are neither relevant to the analysis of the computational complexity, nor essential to the completeness of the presentation.
First, we present an overview of such subroutines. 
FindEPS identifies all the EPS-J's and the EPS-I's in a feasible schedule $\mathcal{S}$.
Algorithm \ref{algorithm:ES} uses FindEPS to compute the maps $\mathcal{M}^\text{J}$ and $\mathcal{M}^\text{I}$, so as to iterate over distinct EPS moves in the main loop of the algorithm.
EvaluateEPSMove efficiently evaluates if an EPS move entails an improvement in the TEC by disregarding some of the moves that cannot lead to improvement by exploiting a bounding condition. 
If the considered EPS move improves the TEC, it is performed.
However, the maps $\mathcal{M}^\text{J}$ and $\mathcal{M}^\text{I}$ may become inconsistent after such an EPS move.
In this case, UpdateEPS updates $\mathcal{M}^\text{J}$ and $\mathcal{M}^\text{I}$ by calling FindEPS so as to identify the new EPS's in the schedule.
Let us now delve into a formal explanation of each subroutine.
\newline

FindEPS takes a feasible schedule $\mathcal{S}$ for $\mathcal{D}(\hat{K})$ as in \eqref{eq:instance-hat}, 
the subsets $\hat{\mathcal{H}} \subseteq \mathcal{H}$ and $\hat{\mathcal{T}} \subseteq \mathcal{T}$, and
$\mathcal{P}_\mathcal{J}$ as an input, 
and returns the two sets of subschedules $\hat{\mathcal{L}}^\text{J}(\mathcal{S})$ and $\hat{\mathcal{L}}^\text{I}(\mathcal{S})$ of the EPS-J's and the EPS-I's in $\mathcal{S}$, respectively.

The identification of the EPS's in a schedule is conceptually similar to the task of finding free locations performed in line $6$ of Algorithm \ref{alg:SGH}.
Formally, for any $p \in \mathcal{P}_\mathcal{J}$ and $h \in \hat{\mathcal{H}}$, two variables $s$ and $e \ge s$ are first set to their initial values $\min_{t \in \hat{\mathcal{T}}}t$ and $\min_{t \in \hat{\mathcal{T}}}t + p - 1$, respectively.
If $e > \max_{t \in \hat{\mathcal{T}}}{t}$, there is no EPS in $\mathcal{T}$ on machine $h$ in schedule $\mathcal{S}$.
Otherwise, FindEPS starts a loop as follows. At each iteration, it considers the set of slots $\mathcal{E} = \{s, s + 1, \ldots, e\}$, and it checks if it is an EPS.
If that is the case, it adds $\mathcal{S}_\mathcal{E}$ to $\hat{\mathcal{L}}^\text{J}(\mathcal{S})$ if $\mathcal{E}$ is an EPS-J, or to $\hat{\mathcal{L}}^\text{I}(\mathcal{S})$ if $\mathcal{E}$ is an EPS-I. 
At the end of the iteration, $s$ is incremented by $1$, or by the processing time of the job starting at $s$, if there is one. Then, $e$ is incremented by the same quantity, so as to preserve the necessary condition $p = e - s + 1$ for $\mathcal{E}$ to be an EPS.
If $e$ reaches the value $\max_{t \in \hat{T}}{t} + 1$, i.e., $e$ cannot be the largest slot in an EPS, and the loop ends.
Otherwise, FindEPS proceeds with the next iteration.

In order for $\mathcal{E}$ to be an EPS-I, or an EPS-J, of cardinality $p$, $e - s + 1$ has to be equal to $p$. Furthermore, $s$ has to be free, or be the start time of a job, and $e$ must be free, or be the end time of a job.
In particular, $\mathcal{E}$ is an EPS-J if the slots in $\mathcal{E}$ are assigned to a single job. Instead, if at least one of the slots in $\mathcal{E}$ if free, then $\mathcal{E}$ is an EPS-I. Otherwise, $\mathcal{E}$ is still an EPS, but it is neither an EPS-J, nor an EPS-I.\\
Checking these conditions requires multiple queries to the LLRB's used to represent schedules, which yield a worst-case 
$O(\log_2{\hat{K}})$ 
complexity for each of such queries. 
Each of them is carried out at most twice for each $t \in \hat{T}$, when $s$ or $e$ reach $t$. 
As a side, but significant, implementation note, for a given EPS $\mathcal{E} = \{t, t + 1, \ldots, t + p - 1\} \subseteq \mathcal{T}$ on a machine $h \in \mathcal{H}$, 
the actual implementation of $\mathcal{M}_p^\text{J}$ and $\mathcal{M}_p^\text{I}$ associates the key $\hat{K} h + t_i$ to $\mathcal{S}_\mathcal{E}$. This allows \addRR{us }to uniquely address each EPS by using a single key. 

FindEPS also stores two additional information for further use: the number of the assigned slots in $\mathcal{E}$ and the cumulative energy cost of the list of slots contained in $\mathcal{E}$, sorted by energy cost in non-decreasing order.
This information is necessary for EvaluateEPSMove in the evaluation of its bounding condition.

Finally, the computational complexity of FindEPS is
\begin{align}\label{eq:complexity-FindEPS}
    O(|\mathcal{P}_\mathcal{J}| |\hat{\mathcal{H}}| |\hat{\mathcal{T}}| \max\{\log_2{\hat{K}},\, p_\text{max} \log_2{p_\text{max}}\})
\end{align} 
The term 
$\max\{\log_2{\hat{K}},\, p_\text{max} \log_2{p_\text{max}}\}$ 
is due to the two distinct operations carried out for each $p \in \mathcal{P}_\mathcal{J}$, $h \in \hat{\mathcal{H}}$, and 
$t \in \{\min_{k \in \hat{\mathcal{T}}}k,
\min_{k \in \hat{\mathcal{T}}}k + 1,
\ldots,
\max_{k \in \hat{\mathcal{T}}}k - p + 1\}
\subseteq \hat{\mathcal{T}}$: 
querying the LLRB to check whether $t$ is free or not on $h$ at most twice, and sorting the slot costs of $\{t, t + 1, \ldots, t + p - 1\}$, if it is an EPS.
\newline

EvaluateEPSMove takes a schedule $\mathcal{S}$ for $\mathcal{D}(\hat{K})$ as in \eqref{eq:instance-hat}, and an EPS-I $\mathcal{E}^{(0)}$ and an EPS-J $\mathcal{E}^{(1)}$ of cardinality $p \in \mathcal{P}_\mathcal{J}$ on machines $h^{(0)}$ and $h^{(1)}$, respectively, as input.
As an output, it returns the schedule $\mathcal{S}'$ that results from the EPS move involving $\mathcal{E}^{(0)}$ and $\mathcal{E}^{(1)}$ on machines $h^{(0)}$ and $h^{(1)}$, respectively, as well as the difference $E(\mathcal{S}') - E(\mathcal{S})$.
The purpose of EvaluateEPSMove is to evaluate the improvement in TEC entailed by the EPS move involving $\mathcal{E}^{(0)}$ and $\mathcal{E}^{(1)}$.

First, EvaluateEPSMove checks a necessary condition for TEC improvement, so as to possibly avoid to unnecessarily evaluate the EPS move.
EvaluateEPSMove achieves so by exploiting the additional information stored with each EPS by FindEPS.
Towards the end of describing such condition, let us consider an EPS 
\begin{align*}
    \mathcal{E} = \{i, i + 1, \ldots, i + p - 1\}, \quad p \in \mathcal{P}_\mathcal{J}, i \in \{ 1, 2, \ldots, K - p + 1 \},
\end{align*}
on machine $h \in \mathcal{H}$ in $\mathcal{S}$. Then, let us refer as $Q_\mathcal{E}$ to the list containing the time slots in $\mathcal{E}$, sorted in non-decreasing order of their cost, i.e., 
\begin{align*}
    Q_\mathcal{E} = (t^{(i)}, t^{(i + 1)}, \ldots, t^{(i + p - 1)}),
    \quad c_{t^{(i + n - 1)}} < c_{t^{(i + n)}}, n = 1, 2, \ldots, p - 1,
\end{align*} 
and such that $Q_\mathcal{E}$ is a permutation of $(i, i + 1, \ldots, i + p - 1)$.

For $n$ such that $0 \le n \le p - 1$, let us denote as
\begin{align*}
    \eta^\text{lb}_{h, n}(Q_\mathcal{E}) = u_h \sum_{r = 0}^{n - 1} c_{t^{(i + r)}}, 
    \qquad
    \eta^\text{ub}_h(\mathcal{E}) = u_h \sum_{r = 0}^{p - 1} c_{i + r},
\end{align*}
the cumulative cost of the first $n$ slots in the list $Q_\mathcal{E}$, and
the sum of the costs of the slots in $\mathcal{E}$, respectively.
Then, the inequality 
\begin{align*}
    \eta^\text{lb}_{h, n}(E) \le E(\mathcal{S}_\mathcal{E}) \le \eta^\text{ub}_h(\mathcal{E})
\end{align*}
holds for $0 \le n \le p - 1$.
Finally, let $\alpha(\mathcal{S}_\mathcal{E}) \coloneqq \sum_{j \in \mathcal{J}(\mathcal{S}_\mathcal{E})}{p_j}$ be the number of slots assigned in $\mathcal{S}_\mathcal{E}$.

Formally, the EPS move involving the EPS-I $\mathcal{E}^{(0)}$ and the EPS-J $\mathcal{E}^{(1)}$
results in two subschedules, say,
$\mathcal{S}_{\mathcal{E}^{(0)}}'$ and
$\mathcal{S}_{\mathcal{E}^{(1)}}'$.
Then, $E(\mathcal{S}_{\mathcal{E}^{(0)}}') + E(\mathcal{S}_{\mathcal{E}^{(1)}}')$ is lower bounded by
$
    \eta^\text{ub}_h(\mathcal{E}^{(0)}) + 
    \eta_{h, \alpha(\mathcal{S}_{\mathcal{E}^{(0)}})}^\text{lb}(Q_{\mathcal{E}^{(1)}})
$,
since, after the EPS move, the job initially in $\mathcal{E}^{(1)}$ is then scheduled entirely in $\mathcal{E}^{(0)}$ with cost $\eta^\text{ub}_h(Q_{\mathcal{E}^{(0)}})$, and the cost of the jobs initially in $\mathcal{E}^{(0)}$, and then scheduled in $\mathcal{E}^{(1)}$, cannot exceed the sum of the $\alpha(\mathcal{S}_{\mathcal{E}^{0}})$ smallest costs of the slots in $\mathcal{E}^{(1)}$, i.e., 
$\eta_{h, \alpha(\mathcal{S}_{\mathcal{E}^{(0)}})}^\text{lb}(Q_{\mathcal{E}^{(1)}})$.
If 
\begin{align}\label{eq:EPS-move-lb}
    E(\mathcal{S}_{\mathcal{E}^{(0)}}) + E(\mathcal{S}_{\mathcal{E}^{(1)}}) \le
    \eta^\text{ub}_h(\mathcal{E}^{(0)}) + 
    \eta_{h, \alpha(\mathcal{S}_{\mathcal{E}^{(0)}})}^\text{lb}(Q_{\mathcal{E}^{(1)}})
\end{align} 
then the EPS move is disregarded, as it cannot entail an improvement in TEC, and EvaluateEPSMove returns $\mathcal{S}$ and $0$.
Otherwise, ES carries out the EPS move by first applying the EPS swap, and then performing the EPS rearrangement of $\mathcal{E}^{(0)}$ with SGH.
As a consequence, ES generates a new schedule $\mathcal{S}'$.
Afterwards, EvaluateEPSMove sets an inner variable $\delta$ as
\begin{align*}
    \delta \leftarrow E(\mathcal{S}_{\mathcal{E}^{(0)}}') + E(\mathcal{S}_{\mathcal{E}^{(1)}}') - 
    E(\mathcal{S}_{\mathcal{E}^{(0)}}) - E(\mathcal{S}_{\mathcal{E}^{(1)}}),
\end{align*}
that is negative if the move is improving, and non-negative otherwise.
Finally, EvaluateEPSMove returns $\mathcal{S}'$ and $\delta$.\\
The computational complexity of EvaluateEPSMove is dominated by the one of SGH, given in \eqref{eq:SGH-complexity-compact}, which is called once for each of the two EPS's involved. Therefore, the complexity of EvaluateEPSMove is 
\begin{align*}
    & O(p_\text{max}(2p_\text{max} + 2p_\text{max})\log_2{\Omega} + 
    4p_\text{max} +
    p_\text{max}\log_2{p_\text{max}}),
\end{align*}
which is equivalent to
\begin{align}\label{eq:complexity-EvaluateEPSMove}
    O(p_\text{max}^2 \log_2{\Omega}).
\end{align}
\newline
UpdateEPS takes a schedule $\mathcal{S}$ for $\mathcal{D}(\hat{K})$ as in \eqref{eq:instance-hat}, 
the two sets $\mathcal{L}^\text{J}$ and $\mathcal{L}^\text{I}$, 
the two sets of maps $\mathcal{M}^\text{J}$ and $\mathcal{M}^\text{I}$, 
a machine $h \in \mathcal{H}$, 
a subset $\hat{\mathcal{T}} \subseteq \mathcal{T}$, and
$\mathcal{P}_\mathcal{J}$ as input, 
and returns the updated sets $\mathcal{L}^\text{J}$ and $\mathcal{L}^\text{I}$ of EPS's in $\hat{\mathcal{T}}$ on $h$, 
as well as the related maps $\mathcal{M}^\text{J}$ and $\mathcal{M}^\text{I}$. 
The purpose of UpdateEPS is to update $\mathcal{L}^\text{J}$, $\mathcal{L}^\text{I}$, $\mathcal{M}^\text{J}$, and $\mathcal{M}^\text{I}$, that may be inconsistent with $\mathcal{S}$ after an improving EPS move performed by EvaluateEPSMove.

In order to achieve this, UpdateEPS first iterates over each $p \in \mathcal{P}_\mathcal{J}$, and for each 
$t$ such that 
\begin{align*}
    \max\{0,\, \min_{t' \in \hat{\mathcal{T}}}t' - p + 1\} \le t \le
    \min\{\hat{K} - p + 1,\, \max_{t' \in \hat{\mathcal{T}}}{t'}\},
\end{align*}
it removes the entry, if present, associated with $(h, t)$ from both $\mathcal{L}_p^\text{J}$ and $\mathcal{L}_p^\text{I}$.
The complexity of such removal operations is $O(|\mathcal{P}_\mathcal{J}| |\hat{\mathcal{T}}|) = O(|\mathcal{P}_\mathcal{J}| p_\text{max})$, as $|\hat{\mathcal{T}}| \le 3p_\text{max} - 2$.
In order to show this, let us call the entry associated with $(h, t)$ as $\mathcal{E}$.
Then, observe that most the $p_\text{max} - 1$ slots before $\mathcal{E}$, the $p_\text{max}$ slots in $\mathcal{E}$, and the $p_\text{max} - 1$ slots after $\mathcal{E}$ are affected by the EPS move prior to the call of UpdateEPS.

Afterwards, UpdateEPS calls FindEPS with parameters $\mathcal{S}$, $\{h\}$, $\hat{\mathcal{T}}$, and $\mathcal{P}_\mathcal{J}$. 
Let us denote its return values as 
\begin{align*}
    \hat{\mathcal{L}}^\text{I} = \{\hat{\mathcal{L}}^\text{I}_p, p \in \mathcal{P}_\mathcal{J}\}, \quad \text{and }
    \hat{\mathcal{L}}^\text{J} = \{\hat{\mathcal{L}}^\text{J}_p, p \in \mathcal{P}_\mathcal{J}\}.
\end{align*}
UpdateEPS updates $\mathcal{L}^\text{I}$ and $\mathcal{L}^\text{J}$ with the newfound EPS's by performing the assignments 
\begin{align*}
    \mathcal{L}^\text{I} \leftarrow \mathcal{L}^\text{I}_p \cup \hat{\mathcal{L}}^\text{I}_p, \quad \text{and }
    \mathcal{L}^\text{J} \leftarrow \mathcal{L}^\text{J}_p \cup \hat{\mathcal{L}}^\text{J}_p
\end{align*}
for each $p \in \mathcal{P}_\mathcal{J}$ 
in $O(|\mathcal{P}_\mathcal{J}| p_\text{max})$ time.
The maps $\mathcal{M}^\text{J}$ and $\mathcal{M}^\text{I}$ are then updated accordingly, again in 
$O(|\mathcal{P}_\mathcal{J}| p_\text{max})$ time. 
Therefore, the complexity of UpdateEPS is 
\begin{align*}
    O(|\mathcal{P}_\mathcal{J}| p_\text{max} 
    \max\{\log_2{\Omega},\, p_\text{max} \log_2{p_\text{max}}\} +
    |\mathcal{P}_\mathcal{J}| p_\text{max}),
\end{align*}
that can be compacted as
\begin{align}\label{eq:complexity-UpdateEPS}
    O(|\mathcal{P}_\mathcal{J}| p_\text{max} \max\{\log_2{\Omega},\, p_\text{max} \log_2{p_\text{max}})\}.
\end{align}

\begin{algorithm}[!t]
\begin{tabularx}{\textwidth}{ c X }
    \textbf{Input} & A feasible schedule $\mathcal{S}$ for a BPMSTP instance $\mathcal{D}(\hat{K})$ as in \eqref{eq:instance-hat}. \\
    \textbf{Output} & A feasible schedule $\mathcal{S}'$ for $\mathcal{D}(\hat{K})$, with 
    $C^\text{max}(\mathcal{S}') \le C^\text{max}(\mathcal{S})$
    and $E(\mathcal{S}') \le E(\mathcal{S})$.\\
\hline
\end{tabularx}
\begin{algorithmic}[1]
\State Let $\mathcal{L}^\text{J}, \mathcal{L}^\text{I} \leftarrow$ FindEPS$(\mathcal{S}, \mathcal{H}, \mathcal{T}, \mathcal{P}_\mathcal{J})$ 
\For{each $p \in \mathcal{P}_\mathcal{J}$}
    \State Build the maps 
    $\mathcal{M}^\text{J}_p : \mathcal{H} \times \mathcal{T} \rightarrow \mathcal{L}^\text{J}_p$ and
    $\mathcal{M}^\text{I}_p : \mathcal{H} \times \mathcal{T} \rightarrow \mathcal{L}^\text{I}_p$
\EndFor
\State Let $\mathcal{M}^\text{J} \leftarrow \{\mathcal{M}_p^\text{J}, p \in \mathcal{P}_\mathcal{J}\}$, 
and $\mathcal{M}^\text{I} \leftarrow \{\mathcal{M}_p^\text{I}, p \in \mathcal{P}_\mathcal{J}\}$
\State {Let $\hat{P}$ be the list of the elements $\mathcal{P}_\mathcal{J}$ sorted in non-increasing order}
\Repeat
    \State Let Improvement $\leftarrow$ false
    \For{$p \in \hat{P}$}
        \For{$\mathcal{E}^\text{J} : ((h^\text{J}, \min_{t \in \mathcal{E}^\text{J}}{t}), \mathcal{S}_{\mathcal{E}^\text{J}})\in \mathcal{M}^\text{J}_p$}
            \For{$\mathcal{E}^\text{I} : ((h^\text{I}, \min_{t \in \mathcal{E}^\text{I}}{t}), \mathcal{S}_{\mathcal{E}^\text{I}}) \in \mathcal{M}^\text{I}_p$}
                \State $\mathcal{S}', \delta $\,\,\,$\leftarrow$ EvaluateEPSMove$(\mathcal{S}, \mathcal{E}^\text{J}, \mathcal{E}^\text{I}, h^\text{J}, h^\text{I})$
                \If{$\delta < 0$}
                    \State Let $\mathcal{S} \leftarrow \mathcal{S}'$
                    \State $\mathcal{L}^\text{J}, \mathcal{L}^\text{I}, \mathcal{M}^\text{J}, \mathcal{M}^\text{I} \leftarrow$
                    UpdateEPS$(\mathcal{S}, \mathcal{L}^\text{J}, \mathcal{L}^\text{I}, \mathcal{M}^\text{J}, \mathcal{M}^\text{I}, h^\text{J}, \hat{\mathcal{T}}^\text{J}, 
                    \mathcal{P}_\mathcal{J})$
                    \State $\mathcal{L}^\text{J}, \mathcal{L}^\text{I}, \mathcal{M}^\text{J}, \mathcal{M}^\text{I} \leftarrow$
                    UpdateEPS$(\mathcal{S}, \mathcal{L}^\text{J}, \mathcal{L}^\text{I}, \mathcal{M}^\text{J}, \mathcal{M}^\text{I}, h^\text{I}, \hat{\mathcal{T}}^\text{I}, 
                    \mathcal{P}_\mathcal{J})$
                    \State Improvement $\leftarrow$ true
                    \State \textbf{break}
                \EndIf
            \EndFor
        \EndFor
    \EndFor
\Until{\textbf{not} Improvement}
\State $\mathcal{S}' \leftarrow \mathcal{S}$
\State \textbf{return} $\mathcal{S}'$
\caption{Exchange-Search (ES)}
\label{algorithm:ES}
\end{algorithmic}
\end{algorithm}

Let us finally describe the pseudo-code of Algorithm \ref{algorithm:ES}.
Line $1$ uses FindEPS in order to identify the EPS's in $\mathcal{S}$. 
Then, lines $2$--$4$, for each $p \in \mathcal{P}_\mathcal{J}$, compute the maps $\mathcal{M}_p^\text{J}$ and $\mathcal{M}_p^\text{I}$.  
Line $5$ instead computes $\mathcal{M}^\text{J}$ and $\mathcal{M}^\text{I}$, and
line $6$ builds the list $\hat{P}$ from $\mathcal{P}_\mathcal{J}$ as in Algorithm \ref{alg:SGH}.
Afterwards, lines $8$--$22$ are iterated as long as they result in at least an improving EPS move.
Specifically, for each possible move involving an EPS-J and an EPS-I in $\mathcal{S}$ (lines $9$--$11$), ES evaluates it with EvaluateEPSMove at line $12$.
If such move is improving, i.e., $\delta < 0$, line $14$ assigns the resulting schedule $\mathcal{S}'$ to $\mathcal{S}$.
Then, lines $15$ and $16$ update the lists of EPS's and the maps with UpdateEPS.
Line $17$ sets the variable Improvement to true. As a result, at the end of the current iteration of lines $8$--$22$, they are executed again.
Line $18$ interrupts the innermost loop according to the first improvement strategy, so as to search for further improving EPS moves that involve other EPS-J's.
If the last iteration of lines $8$--$22$ is unsuccessful in improving the TEC of $\mathcal{S}$, the loop ends.
Finally, line $25$ returns a feasible schedule $\mathcal{S}'$ that improves, or does not worsen, the TEC and the makespan of $\mathcal{S}$.

Let us discuss the computational complexity of Algorithm \ref{algorithm:ES}.
Since there is a total of $M \hat{K}$ keys, 
lines $2$--$4$ take $O(|\mathcal{P}_\mathcal{J}| M \hat{K})$.
Line $5$ takes $O(\mathcal{P}_\mathcal{J})$, while line $6$ takes $O(|\mathcal{P}_\mathcal{J}| \log_2{|\mathcal{P}_\mathcal{J}|})$.
Let us consider lines $7$--$23$.
First, observe that an EPS-J of cardinality $p \in \mathcal{P}_\mathcal{J}$ can be involved in an EPS move with up to $M (\hat{K} - p + 1)$ EPS-I's. 
Therefore, since there are $N$ distinct EPS-J's, one for each job, the maximum number of EPS moves to be evaluated is $O(N M \hat{K})$, that is, lines $12$--$19$ are executed $O(N M \hat{K})$ times.  
However, observe that, generally the higher $N$ and $\sum_{p \in \mathcal{P}_\mathcal{J}}{p}$, the lower the number of idle slots, hence the smaller the set of EPS-I's. 
Moreover, for each EPS-J $\mathcal{E}$, ES interrupts the innermost loop (lines $11$--$20$) as soon as an improving EPS move for an EPS-J $\mathcal{E}$, and then does not generally consider each EPS move involving $\mathcal{E}$.
Hence, $O(N M \hat{K})$ is not the tightest bound for the number of EPS-I's, and the actual performances may be better than the computational complexity suggests.

The number of iterations of lines $7$--$23$ depends on the structure of the solution $\mathcal{S}$, and on the instance $\mathcal{D}(\hat{K})$ as well.
In order to simplify the analysis without losing generality, we suppose that lines $7$--$23$ can be executed at most $R > 0$ times. 
The value of $R$ depends both on the instance of the problem, and on the schedule $\mathcal{S}$ given as input.
\newline
The determination of $R$ is out of the scope of this thesis. However, $R$ has a very small upper bound in practice. 
In fact, experimental observations related to the numerical results presented in Chapter \ref{chap:exp-tests} suggest that $R$ is generally two orders of magnitude lower than $\hat{K}$ on the considered benchmark instances.

\noindent
Finally, 
in light of these observations, the computational complexity of ES is
\begin{align}\label{eq:complexity-ES-uncompacted}
\begin{split}
    & O(|\mathcal{P}_\mathcal{J}|| M \hat{K} \max\{\log_2{\Omega}, p_\text{max} \log_2{p_\text{max}}\} +
    O(|\mathcal{P}_\mathcal{J}| M \hat{K}) + O(|\mathcal{P}_\mathcal{J}| \log_2{|\mathcal{P}_\mathcal{J}|}) + \\
    & O(R N M \hat{K} (p_\text{max}^2 \log_2{\Omega} + 
    |\mathcal{P}_\mathcal{J}| p_\text{max} \max\{\log_2{\Omega}, p_\text{max} \log_2{p_\text{max}}\})).
\end{split}
\end{align} 
The terms in the sum correspond to the complexity of line $1$, as in \eqref{eq:complexity-FindEPS}, lines $2$--$4$, lines $5$--$6$, and to the main loop at lines $7$--$23$, respectively. 
Specifically, as regards the complexity of lines $7$--$23$, the first summand is due to line $12$, and given by \eqref{eq:complexity-EvaluateEPSMove}, while the second summand is due to lines $15$ and $16$, and given by \eqref{eq:complexity-UpdateEPS}.
Expression \eqref{eq:complexity-ES-uncompacted} can be equivalently and more compactly expressed as
\begin{align*}
    O(R N M \hat{K} p_\text{max}^2 (\log_2{\Omega} + p_\text{max}\log_2{p_\text{max}})
    )
\end{align*}
As a computational remark, let us observe that if the schedules were implemented as ordered lists instead of ordered sets, the computational complexity for scheduling each of the $O(p_\text{max})$ jobs in an EPS would be $O(\Omega)$ instead of $O(\log_2{\Omega})$. 
In fact, in this case, each job would require $O(\log_2{\Omega})$ for binary search to identify the insertion point, and then take $O(\Omega)$ for shifting the elements after it.

An example may be useful to illustrate the update operations performed by ES.
\begin{example}\normalfont
    \begin{figure}[b]
    \captionsetup{font=footnotesize}
    \centering
    \includegraphics[scale=0.476]{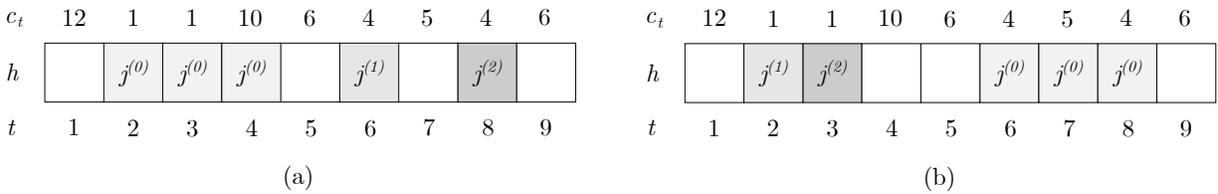}
    \caption{An example of an EPS move performed by ES.
    Such move involves the EPS-J \{2, 3, 4\} and the EPS-I \{6, 7, 8\} in the schedule in Figure \ref{fig:EPS-move-ES}(a).
    The resulting schedule is shown in Figure \ref{fig:EPS-move-ES}(b).
    }
    \label{fig:EPS-move-ES}
    \end{figure}
    Let us consider an example of the application of ES on a schedule, say, $\mathcal{S}$, on a single machine $h$ with $u_h = 1$, depicted in Figure \ref{fig:EPS-move-ES}(a).
    Figure \ref{fig:EPS-move-ES}(b) shows the resulting schedule $\mathcal{S}'$.
    ES performs only one improving EPS move, that leads to a TEC improvement from $E(\mathcal{S}) = 20$ to $E(\mathcal{S}') = 15$.
    
    Let us describe the distinct, possibly improving, EPS moves for $\mathcal{S}$.
    First, the EPS-J's in $\mathcal{S}$ are the ones in
    $\mathcal{L}_3^\text{J}(\mathcal{S}) = 
    \{\{2, 3, 4\}\}$ and
    $\mathcal{L}_1^\text{J}(\mathcal{S}) = \{\{6\}, \{8\}\}$.
    Similarly, 
    \begin{gather*}
        \mathcal{L}_1^\text{I}(\mathcal{S}) = \{\{1\}, \{5\}, \{7\}, \{9\}\}, \quad
        \mathcal{L}_2^\text{I}(\mathcal{S}) = \{\{5, 6\}, \{6, 7\}, \{7, 8\}, \{8, 9\}\}, \\
        \mathcal{L}_3^\text{I}(\mathcal{S}) = \{\{5, 6, 7\}, \{6, 7, 8\}, \{7, 8, 9\}\}.
    \end{gather*}
    Let $\mathcal{E}^{(0)}$ be the EPS-J $\{2, 3, 4\}$ in $\mathcal{L}_3^\text{J}(\mathcal{S})$, and
    let also 
    \begin{align*}
        \mathcal{E}^{(1)} = \{5,6,7\}, \quad
        \mathcal{E}^{(2)} = \{6,7,8\}, \quad 
        \mathcal{E}^{(3)} = \{7,8,9\}
    \end{align*}
    be the EPS-I's in $\mathcal{L}_3^\text{I}(\mathcal{S})$. 
    
    Let us consider the possible EPS moves involving $\mathcal{E}^{(0)}$.
    First,
    \begin{align*}
        E(\mathcal{S}_{\mathcal{E}^{(0)}}) = 12, \quad
        E(\mathcal{S}_{\mathcal{E}^{(1)}}) = 4, \quad
        E(\mathcal{S}_{\mathcal{E}^{(2)}}) = 8, \quad
        E(\mathcal{S}_{\mathcal{E}^{(3)}}) = 4.
    \end{align*}
    Moreover, 
    \begin{align*}
        \eta^\text{ub}_h(\mathcal{E}^{(1)})=15, \quad
        \eta^\text{ub}_h(\mathcal{E}^{(2)})=13,  \quad
        \eta^\text{ub}_h(\mathcal{E}^{(3)})=15,
    \end{align*}
    and the cumulative costs of the slots in $\mathcal{E}^{(0)}$ in non-decreasing order of their costs are
    \begin{align*}
        \eta^\text{lb}_{h, 1}(Q_{\mathcal{E}^{(0)}}) = 1, \quad 
        \eta^\text{lb}_{h, 2}(Q_{\mathcal{E}^{(0)}}) = 2, \quad
        \eta^\text{lb}_{h, 3}(Q_{\mathcal{E}^{(0)}}) = 12.
    \end{align*}
    The necessary condition \eqref{eq:EPS-move-lb} for an improving move with $\mathcal{E}^{(0)}$ is only satisfied by $\mathcal{E}^{(2)}$ among the EPS's in $\mathcal{L}_3^\text{I}(\mathcal{S})$, since
    \begin{align*}
        \eta^\text{ub}_h(\mathcal{E}^{(2)}) + \eta_{h, 2}^\text{lb}(Q_{\mathcal{E}^{(0)}}) = 15 < E(\mathcal{S}_{\mathcal{E}^{(2)}}) + E(\mathcal{S}_{\mathcal{E}^{(0)}}) = 20,
    \end{align*}
    but 
    \begin{gather*}
        \eta^\text{ub}_h(\mathcal{E}^{(1)}) + \eta_{h, 1}^\text{lb}(Q_{\mathcal{E}^{(0)}}) = 16, \quad
        \eta^\text{ub}_h(\mathcal{E}^{(3)}) + \eta_{h, 1}^\text{lb}(Q_{\mathcal{E}^{(0)}}) = 16
    \end{gather*}
    are no greater than 
    \begin{align*}
        E(\mathcal{S}_{\mathcal{E}^{(1)}}) + E(\mathcal{S}_{\mathcal{E}^{(0)}}) = 16,
        \quad
        E(\mathcal{S}_{\mathcal{E}^{(3)}}) + E(\mathcal{S}_{\mathcal{E}^{(0)}}) = 16,
    \end{align*}
    respectively.
    The EPS move involving $\mathcal{E}^{(0)}$ and $\mathcal{E}^{(2)}$ is indeed the only one that entails an improvement in $TEC$.
    \newline
\end{example}

\begin{figure}[!t]
    \captionsetup{font=footnotesize}
    \centering
    \includegraphics[scale=0.525]{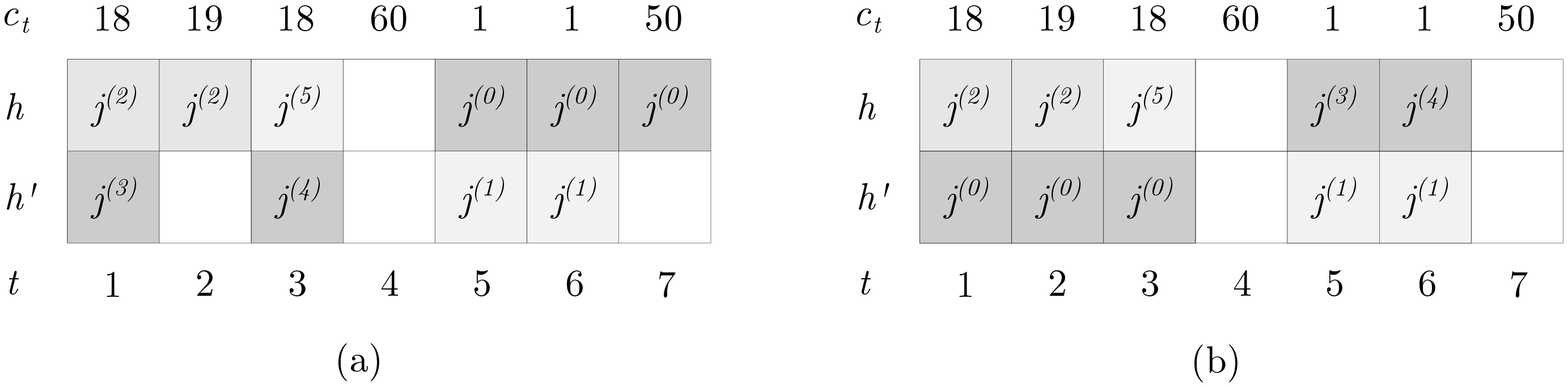}
    \caption{An example of the application of ES.
    There is only one improving EPS move, that involves the EPS-J $\{5, 6, 7\}$ on machine $h$ and the EPS-I $\{1, 2, 3\}$ on $h'$ in Figure~\ref{fig:ES-example}(a).
    Figure~\ref{fig:ES-example}(b) shows the resulting schedule.
    }
    \label{fig:ES-example}
\end{figure}

Another example may be helpful in highlighting why ES can be effective in improving solutions. 
\begin{example}\normalfont
    Figure \ref{fig:ES-example}(a) shows a feasible schedule, say, $\mathcal{S}$, possibly generated by SGH, for two machines $h$ and $h'$, with $u_h = u_{h'} = 1$, 
    and six jobs $j^{(0)}$, $j^{(1)}$, $j^{(2)}$, $j^{(3)}$, $j^{(4)}$ and $j^{(5)}$, with $p_{j^{(0)}} = 3$, $p_{j^{(1)}} = p_{j^{(2)}} = 2$, and $p_{j^{(3)}} = p_{j^{(4)}} = p_{j^{(5)}} = 1$. 
    The makespan and the TEC of $\mathcal{S}$ are $C^\text{max}(\mathcal{S}) = 7$ and $E(\mathcal{S}) = 145$, respectively.
    Figure \ref{fig:ES-example}(b) shows the result of the application of ES to $\mathcal{S}$. 
    ES performs only one improving EPS move, i.e., the one that involves the EPS-J $\mathcal{E}^\text{J} = \{5, 6, 7\}$ on machine $h$, and the EPS-I $\mathcal{E}^\text{I} = \{1, 2, 3\}$ on machine $h'$. 
    After the EPS swap that is part of such move, ES performs the EPS rearrangement of $\mathcal{E}^\text{J}$ by scheduling $j^{(3)}$, $j^{(4)}$ with SGH. 
    The resulting schedule $\mathcal{S}'$ improves the TEC, as $E(\mathcal{S}') = 114 < E(\mathcal{S})$. 
    Incidentally, $\mathcal{S}$ also improves the makespan, as
    $C^\text{max}(\mathcal{S}') = 6 < C^\text{max}(\mathcal{S})$.
\end{example}

\section{Split-Greedy Scheduler}
\label{sec:SGS}

This section describes SGS, the first complete heuristic scheme for the BPMSTP presented in this thesis.

SGS exploits the $\epsilon$-constraint paradigm \citep{Haimes_epsilon} for multi-objective optimization, similarly to CH.
First, let us observe that, for a given BPMSTP instance $\mathcal{I}$, each Pareto-optimal solution $\mathcal{S}^\star$ to $\mathcal{I}$ corresponds to a non-dominated point $(C^\text{max}(\mathcal{S}^\star),$ $E(\mathcal{S}^\star))$ in the optimal Pareto front.
In particular, there exist at most $K - \underline{K}(\mathcal{I}) + 1$ different Pareto-optimal solutions, where
\begin{align}
    & \underline{K}(\mathcal{I}) = \max\left\{ \lfloor \: \sum_{j \in \mathcal{J}} p_j / M \rfloor,
    \max_{j \in \mathcal{J}}\{p_j\}\right\}. \label{eq:K-lb}
\end{align}
Indeed, since the processing times $p_j$, $j \in \mathcal{J}$, are integer numbers, $C^\text{max}(\mathcal{S}^\star)$ ranges between the lower bound $\underline{K}(\mathcal{I})$ and the upper bound $K$.
Observe that the lower bound \eqref{eq:K-lb} used by SGS is stronger than the bound $K^{min} = \sum_{j \in J}p_j / M$ in CH.

SGS computes a set of non-dominated, heuristic solutions by solving, for each $\hat{K}$ such that $\underline{K}(\mathcal{I}) \le \hat{K} \le K$, the BPMSTP instance $\mathcal{D}(\hat{K})$ as in \eqref{eq:instance-hat} with SGH.
The pseudo-code of SGS is reported in Algorithm \ref{alg:SGS}.
At line $1$, Algorithm \ref{alg:SGS} initializes $\underline{K}(\mathcal{I})$ as in \eqref{eq:K-lb}.
Then, at line $2$, it initializes the set of solutions $\mathcal{F}$ as an empty set. It also initializes the iteration variable $\hat{K}$, used as the upper bound for the makespan in the loop at lines $3$--$11$, as $K$.
At lines $3$--$11$, Algorithm \ref{alg:SGS} computes the set of solutions $\mathcal{F}$.
Algorithm \ref{alg:SGS} first evaluates the condition $\hat{K} \ge \underline{K}$ at line $3$. If it does not hold, i.e., the current maximum makespan $\hat{K}$ is not greater or equal than the lower bound $\underline{K}$, then there is no feasible solution for $\hat{K}$.
Afterwards, at line $5$, it solves an instance $\mathcal{D}(\hat{K})$ as in \eqref{eq:instance-hat} for the BPMSTP by means of SGH.
If $\mathcal{S}$ is \addRR{infeasible}, then the loop ends at line $7$. Observe that, as opposed to CH, Algorithm \ref{alg:SGS} verifies the feasibility of the computed solution before proceeding.
If $\mathcal{S}$ is feasible, then \addRR{$\mathcal{S}$} is added to $\mathcal{F}$ at line $9$, and $\hat{K}$ is updated at line $10$ for the next iteration.
At line $12$, Algorithm \ref{alg:SGS} finally returns the set of non-dominated heuristic solutions in $\mathcal{F}$.

The computational complexity of Algorithm \ref{alg:SGS} is dominated by SGH at line $5$, which occurs $O(K)$ times during the execution of the algorithm.
Hence, the complexity of Algorithm \ref{alg:SGS} is
\begin{align*}
    O(  K(
        \Omega M |\mathcal{P}_\mathcal{J}| \log_2{\Omega} +
        \Omega N \log_2{\Omega} + 
        NM +
        |\mathcal{P}_\mathcal{J}| \log_2{\mathcal{P}_\mathcal{J}}
        )
    ).
\end{align*}

\begin{algorithm}[t]
\begin{tabularx}{\textwidth}{ c X }
    \textbf{Input} 
    & An instance $\mathcal{I}$ of the BPMSTP.\\
    \textbf{Output} & A set of non-dominated heuristic solutions \addRR{for $\mathcal{I}$}.\\
    \hline
\end{tabularx}
\begin{algorithmic}[1]
    \State $\underline{K} \leftarrow max\{\floor{\sum_{j \in J}~{p_j / M}}, max_{j \in J}\{p_j\}\}$
    \State Let $\mathcal{F} \leftarrow \emptyset$, and $\hat{K} \leftarrow K$
    \While{$\hat{K} \ge \underline{K}$}
        \State Let $\mathcal{D}(\hat{K})$  be an instance as in \eqref{eq:instance-hat}
        \State $\mathcal{S} \leftarrow SGH(\mathcal{D}(\hat{K}))$
        \If{$\mathcal{S}$ is an empty schedule}~{\quad // Checks if $\mathcal{S}$ is \addRR{infeasible}}
            \State \textbf{break}
        \EndIf
        \State \addRR{Set $\mathcal{F} \leftarrow \mathcal{F} \cup \{\mathcal{S}\}$}
        \State $\hat{K} \leftarrow \hat{K} - 1$
    \EndWhile
    \State \textbf{return} the set of non-dominated solutions in $\mathcal{F}$
    \caption{Split-Greedy Scheduler}
    \label{alg:SGS}
    \end{algorithmic}
\end{algorithm}

As a final remark, observe that the lower bound $\underline{K}(\mathcal{I})$ given by \eqref{eq:K-lb} is not tight for all the instances of the BPMSTP, and generally for the problem $Pm | | C^\text{max}$ as well.

\begin{example}
\normalfont
Let us consider a BPMSTP instance with a set of $N = 4$ jobs with processing times $p_1 = 2$, $p_2 = p_3 = 9$, and $p_4 = 10$, to be scheduled on $M = 3$ machines with $u_1 = u_2 = u_3 = 1$, and a number $K = 10$ of time slots.
The lower bound for $C^\text{max}$ given by \eqref{eq:K-lb} is equal to $10$, but there is no feasible solution with such makespan. In fact, jobs $2$, $3$, and $4$ have to be scheduled on three different machines, without leaving two adjacent time slots for job $1$.
\end{example}


SGS can be enhanced by improving the solution $\mathcal{S}$, computed at line $5$ with SGH, by means of ES before line $9$.
Let us refer to such algorithm as Split-Greedy Scheduler with Exchange Search (SGS-ES).

\section{Exact Algorithm}

The exact algorithm heavily relies on the mathematical models described in Chapter \ref{chap:problem}, and it exploits the $\epsilon$-constraint paradigm similarly to SGS (Section \ref{sec:SGS}).
For the sake of compactness, this section describes the exact algorithm with \hyperlink{F2}{Formulation 2}. The algorithm can be modified to work with \hyperlink{F1}{Formulation 1} with minor changes.

First, let us define the \emph{reduced formulation} of the BPMSTP as the optimization of \eqref{eq:model2:TEC} subject to constraints \eqref{eq:model2:E}-\eqref{eq:model2:NoOverl} and \eqref{eq:model2:belong}.
In other words, the reduced formulation only requires the minimization of the TEC (the makespan is discarded) without considering constraints \eqref{eq:model2:Cj} and \eqref{eq:model2:Cmax-ub} that are related to the makespan.

\begin{algorithm}[tb]
\begin{tabularx}{\textwidth}{lX}
\textbf{Input:} & A BPMSTP instance $\mathcal{I}$.
\\
\textbf{Output:} & The set of Pareto-optimal solutions for $\mathcal{I}$.
\\
\hline
\end{tabularx}
\begin{algorithmic}[1]
    \State Let $\mathcal{O} \leftarrow \emptyset$ \label{alg:optimal-algorithm:O-decl}
    \State Let $\hat{K} \leftarrow K$
    \label{alg:optimal-algorithm:Khat-decl}
    \While{$\hat{K} \ge \underline{K}(\mathcal{I})$}
    \label{alg:optimal-algorithm:start-while}
        \State Solve the reduced formulation of $\mathcal{D}(\mathcal{I}, \hat{K})$ with MILP \label{alg:optimal-algorithm:reduced-formulation-solution}
        \If{no feasible solution exists}
            \State \textbf{break} \label{alg:optimal-algorithm:break}
        \EndIf
        \State Let $\mathcal{S}^\star$ be the schedule computed with Algorithm \ref{alg:fromYtoS} from the solution of $\mathcal{D}(\mathcal{I}, \hat{K})$
        \label{alg:optimal-algorithm:fromYtoS}
        \State Update $\mathcal{O} \leftarrow \mathcal{O} \,\cup\, $\addRR{$\{\mathcal{S}^\star\}$}
        \label{alg:optimal-algorithm:solution-set-update}
        \State $\hat{K} \leftarrow C^\text{max}(\mathcal{S}^\star) - 1$
        \label{alg:optimal-algorithm:hatk-update}
    \EndWhile
    \label{alg:optimal-algorithm:end-while}
    \State \textbf{return} 
    $\mathcal{O}$
\caption{Exact algorithm for the BPMSTP}
\label{alg:optimal-algorithm}
\end{algorithmic}
\end{algorithm}

Algorithm \ref{alg:optimal-algorithm} reports the pseudo-code for an exact algorithm for the BPMSTP.
Algorithm \ref{alg:optimal-algorithm} takes a BPMSTP instance $\mathcal{I}$ as input, and returns the set of the optimal solutions for $\mathcal{I}$.
The algorithm first initializes the solution set $\mathcal{O}$ at line \ref{alg:optimal-algorithm:O-decl} and the parameter $\hat{K}$ at line \ref{alg:optimal-algorithm:Khat-decl}. The latter is used in the downsized instances within the subsequent loop.
Then, it repeats lines \ref{alg:optimal-algorithm:start-while}--\ref{alg:optimal-algorithm:end-while} until either $\hat{K}$ is lower than the lower bound $\underline{K}(\mathcal{I})$ or an \addRR{infeasible} solution is obtained before reaching $\underline{K}(\mathcal{I})$.
In more detail, Algorithm \ref{alg:optimal-algorithm} solves the reduced formulation associated with the downsized instance $\mathcal{D}(\mathcal{I}, \hat{K})$ with MILP at line \ref{alg:optimal-algorithm:reduced-formulation-solution}.
Then, if no feasible solution exists, the loop is stopped at line \ref{alg:optimal-algorithm:break}.
Otherwise, Algorithm \ref{alg:optimal-algorithm} calls Algorithm \ref{alg:fromYtoS} to obtain a representation of the optimal solution of $\mathcal{D}(\mathcal{I}, \hat{K})$ as a feasible schedule at line \ref{alg:optimal-algorithm:fromYtoS}.
Afterwards, it adds the new solution \addRR{$\mathcal{S}^\star$} to the \addRR{set of solutions $\mathcal{O}$} at line \ref{alg:optimal-algorithm:solution-set-update}.
\addRR{O}bserve that, after line \ref{alg:optimal-algorithm:solution-set-update}, any solution $\mathcal{S}'$ to $\mathcal{D}(\mathcal{I}, \hat{K})$ with makespan $C^\text{max}(\mathcal{S}')$ such that $C^\text{max}(\addRR{\mathcal{S}^\star}) \le C^\text{max}(\mathcal{S}') \le \hat{K}$ is either equivalent to or dominated by $\addRR{\mathcal{S}^\star}$.
Hence, at line \ref{alg:optimal-algorithm:hatk-update}, Algorithm \ref{alg:optimal-algorithm} updates the number of slots $\hat{K}$ for the next iteration as $C^\text{max}(\addRR{\mathcal{S}^\star}) - 1$.
Finally, it returns the set \addRR{$\mathcal{O}$} of Pareto optimal solutions at the end of the loop.

The computational efficiency of step \ref{alg:optimal-algorithm:reduced-formulation-solution} in Algorithm \ref{alg:optimal-algorithm} can be enhanced by providing an initial feasible solution for the MILP solver computed by an ad-hoc heuristic, such as SGS or SGS-ES, described in the previous section.
The advantages of such choice will be investigated in Chapter \ref{chap:exp-tests} in comparison to the general-purpose heuristics for initialization used in commercial MILP solvers.

%% file: chap4.tex
\chapter{Numerical results}
\label{chap:exp-tests}

This \addRR{chapter} reports the results of the tests aimed at experimentally evaluating the algorithms described in Chapter \ref{chap:algorithms}.
The tests were motivated by the \addRR{goals of assessing}:
\begin{enumerate}
    \item the performances of SGH with respect to CH;
    \item the impact of ES on the quality of the solutions computed by SGH;
    \item the performances of SGS-ES with respect to other state-of-the-art algorithms;
    \item the relative performances of the exact algorithm with the two mathematical models;
    \item the impact of the initial solution provided by SGH on the computational efficiency of the exact algorithm;
    \item the performances of SGS-ES with respect to the exact algorithm with \hyperlink{F2}{Formulation 2}.
\end{enumerate}

The first goal is essential in experimentally evaluate SGH with respect to the state-of-the-art for the BPMSTP.
The second goal \addRR{is important} to check whether the improvement in solutions quality enabled by the application of ES justifies its computational times.
The third goal is a fundamental step towards establishing, from an experimental standpoint, SGS-ES as a state-of-the-art heuristic scheme for the BPMSTP.
The fourth goal, instead, deals with the the exact algorithm. Specifically, it is focused on highlighting the higher computational efficiency of \hyperlink{F2}{Formulation 2} with respect to \hyperlink{F2}{Formulation 1} as a part of the MILP solving step.
The fifth goal again focuses again on the exact algorithm, by further enhancing its computational efficiency with the use of SGH to find the initial feasible solution.
Finally, the sixth goal is \addRR{fundamental in} evaluating the optimality gap of the solutions provided by SGS-ES.
Moreover, a comparison of the computational times achieved by SGS-ES and the exact algorithm allows \addRR{us }to investigate the impact of the trade-off between the solutions quality and the computational efficiency.

This chapter is organized as follows.
Section~\ref{sec:instances} describes the benchmark instances.
Then, Section~\ref{sec:implementation} provides details on the algorithms implementation, while Section~\ref{sec:performance-metrics} describes the performance metrics used to evaluate the algorithms performances.
Section~\ref{sec:CH-evaluation} experimentally evaluates the impact of the shortcomings of CH, described in Section \ref{sec:SGH} of Chapter \ref{chap:algorithms}.
Section~\ref{sec:ES-evaluation} compares SGS with SGS-ES in order to evaluate the introduction of ES in the heuristic scheme. This section concludes that the quality--computational time trade-off of introducing ES in the heuristic scheme is very favorable for all the instances.
Then, Section~\ref{sec:SGSESvsNSGAMOEADCH} compares the performances of SGS-ES with the ones achieved by NSGA-III, MOEA/D, and CH. 
Section~\ref{sec:formulations-comparison} compares the two mathematical models described in Chapter~\ref{chap:problem}.
Afterwards, Section~\ref{sec:warmstart} evaluates the impact of the heuristic scheme in speeding up the computational process the exact algorithm by providing an initial feasible solution.
Finally, Section~\ref{sec:SGS-ES-vs-exact-algorithm} shows the relative performances of SGS-ES and the exact algorithm.

The experimental tests in Sections~\ref{sec:CH-evaluation}, \ref{sec:ES-evaluation}, and \ref{sec:SGSESvsNSGAMOEADCH} were carried out on a Windows 10 system equipped with a Intel(R) Core i7-8750H CPU @ 2.20GHz, 6 cores processor, and 16 gigabytes of RAM.
Due to machine unavailability, the tests in Section~\ref{sec:formulations-comparison}, Section~\ref{sec:warmstart}, and Section~\ref{sec:SGS-ES-vs-exact-algorithm} were performed on another machine with a Windows 10 system, equipped with a Intel Core i9-9900K Octa-core 3.6 GHz processor as well as 16 GB of RAM.

\section{Instances}
\label{sec:instances}

\begin{table}[!t]
\captionsetup{font=footnotesize}
\centering
    \scalebox{0.75}{
    \begin{tabular}{cccccccccccccccccc}
    \toprule
    
    \textbf{Instance} & \multicolumn{4}{c}{\textbf{Problem data}}    &  & \textbf{Instance} & \multicolumn{4}{c}{\textbf{Problem data}}        &  & \textbf{Instance} & \multicolumn{4}{c}{\textbf{Problem data}}         \\ 
    
    \midrule
    
             & $N$  & $M$ & $K$  & $|\mathcal{P}_\mathcal{J}|$ &  &          & $N$   & $M$  & $K$   & $|\mathcal{P}_\mathcal{J}|$ &  &          & $N$   & $M$  & $K$   & $|\mathcal{P}_\mathcal{J}|$  \\ 
    
    \midrule
    
    1        & 6  & 3 & 50 & 4                     &  & 31       & 30  & 8  & 100 & 3                     &  & 61       & 250 & 25 & 350 & 12                     \\
    2        & 6  & 3 & 80 & 4                     &  & 32       & 60  & 8  & 100 & 3                     &  & 62       & 250 & 25 & 500 & 12                     \\
    3        & 6  & 5 & 50 & 4                     &  & 33       & 100 & 8  & 100 & 3                     &  & 63       & 300 & 25 & 350 & 12                     \\
    4        & 6  & 5 & 80 & 3                     &  & 34       & 150 & 8  & 100 & 3                     &  & 64       & 300 & 25 & 500 & 12                     \\
    5        & 6  & 7 & 50 & 5                     &  & 35       & 200 & 8  & 100 & 4                     &  & 65       & 350 & 25 & 350 & 12                     \\
    6        & 6  & 7 & 80 & 3                     &  & 36       & 30  & 16 & 100 & 3                     &  & 66       & 350 & 25 & 500 & 12                     \\
    7        & 10 & 3 & 50 & 5                     &  & 37       & 60  & 16 & 100 & 3                     &  & 67       & 400 & 25 & 350 & 12                     \\
    8        & 10 & 3 & 80 & 5                     &  & 38       & 100 & 16 & 100 & 3                     &  & 68       & 400 & 25 & 500 & 12                     \\
    9        & 10 & 5 & 50 & 4                     &  & 39       & 150 & 16 & 100 & 3                     &  & 69       & 500 & 25 & 350 & 12                     \\
    10       & 10 & 5 & 80 & 5                     &  & 40       & 200 & 16 & 100 & 4                     &  & 70       & 500 & 25 & 500 & 12                     \\
    11       & 10 & 7 & 50 & 5                     &  & 41       & 30  & 20 & 100 & 3                     &  & 71       & 250 & 30 & 350 & 12                     \\
    12       & 10 & 7 & 80 & 4                     &  & 42       & 60  & 20 & 100 & 3                     &  & 72       & 250 & 30 & 500 & 12                     \\
    13       & 15 & 3 & 50 & 5                     &  & 43       & 100 & 20 & 100 & 3                     &  & 73       & 300 & 30 & 350 & 12                     \\
    14       & 15 & 3 & 80 & 5                     &  & 44       & 150 & 20 & 100 & 3                     &  & 74       & 300 & 30 & 500 & 12                     \\
    15       & 15 & 5 & 50 & 4                     &  & 45       & 200 & 25 & 100 & 4                     &  & 75       & 350 & 30 & 350 & 12                     \\
    16       & 15 & 5 & 80 & 5                     &  & 46       & 30  & 8  & 300 & 4                     &  & 76       & 350 & 30 & 500 & 12                     \\
    17       & 15 & 7 & 50 & 5                     &  & 47       & 60  & 8  & 300 & 4                     &  & 77       & 400 & 30 & 350 & 12                     \\
    18       & 15 & 7 & 80 & 5                     &  & 48       & 100 & 8  & 300 & 4                     &  & 78       & 400 & 30 & 500 & 12                     \\
    19       & 20 & 3 & 50 & 5                     &  & 49       & 150 & 8  & 300 & 4                     &  & 79       & 500 & 30 & 350 & 12                     \\
    20       & 20 & 3 & 80 & 5                     &  & 50       & 200 & 8  & 300 & 4                     &  & 80       & 500 & 30 & 500 & 12                     \\
    21       & 20 & 5 & 50 & 5                     &  & 51       & 30  & 16 & 300 & 4                     &  & 81       & 250 & 40 & 350 & 12                     \\
    22       & 20 & 5 & 80 & 5                     &  & 52       & 60  & 16 & 300 & 4                     &  & 82       & 250 & 40 & 500 & 12                     \\
    23       & 20 & 7 & 50 & 5                     &  & 53       & 100 & 16 & 300 & 4                     &  & 83       & 300 & 40 & 350 & 12                     \\
    24       & 20 & 7 & 80 & 5                     &  & 54       & 150 & 16 & 300 & 4                     &  & 84       & 300 & 40 & 500 & 12                     \\
    25       & 25 & 3 & 50 & 5                     &  & 55       & 200 & 16 & 300 & 4                     &  & 85       & 350 & 40 & 350 & 12                     \\
    26       & 25 & 3 & 80 & 5                     &  & 56       & 30  & 25 & 300 & 4                     &  & 86       & 350 & 40 & 500 & 12                     \\
    27       & 25 & 5 & 50 & 5                     &  & 57       & 60  & 25 & 300 & 4                     &  & 87       & 400 & 40 & 350 & 12                     \\
    28       & 25 & 5 & 80 & 5                     &  & 58       & 100 & 25 & 300 & 4                     &  & 88       & 400 & 40 & 500 & 12                     \\
    29       & 25 & 7 & 50 & 5                     &  & 59       & 150 & 25 & 300 & 4                     &  & 89       & 500 & 40 & 350 & 12                     \\
    30       & 25 & 7 & 80 & 5                     &  & 60       & 200 & 25 & 300 & 4                     &  & 90       & 500 & 40 & 500 & 12         \\
    
    \bottomrule
    \end{tabular}
    }
\caption{\addRR{For each instance, this table reports the number of jobs $N$, the number of machines $M$, the number of time slots $K$, and the number of distinct processing times $|\mathcal{P}_\mathcal{J}|$.
The maximum processing time $p_\text{max}$ is equal to $|\mathcal{P}_\mathcal{J}|$ for all the instances except for instances 
$1$, $3$, $4$, $6$, $9$, $12$ and $15$, 
where $p_\text{max}$ is equal to $5$.}
\label{table:instances-parameters}
}
\end{table}

The tests were performed on a benchmark that consists of a set of 60 instances proposed by \citet{TimeSch33_wang_bi-objective_2018}, and another set of 30 instances proposed in our work \citep{TimeSch11_Anghinolfi2021ABH}. 

The first set of instances consists of 30 small-scale instances (numbered from 1 to 30), and 30 medium-scale and large-scale instances (numbered from 31 to 60). The latter set is hereinafter referred to MLS instances for short.

The second set of instances contains 30 very large-scale instances, or VLS instances for short (numbered from 61 to 90).
The value of the number of machines $M$, 
the number of jobs $N$, 
and the number of time slots $K$ 
in each instance of the VLS set is given by an element $(M, N, K)$ of the cartesian product between 
$\{25, 30, 40\}$,
$\{250, 300, 350, 400, 500\}$, and
$\{350, 500\}$. 
The processing times $p_j$, $j \in \mathcal{J}$, were randomly drawn from the uniform distribution $U[1,12]$. 
In order to reflect the tendencies of a highly volatile electricity market, 
the values of the consumption rate $u_h$, $h \in \mathcal{H}$,
and the values of the time slot costs $c_k$, $k \in \mathcal{T}$, 
were randomly drawn from the uniform distributions $U[1, 6]$ and $U[1, 8]$, respectively. 
Such choice fostered the generation of smaller time intervals with respect to the first 60 instances.

Table \ref{table:instances-parameters} shows the parameters $N$, $M$, $K$, and \addRR{$|\mathcal{P}_\mathcal{J}|$} for each of the $90$ instances in the benchmark.
For compactness, $p_\text{max}$ is not shown since it is equal to \addRR{$|\mathcal{P}_\mathcal{J}|$} for all the instances, except for instance\addRR{s} 
$1$, $3$, $4$, $6$, $9$, $12$ and $15$,
where $p_\text{max}$ is equal to $5$.
The whole benchmark is available at \textit{https://github.com/ORresearcher/PhD-thesis}.

\section{Implementation}
\label{sec:implementation}

All the algorithms were implemented in Java 16. 
\addRR{In particular, t}he implementation of the exact algorithm also employs the Java CPLEX 20.1.0 API. 
Furthermore, the exact algorithm accepts solutions with an optimal gap that does not exceed $10^{-4}$.

SGS-ES was compared with three distinct state-of-the-art solution approaches: CH, NSGA-III \citep{Deb2014}, and MOEA/D \citep{zhang2007moead}.
CH was first implemented by strictly following the description of \citet{TimeSch33_wang_bi-objective_2018}. Such implementation is referred to as \emph{CH-M}, since it is the Java version of the MATLAB implementation kindly shared by the authors.
Then, CH was reimplemented by building upon CH-M, and correcting the identified shortcomings. 
Such second implementation, called \emph{CH-J}, is the reference CH implementation for the comparisons in this thesis. 

NSGA-III is a genetic algorithm that builds upon NSGA-II \citep{Deb2002}.
NSGA-II was proposed by \citet{Deb2002} as a further step towards a computationally efficient genetic algorithm for multi-objective optimization.
It was used by \citet{TimeSch33_wang_bi-objective_2018} as a reference state-of-the-art algorithm as the benchmark for CH performances.
NSGA-III differs from NSGA-II in the operator used to select the solutions to be included in the population. In fact, in order to ensure the diversity of the solutions, NSGA-III adopts a predefined set of reference points instead of the niche-preservation operator based on the crowding distance exploited by NSGA-II.
The Java implementation of NSGA-III is based on the MATLAB implementation of NSGA-II of \citet{TimeSch33_wang_bi-objective_2018}.
The design of NSGA-II was preserved, i.e., the Java implementation of NSGA-III used the same solution representation, initialization procedure, crossover and mutation operators, and the same parameters setting as well.

MOEA/D is an evolutionary multi-objective algorithm based on a similar idea.
In fact, it defines a set of weight vectors a priori, so that each generated solution is uniquely associated with a single vector to guarantee solutions diversity.
The MOEA/D algorithm presented by \citet{zhang2007moead} was implemented for the BPMSTP by exploiting some parts of NSGA-III design, i.e., the same solution representation, initialization procedure, crossover operator, and termination condition.
In addition, at each iteration, MOEA/D generates the candidate solutions to possibly update the ones currently associated with each weight vector by means of the crossover operator.
This operator is applied to a pair of parent solutions randomly drawn from the set including the $CW$ solutions associated with the closest weight vectors, where the parameter $CW$ is fixed to $10$ as in \citep{zhang2007moead}. 

\section{Performance metrics}
\label{sec:performance-metrics}

This chapter uses two different classes of state-of-the-art performance metrics to evaluate the performances of multi-objective algorithms.
The metrics in the first class evaluate the quality of the Pareto front, while the ones in the second class provide a measure of the distribution uniformity of the non-dominated points in the front \citep{Bandyopadhyay2004}.


Let $\mathcal{F} \subseteq \mathbb{R}^n$ be an $n$-dimensional Pareto front, and 
$\mathcal{F}^\star$ be its reference optimal Pareto front.
In practice, if $\mathcal{F}^\star$ is not known, then it is approximated as the set of the non-dominated solutions in the union of the fronts computed by the algorithms under comparison.
The first class of metrics is constituted by the three following ones:
\begin{itemize}
    \item $D_R$ \citep{Ishibuchi2003}, given by
    \begin{equation*}
        D_R(\mathcal{F}) = \frac{1}{|\mathcal{F}^\star|}
        \sum_{y \in \mathcal{F}^\star} \min\{d_{x, y}:x \in \mathcal{F}\}  
    \end{equation*}
    where $\mathcal{F}^\star$ is the optimal Pareto front, and $d_{x, y}$ is the Euclidean distance between the two points $x$ and $y$ in $\mathcal{F}$ and $\mathcal{F}^\star$, respectively.
    Smaller values of $D_R$ denote higher quality fronts;
    \item \textit{Purity} \citep{Bandyopadhyay2004}, given by 
    \begin{equation*}
        P(\mathcal{F}) = \frac{N_d(\mathcal{F}^\star \cap \mathcal{F})}{N_d(\mathcal{F})},
    \end{equation*}
    where $N_d(\mathcal{F})$ is the number of efficient solutions in $\mathcal{F}$. 
    The value of $P(\mathcal{F})$ is the ratio between the number of non-dominated points in the intersection of the reference Pareto front $\mathcal{F}^\star$ and the front $\mathcal{F}$, and the number of non-dominated points in $\mathcal{F}$. As a consequence, larger values of Purity denote higher quality fronts.
    \item \emph{Hypervolume} \citep{zitzler1999multiobjective,Guerreiro2020TheHI}, that measures the hypervolume covered by $\mathcal{F}$ with respect to a reference point in the objectives space. As such, it can be used to compare two or more fronts by assuming a common reference point. 
    Formally, let $r$ be a reference point in $\mathbb{R}^n$. Then, the Hypervolume value of $\mathcal{F}$ is the measure of the region weakly dominated by $\mathcal{F}$ and bounded above by $r$, i.e.,
    \begin{align*}
        H(\mathcal{F}) = \Lambda(\{
        q \in \mathbb{R}^n : \exists p \in \mathcal{F} : p \le q \le r\}
        \}),
    \end{align*}
    where $\Lambda(\cdot)$ is the Lebesgue measure.
    Larger Hypervolume values denote a better approximation of the optimal Pareto front.
\end{itemize}

Let also $f_k(x)$, $k = 1, \ldots, n$, be the the value of the $k$-th objective function in $\mathcal{F}$ computed for some $x \in \mathcal{F}$.
The second class of metrics is constituted by the two following ones:
\begin{itemize}
    \item \emph{Spacing} \citep{Bandyopadhyay2004}, given by
    \begin{equation*}
        SP(\mathcal{F}) = \sqrt{\frac{1}{N_d(\mathcal{F})}\sum_{i = 1}^{N_d(\mathcal{F})}(\delta_i - \overline{\delta})^2}
    \end{equation*}
    where
    \begin{equation*}
        \delta_i = \min_{j \neq i} \sum_{k = 1}^n |f_k(x_i) - f_k(x_j)|,
    \end{equation*}
    and
    \begin{equation*}
        \overline{\delta} = \frac{1}{N_d(\mathcal{F})-1} \sum_{i = 1}^{N_d(\mathcal{F}) - 1} \delta_i,
    \end{equation*}
    The smaller the values of Spacing, the better the uniformity of the distribution of the solutions in $\mathcal{F}$.
    
    \item \textit{Spread} \citep{Deb2002}, also denoted as $D_2$ \citep{TimeSch33_wang_bi-objective_2018}, given by
    \begin{equation*}
        D_2(\mathcal{F}) = \frac{d_f + d_l + \sum_{i = 1}^{N_d(\mathcal{F}) - 1}|d_i - \overline{d}|}{d_f + d_l + (N_d(\mathcal{F}) - 1) \overline{d}},
    \end{equation*}
    where 
    \begin{equation*}
        d_i = \sqrt{\sum_{k=1}^{n}\big(f_k(x_i)-f_k(x_{i+1})\big)^2},
    \end{equation*}
    being $x_i$ and $x_{i+1}$ two neighboring points in $\mathcal{F}$, and 
    \begin{equation*}
        \overline{d} = \frac{1}{N_d(\mathcal{F})-1}\sum_{k=1}^{n} d_i.
    \end{equation*}
    Finally, $d_f$ and $d_l$ are the Euclidean distances between the extreme solutions and the boundary solutions of the Pareto-optimal solutions in $\mathcal{F}$ \citep{Deb2002}.
    The smaller the values of Spread, the better the uniformity of the distribution of the solutions in $\mathcal{F}$. 
\end{itemize}

Observe that distribution metrics are secondary to quality metrics. 
In fact, distribution metrics should be used as a complementary evaluation metrics for fronts with a comparable degree of quality. 
As an example, let us consider a well distributed front $\mathcal{F}$ that is totally dominated by another front $\mathcal{F}'$ which is instead poorly distributed, but qualitatively higher than $\mathcal{F}$.
In such a case, distribution indices allow \addRR{one }to perform a more accurate comparison, thanks to the better insight on the structure of the Pareto fronts.

When the quality metrics allow \addRR{one }to draw clear conclusions, distribution metrics have limited use in enhancing the comparisons.
In fact, the distribution of the Pareto-optimal points in a front depend on the related problem, and problem instance as well. The structure of such a specific optimal Pareto front cannot be captured by a general-purpose metric that aims at measuring the distance between distinct non-dominated points, regardless of the nature of the problem.

This thesis also considers the \textit{Empirical Attainment Function} (EAF) \citep{dafonseca2001} as a further metric to evaluate performances.
EAF provides a graphical representation of the probability of an algorithm of generating solutions that dominate a given point of the objective space in a single run \citep{minella2011restarted}. 
In particular, the \textit{Differential Empirical Attainment Function} (Diff-EAF) proposed in \citep{lopez2006hybrid} visually shows the differences between a pair of EAFs, by highlighting the regions of the objective space in which an algorithm outperforms another in terms of generated solutions. 
However, differently from the other metrics, EAF and Diff-EAF are only a visual tools for comparison, as they do not provide an overall score.

Finally, the thesis uses two further metrics, $FM_1$ and $FM_2$, in order to evaluate the feasibility of a front generated by an algorithm. Such metrics are specifically introduced to measure the number of \addRR{infeasible} solutions in the Pareto fronts generated by CH.

For a given Pareto front $\mathcal{F}$, $FM_1(\mathcal{F})$ is the percentage of \addRR{infeasible} points in $\mathcal{F}$:
\begin{equation}\label{eq:FM1}
    FM_1(\mathcal{F}) = \frac{|\mathcal{U}(\mathcal{F})|}{|\mathcal{F}|},
\end{equation}
where $\mathcal{U}(\mathcal{F})$ denotes the set of \addRR{infeasible} points of $\mathcal{F}$. 
Instead, $FM_2$ is the average percentage of non-scheduled jobs:
\begin{equation}\label{eq:FM2}
    FM_2(\mathcal{F})=
    \begin{cases}
        \frac{|\mathcal{J}^\text{ns}|}
        {|\mathcal{U}(\mathcal{F})|N} & \text{if}~ FM_1(\mathcal{F})\ne 0,\\
        0  & \text{otherwise},
    \end{cases}
\end{equation}
where $\mathcal{J}^\text{ns}$ is the set of non-scheduled jobs in the \addRR{infeasible} solutions in $\mathcal{F}$.

\section{Evaluating the limits of CH}
\label{sec:CH-evaluation}

This section performs an experimental evaluation of the limits of the implementation of the original version of CH, i.e., CH-M.
Indeed, as discussed in Chapter \ref{chap:algorithms}, CH \citep{TimeSch33_wang_bi-objective_2018} may mistakenly add \addRR{infeasible} solutions in the computed Pareto front.

Table \ref{table:CH-M-FMs} reports the feasibility metrics $FM_1$ and $FM_2$ for each of the fronts computed by CH-M for each the last $60$ of instances.
Let us observe that $13.33\%$ of the computed Pareto front for the MLS instances includes \addRR{infeasible} solutions. Moreover, an average of $1\%$ of the solutions are \addRR{infeasible}, since an average of $2.58\%$ of the jobs are not scheduled.
Let us now consider the results obtained on the set of VLS instances. For $90\%$ of the instances, the computed Pareto fronts include solutions with unscheduled jobs, even though the average numbers of solutions and unscheduled jobs are close to the relative ones for the MLS instances.
In light of these results, we conclude that CH-J is indeed the best implementation of CH to ensure fair comparisons.

\begin{table}[!t]
\footnotesize
\captionsetup{font=footnotesize}
\centering
\scalebox{0.8}{
\begin{tabular}{ccccccc}
\toprule
\textbf{Instance} & \textbf{FM1} & \textbf{FM2} & \textbf{} & \textbf{Instance} & \textbf{FM1} & \textbf{FM2} \\ 

\midrule
31 &  &  &  & 61 & 0.81\% & 2.00\% \\
32 &  &  &  & 62 & 0.58\% & 2.80\% \\
33 &  &  &  & 63 & 0.83\% & 1.50\% \\
34 &  &  &  & 64 & 0.85\% & 3.00\% \\
35 &  &  &  & 65 & 0.41\% & 0.57\% \\
36 &  &  &  & 66 &  &  \\
37 &  &  &  & 67 & 1.81\% & 3.50\% \\
38 &  &  &  & 68 & 1.17\% & 3.06\% \\
39 &  &  &  & 69 & 0.51\% & 1.60\% \\
40 &  &  &  & 70 & 1.23\% & 2.40\% \\
41 &  &  &  & 71 & 1.28\% & 2.00\% \\
42 &  &  &  & 72 & 0.78\% & 1.40\% \\
43 & 1.54\% & 3.00\% &  & 73 & 1.72\% & 3.42\% \\
44 &  &  &  & 74 &  &  \\
45 &  &  &  & 75 & 1.73\% & 3.93\% \\
46 &  &  &  & 76 & 1.44\% & 4.14\% \\
47 &  &  &  & 77 & 0.83\% & 3.25\% \\
48 &  &  &  & 78 & 0.55\% & 4.25\% \\
49 & 0.95\% & 2.67\% &  & 79 & 1.30\% & 2.47\% \\
50 &  &  &  & 80 & 1.37\% & 2.00\% \\
51 &  &  &  & 81 & 1.72\% & 3.40\% \\
52 &  &  &  & 82 & 0.31\% & 0.80\% \\
53 &  &  &  & 83 & 0.43\% & 1.00\% \\
54 & 0.79\% & 2.67\% &  & 84 & 0.30\% & 0.33\% \\
55 &  &  &  & 85 & 1.69\% & 3.43\% \\
56 &  &  &  & 86 &  &  \\
57 &  &  &  & 87 & 0.45\% & 0.50\% \\
58 &  &  &  & 88 & 0.61\% & 3.75\% \\
59 & 0.72\% & 2.00\% &  & 89 & 0.43\% & 3.40\% \\
60 &  &  &  & 90 & 0.60\% & 1.80\% \\ 

\midrule

\textbf{Average} & 1.00\% & 2.58\% &  &  & 0.95\% & 2.43\% \\
\textbf{Non-feasible} & 13.33\% &  &  &  & 90.00\% &  \\
\bottomrule
\end{tabular}
}
\caption{The two feasibility metrics $FM_1$ and $FM_2$ applied to results achieved by CH-M on instances $31$--$90$.}
\label{table:CH-M-FMs}
\end{table}

\section{Evaluating Exchange Search}
\label{sec:ES-evaluation}

This section compares SGS and SGS-ES so as to evaluate the computational efficiency of ES with respect to the quality of the computed solutions.

Table \ref{table:SGS-vs-SGS-ES} reports the Hypervolume, Purity, and Spread metrics, as well as the computational times (in seconds), achieved by SGS and SGS-ES on the last $60$ instances, grouped by the MLS and the VLS sets.
In addition, for each metric and each of the two compared algorithms, the third-to-last row shows the average performance, while the second-to-last row reports the number of times the considered algorithm achieved the best performance over the instances in the set.
For each instance, the best values of each metric are highlighted in bold.
The last row reports the probability value obtained with the Friedman non-parametric test. 
If $p$ is smaller than 1\%, then we can reject the null hypothesis that SGS and SGS-ES generate results that do not differ significantly from a statistical standpoint. 

The values reported in Table \ref{table:SGS-vs-SGS-ES} were obtained by averaging the results of the comparisons between each of the 10 runs of SGS with each of the 10 runs of SGS-ES, due to the stochasticity of SGH in both SGS and SGS-ES.
Table \ref{table:SGS-vs-SGS-ES} shows that SGS-ES prevailed in the Hypervolume and Purity metrics on both the sets of instances, whereas SGS got slightly better average values for Spread.
The statistical tests indeed reveal that SGS-ES is significantly better than SGS for Hypervolume and Purity, whereas the results for Spread are statistically comparable.
Furthermore, the comparison of the computational times reveals that SGS-ES is unavoidably more computationally demanding than SGS, since the former algorithm uses ES to improve the quality of the solutions.
However, the average and the maximum computational time required by SGS-ES are acceptable for the MLS instances. As regards the VLS instances, the average time is about 67 seconds, while the maximum time is about 161 seconds. Moreover, the computational times exceeds 100 seconds in only 6 of the VLS instances.
Such computational times can be considered acceptable as well, especially in view of the achieved improvement in the performances.

\begin{landscape}
\begin{table}[t]
\captionsetup{font=footnotesize}
\centering
\caption{Comparison of the results achieved by SGS and SGS-ES on the MLS and the VLS instances based on the Hypervolume, Purity, Spread, and CPU time (s) metrics.}
\label{table:SGS-vs-SGS-ES}
\resizebox{21cm}{!}{
\begin{tabular}{ccccccccccccccccccccccccccc}
\toprule

\textbf{Instance} & & \multicolumn{2}{c}{\textbf{Hypervolume}} &  & \multicolumn{2}{c}{\textbf{Purity}} &  & \multicolumn{2}{c}{\textbf{Spread}} &  & \multicolumn{2}{c}{\textbf{CPU}} &  &
\textbf{Instance} &  & 
\multicolumn{2}{c}{\textbf{Hypervolume}} &  & \multicolumn{2}{c}{\textbf{Purity}} &  & \multicolumn{2}{c}{\textbf{Spread}} &  & \multicolumn{2}{c}{\textbf{CPU}} \\ 

\midrule

&  & \textbf{SGS-ES} & \textbf{SGS} &  & \textbf{SGS-ES} & \textbf{SGS} &  & \textbf{SGS-ES} & \textbf{SGS} &  & \textbf{SGS-ES} & \textbf{SGS} &  & \textbf{} &  & \textbf{SGS-ES} & \textbf{SGS} &  & \textbf{SGS-ES} & \textbf{SGS} &  & \textbf{SGS-ES} & \textbf{SGS} &  & \textbf{SGS-ES} & \textbf{SGS} \\ 

\midrule

31 & & \textbf{0.8813} & 0.8790 & & \textbf{0.9232} & 0.4475 & & \textbf{0.9173} & 0.9292 & \textbf{} & 0.0518 & \textbf{0.0104} & & 61 & & \textbf{0.8198} & 0.8130 & & \textbf{0.9994} & 0.0068 & & 1.0602 & \textbf{1.0461} & & 21.7450 & 0.8068 \\
32 & & \textbf{0.8090} & 0.8082 & & \textbf{0.9742} & 0.8478 & & 0.9482 & \textbf{0.9458} & & 0.0738 & \textbf{0.0181} & & 62 & & \textbf{0.8295} & 0.8220 & & \textbf{1.0000} & 0.0032 & & \textbf{1.0119} & 1.0160 & & 62.3640 & 1.7699 \\
33 & & \textbf{0.7791} & 0.7777 & & \textbf{0.9876} & 0.5517 & & \textbf{0.8728} & 0.8882 & & 0.0937 & \textbf{0.0254} & & 63 & & \textbf{0.7660} & 0.7599 & & \textbf{1.0000} & 0.0078 & & 0.8898 & \textbf{0.8866} & & 25.8207 & 0.8211 \\
34 & & \textbf{0.6960} & 0.6901 & & \textbf{0.9510} & 0.2159 & & 0.7688 & \textbf{0.7570} & & 0.1220 & \textbf{0.0299} & & 64 & & \textbf{0.7862} & 0.7805 & & \textbf{0.9996} & 0.0033 & & 0.9226 & \textbf{0.9076} & & 57.0123 & 1.8385 \\
35 & & \textbf{0.6011} & 0.6009 & & \textbf{1.0000} & 0.9541 & & 0.5066 & \textbf{0.5027} & & 0.0586 & \textbf{0.0251} & & 65 & & \textbf{0.7245} & 0.7116 & & \textbf{1.0000} & 0.0044 & & 0.8744 & \textbf{0.8567} & & 30.4711 & 0.8903 \\
36 & & \textbf{0.8542} & 0.8518 & & \textbf{0.9496} & 0.6613 & & 0.7243 & \textbf{0.7106} & & 0.0810 & \textbf{0.0171} & & 66 & & \textbf{0.8014} & 0.7910 & & \textbf{1.0000} & 0.0075 & & \textbf{0.9823} & 0.9981 & & 75.7831 & 2.0618 \\
37 & & \textbf{0.8709} & 0.8695 & & \textbf{0.9463} & 0.4795 & & 0.8880 & \textbf{0.8757} & & 0.1202 & \textbf{0.0214} & & 67 & & \textbf{0.7581} & 0.7518 & & \textbf{1.0000} & 0.0060 & & 0.8073 & \textbf{0.7831} & & 25.9146 & 0.8793 \\
38 & & \textbf{0.8409} & 0.8399 & & \textbf{0.8638} & 0.4692 & & \textbf{1.1245} & 1.1427 & & 0.1838 & \textbf{0.0317} & & 68 & & \textbf{0.7391} & 0.7266 & & \textbf{0.9998} & 0.0046 & & 0.8607 & \textbf{0.8409} & & 84.8752 & 2.1309 \\
39 & & \textbf{0.7830} & 0.7814 & & \textbf{0.9895} & 0.6211 & & \textbf{0.8684} & 0.8748 & & 0.2712 & \textbf{0.0430} & & 69 & & \textbf{0.7368} & 0.7289 & & \textbf{1.0000} & 0.0188 & & 0.8101 & \textbf{0.8012} & & 25.9769 & 0.9363 \\
40 & & \textbf{0.7827} & 0.7730 & & \textbf{0.9553} & 0.2247 & & \textbf{0.8447} & 0.8456 & & 0.3673 & \textbf{0.0548} & & 70 & & \textbf{0.7508} & 0.7378 & & \textbf{0.9999} & 0.0064 & & 0.8576 & \textbf{0.8521} & & 90.3612 & 2.4123 \\
41 & & \textbf{0.9282} & 0.9282 & & \textbf{1.0000} & 0.9938 & & \textbf{0.8120} & 0.8125 & & 0.0815 & \textbf{0.0161} & & 71 & & \textbf{0.8057} & 0.7987 & & \textbf{1.0000} & 0.0000 & & \textbf{0.9612} & 0.9786 & & 25.2129 & 0.8829 \\
42 & & \textbf{0.8529} & 0.8523 & & \textbf{0.9943} & 0.8179 & & \textbf{0.9458} & 0.9489 & & 0.1289 & \textbf{0.0252} & & 72 & & \textbf{0.8735} & 0.8691 & & \textbf{1.0000} & 0.0000 & & 1.0414 & \textbf{1.0401} & & 65.3719 & 2.0116 \\
43 & & \textbf{0.8613} & 0.8599 & & \textbf{0.8563} & 0.3881 & & \textbf{1.0217} & 1.0239 & & 0.1995 & \textbf{0.0338} & & 73 & & \textbf{0.7955} & 0.7878 & & \textbf{1.0000} & 0.0041 & & 0.9906 & \textbf{0.9669} & & 30.7839 & 0.9326 \\
44 & & \textbf{0.8312} & 0.8301 & & \textbf{0.7902} & 0.4311 & & 1.0712 & \textbf{1.0695} & & 0.3085 & \textbf{0.0481} & & 74 & & \textbf{0.8508} & 0.8456 & & \textbf{0.9992} & 0.0048 & & 1.0319 & \textbf{1.0230} & & 77.8728 & 2.1324 \\
45 & & \textbf{0.7900} & 0.7871 & & \textbf{0.9491} & 0.3861 & & 0.8651 & \textbf{0.8599} & & 0.5005 & \textbf{0.0622} & & 75 & & \textbf{0.7888} & 0.7825 & & \textbf{1.0000} & 0.0043 & & 0.9003 & \textbf{0.8973} & & 34.7806 & 0.9713 \\
46 & & \textbf{0.8242} & 0.8194 & & \textbf{0.9145} & 0.4845 & & \textbf{0.5920} & 0.6045 & & 0.4204 & \textbf{0.1037} & & 76 & & \textbf{0.8508} & 0.8438 & & \textbf{1.0000} & 0.0059 & & 0.9747 & \textbf{0.9655} & & 92.2981 & 2.2562 \\
47 & & \textbf{0.8832} & 0.8811 & & \textbf{0.8538} & 0.4783 & & \textbf{0.7928} & 0.7968 & & 0.7459 & \textbf{0.1384} & & 77 & & \textbf{0.7478} & 0.7404 & & \textbf{0.9999} & 0.0050 & & 0.8656 & \textbf{0.8538} & & 37.2449 & 1.0383 \\
48 & & \textbf{0.8724} & 0.8676 & & \textbf{0.9573} & 0.1964 & & 0.7753 & \textbf{0.7696} & & 0.9804 & \textbf{0.1999} & & 78 & & \textbf{0.7890} & 0.7781 & & \textbf{1.0000} & 0.0058 & & \textbf{0.9621} & 0.9639 & & 107.1587 & 2.4133 \\
49 & & \textbf{0.8025} & 0.7972 & & \textbf{0.9686} & 0.1526 & & 0.8234 & \textbf{0.8118} & & 1.4556 & \textbf{0.2815} & & 79 & & \textbf{0.7464} & 0.7371 & & \textbf{1.0000} & 0.0128 & & 0.8274 & \textbf{0.7982} & & 45.8857 & 1.1171 \\
50 & & \textbf{0.8275} & 0.8226 & & \textbf{0.9864} & 0.3545 & & \textbf{0.6961} & 0.7183 & & 1.8296 & \textbf{0.3628} & & 80 & & \textbf{0.8158} & 0.8099 & & \textbf{1.0000} & 0.0035 & & 0.9324 & \textbf{0.9251} & & 101.0368 & 2.6189 \\
51 & & \textbf{0.8457} & 0.8427 & & \textbf{0.9286} & 0.6523 & & 0.8498 & \textbf{0.8278} & & 0.7754 & \textbf{0.1494} & & 81 & & \textbf{0.8201} & 0.8147 & & \textbf{0.9987} & 0.0062 & & \textbf{0.9468} & 0.9510 & & 37.5365 & 1.0809 \\
52 & & \textbf{0.8657} & 0.8642 & & \textbf{0.8922} & 0.5680 & & \textbf{0.7515} & 0.7907 & & 1.1850 & \textbf{0.1941} & & 82 & & \textbf{0.8646} & 0.8596 & & \textbf{0.9970} & 0.0066 & & \textbf{1.1053} & 1.1441 & & 92.0232 & 2.4982 \\
53 & & \textbf{0.8875} & 0.8864 & & \textbf{0.9411} & 0.4613 & & 0.8235 & \textbf{0.8185} & & 1.6191 & \textbf{0.2666} & & 83 & & \textbf{0.8199} & 0.8135 & & \textbf{0.9992} & 0.0024 & & \textbf{1.0663} & 1.0665 & & 42.3040 & 1.1566 \\
54 & & \textbf{0.8836} & 0.8825 & & \textbf{0.9019} & 0.4204 & & \textbf{0.6813} & 0.7081 & & 2.1862 & \textbf{0.3516} & & 84 & & \textbf{0.8282} & 0.8230 & & \textbf{0.9998} & 0.0041 & & \textbf{0.9831} & 0.9930 & & 113.6161 & 2.5701 \\
55 & & \textbf{0.8445} & 0.8417 & & \textbf{0.9874} & 0.4762 & & 0.8538 & \textbf{0.8484} & & 3.3929 & \textbf{0.4477} & & 85 & & \textbf{0.8237} & 0.8175 & & \textbf{1.0000} & 0.0030 & & \textbf{1.0316} & 1.0450 & & 49.3493 & 1.2331 \\
56 & & \textbf{0.8867} & 0.8852 & & \textbf{0.9792} & 0.7035 & & 0.8028 & \textbf{0.7363} & & 1.1292 & \textbf{0.1762} & & 86 & & \textbf{0.8308} & 0.8248 & & \textbf{0.9999} & 0.0007 & & \textbf{0.9742} & 0.9781 & & 129.1932 & 2.7451 \\
57 & & \textbf{0.7435} & 0.7377 & & \textbf{0.9392} & 0.4743 & & 0.6317 & \textbf{0.6155} & & 1.7694 & \textbf{0.2853} & & 87 & & \textbf{0.8091} & 0.8034 & & \textbf{1.0000} & 0.0003 & & 0.9481 & \textbf{0.9412} & & 52.3694 & 1.2788 \\
58 & & \textbf{0.9057} & 0.9047 & & \textbf{0.9960} & 0.5480 & & \textbf{0.6884} & 0.7099 & & 2.4266 & \textbf{0.3276} & & 88 & & \textbf{0.8372} & 0.8308 & & \textbf{1.0000} & 0.0033 & & \textbf{1.0088} & 1.0124 & & 149.7154 & 2.9976 \\
59 & & \textbf{0.9012} & 0.9002 & & \textbf{0.8873} & 0.4814 & & 0.9665 & \textbf{0.9515} & & 4.6183 & \textbf{0.4209} & & 89 & & \textbf{0.7629} & 0.7485 & & \textbf{1.0000} & 0.0004 & & 0.9555 & \textbf{0.9426} & & 72.6986 & 1.3990 \\
60 & & \textbf{0.8169} & 0.8150 & & \textbf{0.9478} & 0.5063 & & 0.7862 & \textbf{0.7788} & & 4.2915 & \textbf{0.5277} & & 90 & & \textbf{0.8093} & 0.8004 & & \textbf{0.9998} & 0.0066 & & 1.0377 & \textbf{1.0320} & & 161.0501 & 3.2595 \\ \hline
\textbf{Avg} & & \textbf{0.8318} & 0.8292 & & \textbf{0.9404} & 0.5149 & & 0.8231 & \textbf{0.8225} & & 1.0489 & \textbf{0.1565} & & & & \textbf{0.7994} & 0.7918 & & \textbf{0.9997} & 0.0050 & & 0.9541 & \textbf{0.9502} & & 67.2609 & \textbf{1.7047} \\
\textbf{N. best} & & \textbf{30} & 0 & & \textbf{30} & 0 & & 14 & 16 & & 0 & \textbf{30} & & & & \textbf{30} & 0 & & \textbf{30} & 0 & & 11 & 19 & & 0 & 30 \\
\textbf{\textit{p}} & & \multicolumn{2}{c}{7.24E-04} & & \multicolumn{2}{c}{4.32E-04} & & \multicolumn{2}{c}{0.715} & & \multicolumn{2}{c}{4.32E-04} & & & & \multicolumn{2}{c}{4.32E-04} & & \multicolumn{2}{c}{4.32E-04} & & \multicolumn{2}{c}{0.1441} & & \multicolumn{2}{c}{4.32E-04}
 \\ \hline
\end{tabular}
}
\end{table}
\end{landscape}

\section{Comparing SGS-ES with other state-of-the-art heuristics}\label{sec:SGSESvsNSGAMOEADCH}

Table \ref{tab:Tab541}, \ref{tab:Tab542}, \ref{tab:Tab551} and \ref{tab:Tab552} reports the results of the comparisons between SGS-ES, CH-J, NSGA-III and MOEA/D on the MLS and the VLS instances. 
The computational times (CPU) are expressed in seconds.
Furthermore, the last two rows of the tables allow\addRR{ us} to determine, for each metric, the ranks of the four algorithms under comparison.
Specifically, for each metric, the second-to-last row reports the mean ranks achieved by the four algorithms, according to the Friedman test based multiple comparison test. 
For each metric, the last row shows the relative ranking of the four algorithms with a 99\% confidence interval.
A lower rank denotes a better algorithm, and algorithms that did not generate statistically different results have the same rank.

Table \ref{tab:Tab541} and \ref{tab:Tab542} show the results achieved by the four algorithms on the MLS set.
In more detail, Table \ref{tab:Tab541} reports the algorithms performances according to the three quality metrics.
SGS-ES achieves the best value for Hypervolume on each instance. The only instance where SGS-ES achieves an inferior score, according to the Purity and $D_R$ metrics, is instance $41$.
Instead, Table \ref{tab:Tab542} reports the algorithms performances according to the two distribution metrics, as well as the computational times.
As regards the values of the distribution metrics, Table \ref{tab:Tab542} shows that SGS-ES attained the best average results for Spacing, whereas MOEA/D has the best average Spread. 
The $p$ values denote that the results are statistically equivalent. 
In addition, the algorithms have the same rank on Spacing, whereas only NSGA-III has a better rank than SGS-ES on Spread.
SGS-ES and CH-J are also the best algorithms from a computational efficiency standpoint. Indeed, even though SGS-ES needed on average less than half the computational time of CH-J, they share the same rank.

Table \ref{tab:Tab551} and \ref{tab:Tab552} show the results achieved by the four algorithms on the VLS set.
In particular, Table \ref{tab:Tab551} reports the values of the quality metrics attained by the four algorithms. SGS-ES significantly outperforms the other algorithms, since it achieved the best values on all the metrics, and it is always ranked first.
In particular, it is worth observing that, except for instance 74, SGS-ES achieved the best score for Purity, while CH, NSGA-III, and MOEA/D achieved the worst.
SGS-ES also achieved the best score for $D_R$ on the whole VLS set.
Table \ref{tab:Tab552} reports the values of the distribution metrics, and the computational times as well.
CH-J achieved the best result for Spread and Spacing. However, the $p$ values reveal the statistical equivalence of the four algorithms.
Moreover, it is important to observe that SGS-ES attained significantly better computational times with respect to the other algorithms.

\begin{figure}[!t]
    \captionsetup{font=footnotesize}
    \centering
    \includegraphics[scale=0.355]{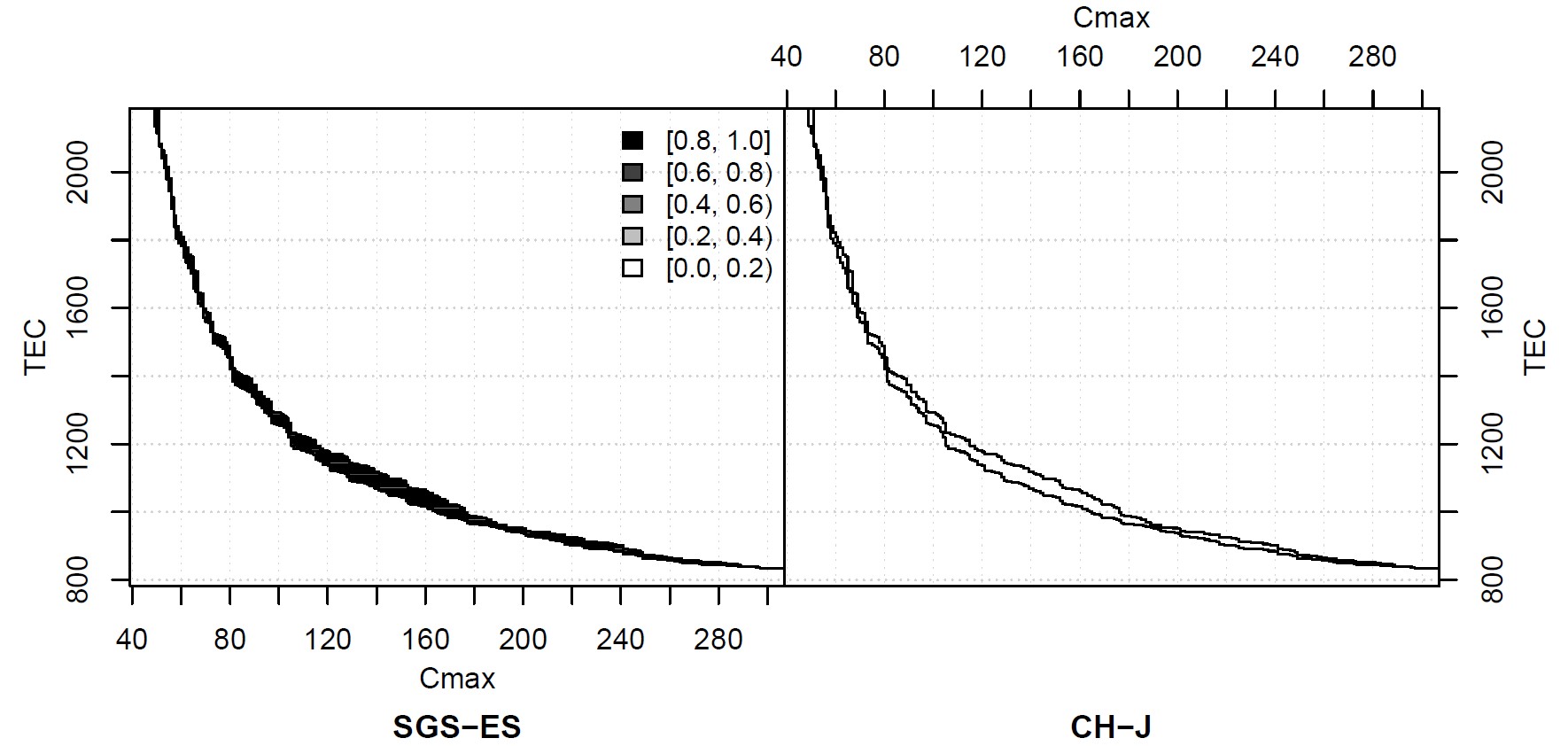}
    \includegraphics[scale=0.28]{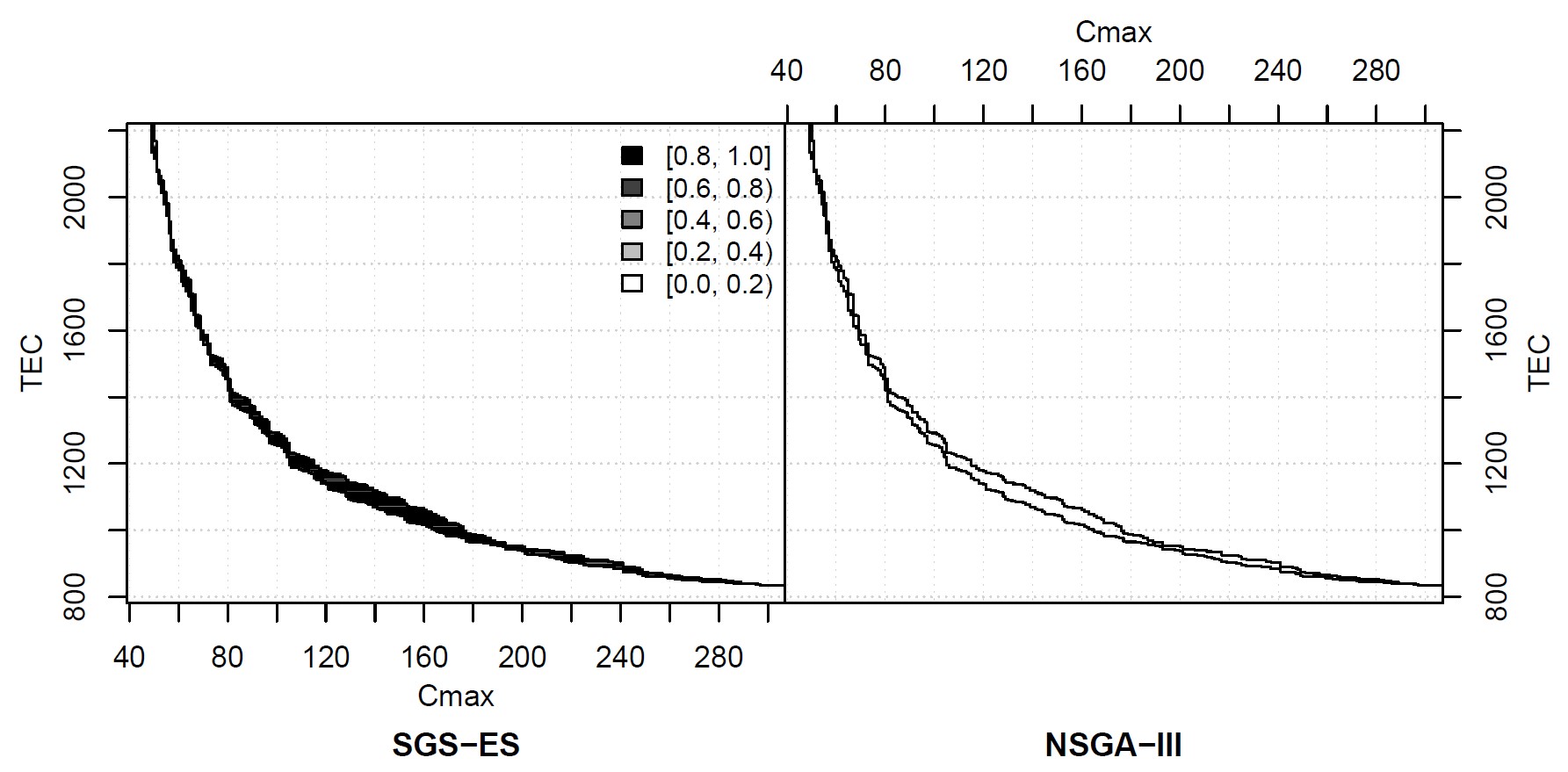}
    \includegraphics[scale=0.28]{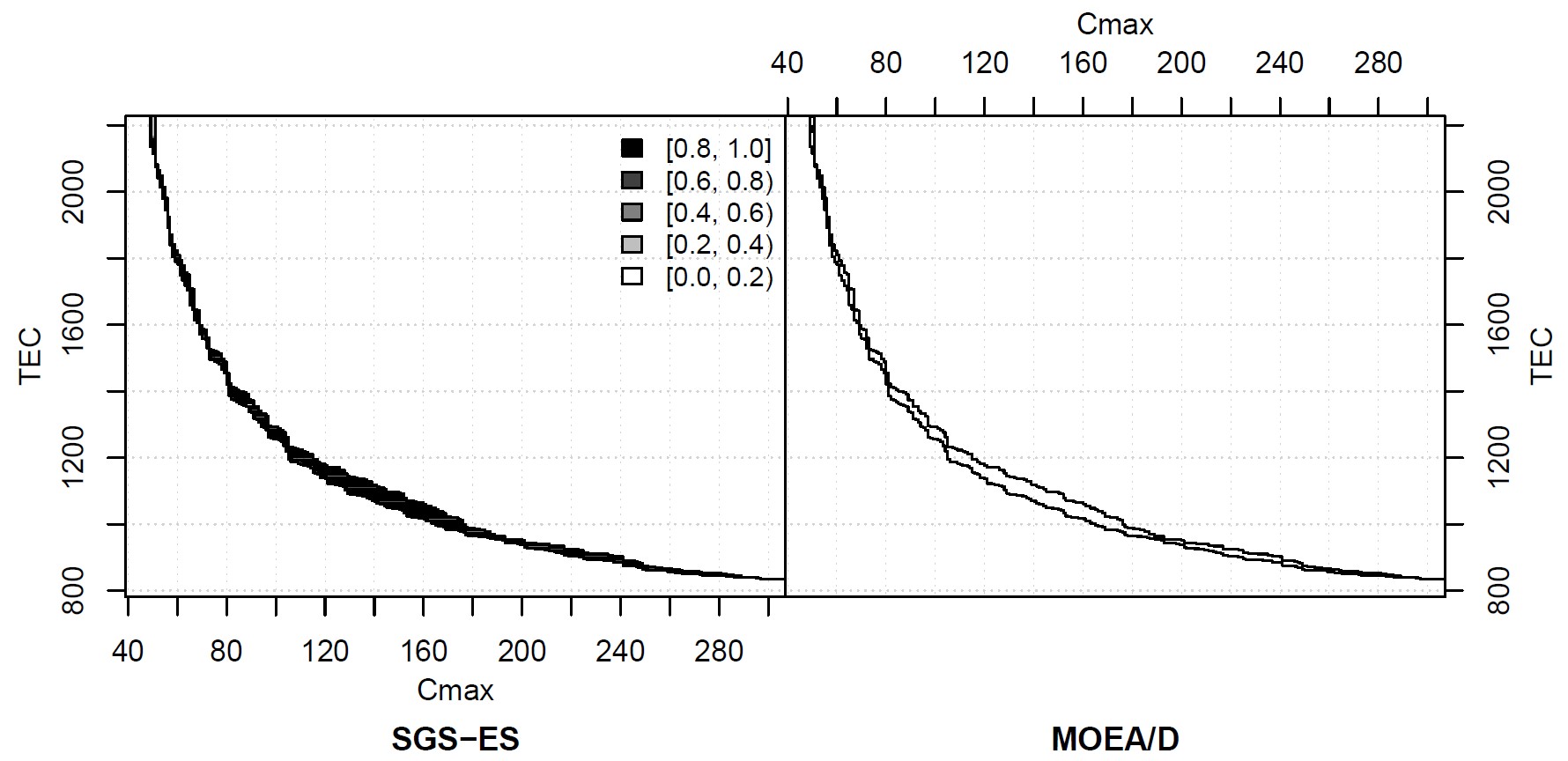}
    \caption{The Diff-EAF plots that compare SGS-ES with CH-J, NSGA-III and MOEA/D for instance 49.}
\label{fig:diff-eafs}
\end{figure}

Finally, Figure \ref{fig:diff-eafs} reports
the Diff-EAF's plots of SGS-ES with respect to the other three algorithms, i.e., CH-J, NSGA-III and MOEA/D for instance $49$.
Such plots were obtained by using the visualization tool proposed by \citet{lopez2010expl}.

The comparisons of the three stochastic algorithms under comparison, that is, SGS-ES, NSGA-III and MOEA/D, were carried by considering the results achieved in 10 distinct runs. 
Each diagram plots a lower and an upper line that connect the best and worst points attained by the compared methods in all the runs, respectively.
The plots also show a dashed line corresponding to the median attained by each algorithm. 
The areas highlighted with different shades of grey show the points where the EAF of an algorithm is larger than the one of the compared algorithm of at least 20\% (the darker is the area, the larger is the difference). 
Observe that, according to all the pairs of diagrams, SGS-ES computed better results than CH-J, NSGA-III and MOEA/D. 

As pointed out in Section \ref{sec:performance-metrics}, a thorough analysis using Diff-EAF's is not practical.
However, the results obtained by considering similar diagrams for several instances were similar to the ones depicted in Figure \ref{fig:diff-eafs}.

\begin{landscape}
\begin{table}[!t]
\captionsetup{font=footnotesize}
\centering
\caption{Comparison of the results achieved by SGS-ES, CH-J, NSGA-III and MOEA/D on the MLS instances based on the Hypervolume, Purity, and $D_R$ metrics.}
\label{tab:Tab541}
\resizebox{21cm}{!}{
\begin{tabular}{ccccclcccclcccc}
\toprule

\textbf{Instance} & \multicolumn{4}{c}{\textbf{Hypervolume}} &  & \multicolumn{4}{c}{\textbf{Purity}} &  & \multicolumn{4}{c}{\textbf{$D_R$}} \\ 

\midrule

& \textbf{SGS-ES} & \textbf{CH-J} & \textbf{NSGA-III} & \textbf{MOEA/D} &  & \textbf{SGS-ES} & \textbf{CH-J} & \textbf{NSGA-III} & \textbf{MOEA/D} &  & \textbf{SGS-ES} & \textbf{CH-J} & \textbf{NSGA-III} & \textbf{MOEA/D} \\ 

\midrule

31 & \textbf{0.8816} & 0.8717 & 0.8717 & 0.8717 &  & \textbf{0.99556} & 0.0540 & 0.0540 & 0.0540 &  & \textbf{0.0000} & 0.0117 & 0.0116 & 0.0117 \\
32 & \textbf{0.8135} & 0.8006 & 0.8006 & 0.8006 &  & \textbf{1.00000} & 0.0981 & 0.0983 & 0.0981 &  & \textbf{0.0000} & 0.0103 & 0.0102 & 0.0103 \\
33 & \textbf{0.7772} & 0.7736 & 0.7736 & 0.7736 &  & \textbf{1.00000} & 0.3702 & 0.3702 & 0.3702 &  & \textbf{0.0000} & 0.0032 & 0.0032 & 0.0032 \\
34 & \textbf{0.6962} & 0.6838 & 0.6839 & 0.6839 &  & \textbf{0.99349} & 0.1082 & 0.1082 & 0.1082 &  & \textbf{0.0000} & 0.0120 & 0.0119 & 0.0119 \\
35 & \textbf{0.6011} & 0.5875 & 0.5875 & 0.5875 &  & \textbf{1.00000} & 0.3889 & 0.3889 & 0.3889 &  & \textbf{0.0000} & 0.0115 & 0.0115 & 0.0115 \\
36 & \textbf{0.8868} & 0.8771 & 0.8773 & 0.8771 &  & \textbf{1.00000} & 0.5043 & 0.5304 & 0.5043 &  & \textbf{0.0000} & 0.0079 & 0.0076 & 0.0079 \\
37 & \textbf{0.8709} & 0.8658 & 0.8658 & 0.8658 &  & \textbf{0.98307} & 0.2972 & 0.2992 & 0.2972 &  & \textbf{0.0001} & 0.0069 & 0.0068 & 0.0069 \\
38 & \textbf{0.8405} & 0.8380 & 0.8380 & 0.8380 &  & \textbf{0.84029} & 0.4113 & 0.4113 & 0.4113 &  & \textbf{0.0006} & 0.0039 & 0.0039 & 0.0039 \\
39 & \textbf{0.7793} & 0.7742 & 0.7742 & 0.7742 &  & \textbf{1.00000} & 0.3333 & 0.3333 & 0.3333 &  & \textbf{0.0000} & 0.0052 & 0.0052 & 0.0052 \\
40 & \textbf{0.7827} & 0.7687 & 0.7687 & 0.7687 &  & \textbf{0.98297} & 0.1188 & 0.1188 & 0.1188 &  & \textbf{0.0001} & 0.0109 & 0.0109 & 0.0109 \\
41 & \textbf{0.9282} & 0.9276 & 0.9280 & 0.9280 &  & 0.93750 & 0.9267 & \textbf{0.9933} & \textbf{0.9933} &  & 0.0016 & 0.0016 & \textbf{0.0000} & \textbf{0.0000} \\
42 & \textbf{0.8569} & 0.8541 & 0.8541 & 0.8541 &  & \textbf{1.00000} & 0.7143 & 0.7143 & 0.7143 &  & \textbf{0.0000} & 0.0031 & 0.0031 & 0.0031 \\
43 & \textbf{0.8612} & 0.8573 & 0.8573 & 0.8573 &  & \textbf{0.96216} & 0.1918 & 0.1918 & 0.1918 &  & \textbf{0.0002} & 0.0045 & 0.0045 & 0.0045 \\
44 & \textbf{0.8312} & 0.8285 & 0.8285 & 0.8285 &  & \textbf{0.80172} & 0.3508 & 0.3508 & 0.3508 &  & \textbf{0.0004} & 0.0029 & 0.0029 & 0.0029 \\
45 & \textbf{0.7900} & 0.7800 & 0.7800 & 0.7800 &  & \textbf{0.86024} & 0.2411 & 0.2411 & 0.2411 &  & \textbf{0.0006} & 0.0084 & 0.0084 & 0.0084 \\
46 & \textbf{0.8173} & 0.8096 & 0.8114 & 0.8096 &  & \textbf{0.89985} & 0.3596 & 0.4207 & 0.3596 &  & \textbf{0.0006} & 0.0072 & 0.0068 & 0.0072 \\
47 & \textbf{0.8949} & 0.8903 & 0.8903 & 0.8903 &  & \textbf{0.94731} & 0.3377 & 0.3377 & 0.3377 &  & \textbf{0.0001} & 0.0134 & 0.0134 & 0.0134 \\
48 & \textbf{0.8724} & 0.8654 & 0.8654 & 0.8654 &  & \textbf{0.87835} & 0.2500 & 0.2500 & 0.2500 &  & \textbf{0.0004} & 0.0076 & 0.0076 & 0.0076 \\
49 & \textbf{0.8059} & 0.7865 & 0.7865 & 0.7865 &  & \textbf{1.00000} & 0.0000 & 0.0000 & 0.0000 &  & \textbf{0.0000} & 0.0158 & 0.0158 & 0.0159 \\
50 & \textbf{0.8275} & 0.8119 & 0.8120 & 0.8119 &  & \textbf{0.99789} & 0.3222 & 0.3222 & 0.3222 &  & \textbf{0.0000} & 0.0134 & 0.0134 & 0.0134 \\
51 & \textbf{0.8492} & 0.8397 & 0.8420 & 0.8420 &  & \textbf{0.81529} & 0.4771 & 0.5600 & 0.5486 &  & \textbf{0.0051} & 0.0107 & 0.0056 & 0.0058 \\
52 & \textbf{0.8681} & 0.8657 & 0.8658 & 0.8658 &  & \textbf{0.79728} & 0.5429 & 0.5429 & 0.5429 &  & \textbf{0.0022} & 0.0101 & 0.0101 & 0.0101 \\
53 & \textbf{0.8875} & 0.8847 & 0.8847 & 0.8847 &  & \textbf{0.93060} & 0.3286 & 0.3286 & 0.3286 &  & \textbf{0.0002} & 0.0058 & 0.0058 & 0.0058 \\
54 & \textbf{0.8836} & 0.8803 & 0.8803 & 0.8803 &  & \textbf{0.97846} & 0.1765 & 0.1765 & 0.1765 &  & \textbf{0.0001} & 0.0056 & 0.0056 & 0.0056 \\
55 & \textbf{0.8445} & 0.8411 & 0.8411 & 0.8411 &  & \textbf{0.97501} & 0.4133 & 0.4133 & 0.4133 &  & \textbf{0.0000} & 0.0026 & 0.0026 & 0.0026 \\
56 & \textbf{0.8867} & 0.8833 & 0.8841 & 0.8841 &  & \textbf{0.87550} & 0.4033 & 0.5278 & 0.5144 &  & \textbf{0.0016} & 0.0082 & 0.0054 & 0.0055 \\
57 & \textbf{0.7435} & 0.7353 & 0.7363 & 0.7363 &  & \textbf{0.85679} & 0.5326 & 0.6011 & 0.5853 &  & \textbf{0.0009} & 0.0040 & 0.0031 & 0.0031 \\
58 & \textbf{0.9057} & 0.9046 & 0.9046 & 0.9046 &  & \textbf{1.00000} & 0.4483 & 0.4483 & 0.4483 &  & \textbf{0.0000} & 0.0039 & 0.0039 & 0.0039 \\
59 & \textbf{0.9017} & 0.8995 & 0.8995 & 0.8995 &  & \textbf{0.96119} & 0.3170 & 0.3170 & 0.3170 &  & \textbf{0.0000} & 0.0040 & 0.0040 & 0.0040 \\
60 & \textbf{0.8184} & 0.8162 & 0.8162 & 0.8162 &  & \textbf{0.93429} & 0.3607 & 0.3607 & 0.3607 &  & \textbf{0.0003} & 0.0037 & 0.0037 & 0.0037 \\ \hline
\textbf{Avg} & \textbf{0.8335} & 0.8267 & 0.8270 & 0.8269 &  & \textbf{0.94016} & 0.3460 & 0.3604 & 0.3560 &  & \textbf{0.0005} & 0.0073 & 0.0070 & 0.0070 \\
\textbf{N. best} & 30 & 0 & 0 & 0 &  & 29 & 0 & 1 & 1 &  & 29 & 0 & 1 & 1 \\
\textbf{p} & \multicolumn{4}{c}{1.87E-17} &  & \multicolumn{4}{c}{4.49E-16} &  & \multicolumn{4}{c}{2.05E-15} \\
\textbf{Mean rank} & 4 & 1.75 & 2.2 & 2.05 &  & 3.9333 & 1.8 & 2.2833 & 1.9833 &  & 1.0833 & 3.2167 & 2.6667 & 3.0333 \\
\textbf{Rank position} & 1 & 2 & 2 & 2 &  & 1 & 2 & 2 & 2 &  & 1 & 2 & 2 & 2 \\ 

\bottomrule
\end{tabular}%
}
\end{table}
\end{landscape}

\begin{landscape}
\begin{table}[!t]
\captionsetup{font=footnotesize}
\centering
\caption{Comparison of the results achieved by SGS-ES, CH-J, NSGA-III and MOEA/D on the MLS instances based on the Spread, Spacing, and CPU time (s) metrics.}
\label{tab:Tab542}
\resizebox{21cm}{!}{
\begin{tabular}{ccccccccccccccc}
\toprule

\textbf{Instance} & \multicolumn{4}{c}{\textbf{Spread}} & \textbf{} & \multicolumn{4}{c}{\textbf{Spacing}} & \textbf{} & \multicolumn{4}{c}{\textbf{CPU}} \\ 

\midrule

\textbf{} & \textbf{SGS-ES} & \textbf{CH-J} & \textbf{NSGA-III} & \textbf{MOEA/D} & \textbf{} & \textbf{SGS-ES} & \textbf{CH-J} & \textbf{NSGA-III} & \textbf{MOEA/D} & \textbf{} & \textbf{SGS-ES} & \textbf{CH-J} & \textbf{NSGA-III} & \textbf{MOEA/D} \\

\midrule

31 & \textbf{0.9173} & 0.9638 & 0.9662 & 0.9638 &  & 0.0453 & \textbf{0.0443} & 0.0445 & \textbf{0.0443} &  & \textbf{0.0518} & 0.2630 & 14.0060 & 2.8346 \\
32 & 0.9482 & 0.8792 & \textbf{0.8782} & 0.8792 &  & 0.0304 & \textbf{0.0294} & 0.0295 & \textbf{0.0294} &  & \textbf{0.0738} & 0.1037 & 14.4933 & 3.3755 \\
33 & 0.8728 & \textbf{0.8562} & \textbf{0.8562} & \textbf{0.8562} &  & \textbf{0.0232} & 0.0238 & 0.0238 & 0.0238 &  & \textbf{0.0937} & 0.1145 & 15.1787 & 4.9112 \\
34 & \textbf{0.7687} & 0.7887 & 0.7887 & 0.7887 &  & \textbf{0.0169} & 0.0171 & 0.0170 & 0.0170 &  & 0.1220 & \textbf{0.0928} & 16.0313 & 6.4969 \\
35 & \textbf{0.5066} & 0.6053 & 0.6053 & 0.6053 &  & \textbf{0.0205} & 0.0232 & 0.0232 & 0.0232 &  & \textbf{0.0586} & 0.0918 & 17.3539 & 8.2203 \\
36 & 0.7243 & \textbf{0.6877} & 0.6927 & \textbf{0.6877} &  & 0.0770 & \textbf{0.0694} & 0.0696 & \textbf{0.0694} &  & \textbf{0.0810} & 0.2132 & 15.1890 & 3.0904 \\
37 & \textbf{0.8880} & 0.9034 & 0.9006 & 0.9034 &  & 0.0515 & \textbf{0.0509} & 0.0509 & \textbf{0.0509} &  & \textbf{0.1202} & 0.1646 & 15.3972 & 4.3526 \\
38 & 1.1245 & 1.0910 & \textbf{1.0910} & 1.0910 &  & 0.0303 & \textbf{0.0298} & 0.0298 & \textbf{0.0298} &  & \textbf{0.1838} & 0.2798 & 16.1686 & 5.9278 \\
39 & \textbf{0.8684} & 0.8764 & 0.8764 & 0.8764 &  & 0.0244 & \textbf{0.0243} & \textbf{0.0243} & \textbf{0.0243} &  & 0.2712 & 0.1295 & 17.0252 & 8.3602 \\
40 & 0.8447 & 0.8119 & \textbf{0.8118} & 0.8119 &  & \textbf{0.0250} & 0.0258 & 0.0258 & 0.0258 &  & 0.3673 & \textbf{0.1424} & 17.6690 & 10.3599 \\
41 & 0.8159 & 0.7885 & \textbf{0.7667} & \textbf{0.7667} &  & 0.1247 & 0.1259 & \textbf{0.1234} & \textbf{0.1234} &  & \textbf{0.0815} & 0.1231 & 14.7031 & 3.4379 \\
42 & \textbf{0.9458} & 0.9669 & 0.9668 & 0.9669 &  & \textbf{0.0682} & 0.0713 & 0.0712 & 0.0713 &  & 0.1289 & \textbf{0.1181} & 15.3369 & 4.8556 \\
43 & 1.0217 & \textbf{0.9456} & \textbf{0.9456} & \textbf{0.9456} &  & \textbf{0.0376} & 0.0389 & 0.0389 & 0.0389 &  & 0.1995 & \textbf{0.1376} & 15.7598 & 6.4588 \\
44 & \textbf{1.0712} & 1.0759 & 1.0759 & 1.0759 &  & 0.0261 & \textbf{0.0261} & 0.0261 & 0.0261 &  & 0.3085 & \textbf{0.1456} & 16.7421 & 8.9836 \\
45 & 0.8651 & \textbf{0.8328} & \textbf{0.8328} & \textbf{0.8328} &  & \textbf{0.0237} & 0.0237 & 0.0237 & 0.0237 &  & 0.5005 & \textbf{0.1736} & 18.4610 & 12.3467 \\
46 & \textbf{0.5925} & 0.6452 & 0.6491 & 0.6452 &  & \textbf{0.0487} & 0.0510 & 0.0509 & 0.0510 &  & \textbf{0.4204} & 2.7928 & 16.9153 & 5.2404 \\
47 & 0.7928 & \textbf{0.7753} & 0.7753 & \textbf{0.7753} &  & 0.0282 & \textbf{0.0264} & 0.0264 & \textbf{0.0264} &  & \textbf{0.7459} & 2.8224 & 17.0613 & 6.1811 \\
48 & 0.7753 & \textbf{0.7632} & 0.7636 & \textbf{0.7632} &  & 0.0223 & \textbf{0.0219} & 0.0220 & \textbf{0.0219} &  & \textbf{0.9804} & 2.9008 & 17.7563 & 7.5760 \\
49 & \textbf{0.8234} & 0.8301 & 0.8435 & 0.8508 &  & \textbf{0.0088} & 0.0089 & 0.0091 & 0.0093 &  & \textbf{1.4556} & 5.8274 & 21.1909 & 11.8035 \\
50 & 0.6961 & \textbf{0.6925} & 0.6925 & \textbf{0.6925} &  & 0.0172 & 0.0155 & \textbf{0.0155} & 0.0155 &  & \textbf{1.8296} & 2.8851 & 19.4386 & 10.7617 \\
51 & 0.8559 & 0.8635 & \textbf{0.8297} & 0.8297 &  & \textbf{0.1124} & 0.1550 & 0.1522 & 0.1524 &  & \textbf{0.7754} & 5.3334 & 19.7377 & 8.4114 \\
52 & 0.7515 & 0.6922 & \textbf{0.6921} & 0.6922 &  & \textbf{0.0848} & 0.0903 & 0.0892 & 0.0892 &  & \textbf{1.1850} & 5.3997 & 20.4408 & 9.6556 \\
53 & 0.8235 & 0.7922 & \textbf{0.7922} & 0.7922 &  & 0.0388 & \textbf{0.0379} & \textbf{0.0379} & \textbf{0.0379} &  & \textbf{1.6191} & 5.4256 & 20.8525 & 11.3704 \\
54 & 0.6813 & 0.6765 & \textbf{0.6764} & 0.6765 &  & \textbf{0.0338} & 0.0341 & 0.0341 & 0.0341 &  & \textbf{2.1862} & 5.4775 & 22.1008 & 12.8782 \\
55 & 0.8538 & \textbf{0.8308} & \textbf{0.8308} & \textbf{0.8308} &  & \textbf{0.0144} & 0.0148 & 0.0148 & 0.0148 &  & \textbf{3.3929} & 5.7765 & 22.7618 & 15.9638 \\
56 & 0.8049 & \textbf{0.7689} & 0.7824 & 0.7821 &  & \textbf{0.1013} & 0.1122 & 0.1116 & 0.1115 &  & \textbf{1.1292} & 8.5075 & 23.0410 & 12.5976 \\
57 & 0.6317 & 0.6023 & \textbf{0.6018} & 0.6023 &  & \textbf{0.0835} & 0.0993 & 0.0996 & 0.0996 &  & \textbf{1.7694} & 8.4910 & 23.5395 & 13.0080 \\
58 & 0.6884 & \textbf{0.6690} & \textbf{0.6690} & \textbf{0.6690} &  & 0.0558 & \textbf{0.0557} & \textbf{0.0557} & \textbf{0.0557} &  & \textbf{2.4266} & 8.4105 & 24.5284 & 15.3027 \\
59 & 0.9664 & \textbf{0.8668} & 0.8668 & \textbf{0.8668} &  & 0.0232 & \textbf{0.0203} & 0.0203 & \textbf{0.0203} &  & \textbf{4.6183} & 8.5497 & 25.4301 & 18.2997 \\
60 & 0.7862 & \textbf{0.7681} & \textbf{0.7681} & \textbf{0.7681} &  & \textbf{0.0228} & 0.0238 & 0.0238 & 0.0238 &  & \textbf{4.2915} & 8.5050 & 26.9437 & 20.8700 \\ 

\midrule

\textbf{Avg} & 0.8236 & 0.8103 & 0.8096 & \textbf{0.8096} &  & \textbf{0.0440} & 0.0464 & 0.0462 & 0.0462 &  & \textbf{1.0489} & 2.9799 & 18.7085 & 9.1311 \\
\textbf{N. best} & 9 & 13 & 15 & 13 &  & 16 & 12 & 5 & 12 &  & 23 & 7 & 0 & 0 \\
\textit{\textbf{p}} & \multicolumn{4}{c}{0.0138} &  & \multicolumn{4}{c}{0.8082} &  & \multicolumn{4}{c}{5.28E-18} \\
\textbf{Mean rank} & 3.0667 & 2.3833 & 2.2 & 2.35 &  & 2.3667 & 2.6333 & 2.5333 & 2.4667 &  & 1.2333 & 1.7667 & 4 & 3 \\
\textbf{Rank position} & 2 & 1, 2 & 1 & 1, 2 &  & 1 & 1 & 1 & 1 &  & 1 & 1 & 3 & 2 \\ 

\bottomrule
\end{tabular}%
}
\end{table}
\end{landscape}


\begin{landscape}
\begin{table}[!t]
\captionsetup{font=footnotesize}
\centering
\caption{Comparison of the results achieved by SGS-ES, CH-J, NSGA-III and MOEA/D on the VLS instances based on the Hypervolume, Purity, and $D_R$ metrics.}
\label{tab:Tab551}
\resizebox{21cm}{!}{
\begin{tabular}{ccccccccccccccc}
\toprule

\textbf{Instance} & \multicolumn{4}{c}{\textbf{Hypervolume}} & \textbf{} & \multicolumn{4}{c}{\textbf{Purity}} & \textbf{} & \multicolumn{4}{c}{\textbf{$D_R$}} \\ 

\midrule

\textbf{} & \textbf{SGS-ES} & \textbf{CH-J} & \textbf{NSGA-III} & \textbf{MOEA/D} & \textbf{} & \textbf{SGS-ES} & \textbf{CH-J} & \textbf{NSGA-III} & \textbf{MOEA/D} & \textbf{} & \textbf{SGS-ES} & \textbf{CH-J} & \textbf{NSGA-III} & \textbf{MOEA/D} \\ 

\midrule

61 & \textbf{0.8197} & 0.8047 & 0.8047 & 0.8047 &  & \textbf{1.0000} & 0.0000 & 0.0000 & 0.0000 &  & \textbf{0.0000} & 0.0118 & 0.0117 & 0.0118 \\
62 & \textbf{0.8295} & 0.8172 & 0.8172 & 0.8172 &  & \textbf{1.0000} & 0.0000 & 0.0000 & 0.0000 &  & \textbf{0.0000} & 0.0097 & 0.0096 & 0.0097 \\
63 & \textbf{0.7660} & 0.7498 & 0.7498 & 0.7498 &  & \textbf{1.0000} & 0.0000 & 0.0000 & 0.0000 &  & \textbf{0.0000} & 0.0113 & 0.0113 & 0.0113 \\
64 & \textbf{0.7862} & 0.7695 & 0.7695 & 0.7695 &  & \textbf{1.0000} & 0.0000 & 0.0000 & 0.0000 &  & \textbf{0.0000} & 0.0110 & 0.0110 & 0.0110 \\
65 & \textbf{0.7245} & 0.7015 & 0.7015 & 0.7015 &  & \textbf{1.0000} & 0.0000 & 0.0000 & 0.0000 &  & \textbf{0.0000} & 0.0171 & 0.0171 & 0.0171 \\
66 & \textbf{0.8015} & 0.7835 & 0.7835 & 0.7835 &  & \textbf{1.0000} & 0.0058 & 0.0058 & 0.0058 &  & \textbf{0.0000} & 0.0153 & 0.0153 & 0.0153 \\
67 & \textbf{0.7581} & 0.7414 & 0.7416 & 0.7415 &  & \textbf{1.0000} & 0.0000 & 0.0000 & 0.0000 &  & \textbf{0.0000} & 0.0105 & 0.0106 & 0.0105 \\
68 & \textbf{0.7390} & 0.7136 & 0.7138 & 0.7137 &  & \textbf{1.0000} & 0.0000 & 0.0000 & 0.0000 &  & \textbf{0.0000} & 0.0176 & 0.0177 & 0.0177 \\
69 & \textbf{0.7368} & 0.7177 & 0.7177 & 0.7177 &  & \textbf{1.0000} & 0.0000 & 0.0000 & 0.0000 &  & \textbf{0.0000} & 0.0131 & 0.0131 & 0.0131 \\
70 & \textbf{0.7525} & 0.7282 & 0.7282 & 0.7282 &  & \textbf{1.0000} & 0.0000 & 0.0000 & 0.0000 &  & \textbf{0.0000} & 0.0172 & 0.0173 & 0.0172 \\
71 & \textbf{0.8098} & 0.7951 & 0.7952 & 0.7951 &  & \textbf{1.0000} & 0.0000 & 0.0000 & 0.0000 &  & \textbf{0.0000} & 0.0106 & 0.0107 & 0.0107 \\
72 & \textbf{0.8734} & 0.8629 & 0.8630 & 0.8630 &  & \textbf{1.0000} & 0.0000 & 0.0000 & 0.0000 &  & \textbf{0.0000} & 0.0075 & 0.0075 & 0.0075 \\
73 & \textbf{0.7979} & 0.7810 & 0.7812 & 0.7811 &  & \textbf{1.0000} & 0.0000 & 0.0000 & 0.0000 &  & \textbf{0.0000} & 0.0111 & 0.0112 & 0.0112 \\
74 & \textbf{0.8509} & 0.8415 & 0.8416 & 0.8416 &  & \textbf{0.9964} & 0.0038 & 0.0038 & 0.0038 &  & \textbf{0.0000} & 0.0067 & 0.0067 & 0.0067 \\
75 & \textbf{0.7879} & 0.7710 & 0.7712 & 0.7711 &  & \textbf{1.0000} & 0.0000 & 0.0000 & 0.0000 &  & \textbf{0.0000} & 0.0114 & 0.0115 & 0.0115 \\
76 & \textbf{0.8505} & 0.8394 & 0.8396 & 0.8395 &  & \textbf{1.0000} & 0.0000 & 0.0000 & 0.0000 &  & \textbf{0.0000} & 0.0072 & 0.0073 & 0.0072 \\
77 & \textbf{0.7478} & 0.7223 & 0.7223 & 0.7223 &  & \textbf{1.0000} & 0.0000 & 0.0000 & 0.0000 &  & \textbf{0.0000} & 0.0176 & 0.0176 & 0.0176 \\
78 & \textbf{0.7890} & 0.7716 & 0.7716 & 0.7716 &  & \textbf{1.0000} & 0.0000 & 0.0000 & 0.0000 &  & \textbf{0.0000} & 0.0136 & 0.0136 & 0.0136 \\
79 & \textbf{0.7465} & 0.7253 & 0.7254 & 0.7253 &  & \textbf{1.0000} & 0.0000 & 0.0000 & 0.0000 &  & \textbf{0.0000} & 0.0141 & 0.0142 & 0.0141 \\
80 & \textbf{0.8157} & 0.8011 & 0.8012 & 0.8011 &  & \textbf{1.0000} & 0.0000 & 0.0000 & 0.0000 &  & \textbf{0.0000} & 0.0096 & 0.0097 & 0.0096 \\
81 & \textbf{0.8205} & 0.8066 & 0.8069 & 0.8069 &  & \textbf{1.0000} & 0.0000 & 0.0000 & 0.0000 &  & \textbf{0.0000} & 0.0090 & 0.0089 & 0.0089 \\
82 & \textbf{0.8650} & 0.8562 & 0.8563 & 0.8563 &  & \textbf{1.0000} & 0.0000 & 0.0000 & 0.0000 &  & \textbf{0.0000} & 0.0084 & 0.0084 & 0.0084 \\
83 & \textbf{0.8200} & 0.8034 & 0.8036 & 0.8036 &  & \textbf{1.0000} & 0.0000 & 0.0000 & 0.0000 &  & \textbf{0.0000} & 0.0111 & 0.0109 & 0.0109 \\
84 & \textbf{0.8292} & 0.8169 & 0.8169 & 0.8169 &  & \textbf{1.0000} & 0.0000 & 0.0000 & 0.0000 &  & \textbf{0.0000} & 0.0090 & 0.0089 & 0.0089 \\
85 & \textbf{0.8239} & 0.8096 & 0.8097 & 0.8096 &  & \textbf{1.0000} & 0.0000 & 0.0000 & 0.0000 &  & \textbf{0.0000} & 0.0102 & 0.0102 & 0.0102 \\
86 & \textbf{0.8317} & 0.8202 & 0.8202 & 0.8202 &  & \textbf{1.0000} & 0.0000 & 0.0000 & 0.0000 &  & \textbf{0.0000} & 0.0082 & 0.0082 & 0.0082 \\
87 & \textbf{0.8090} & 0.7974 & 0.7974 & 0.7974 &  & \textbf{1.0000} & 0.0000 & 0.0000 & 0.0000 &  & \textbf{0.0000} & 0.0081 & 0.0081 & 0.0081 \\
88 & \textbf{0.8372} & 0.8251 & 0.8251 & 0.8251 &  & \textbf{1.0000} & 0.0000 & 0.0000 & 0.0000 &  & \textbf{0.0000} & 0.0089 & 0.0089 & 0.0089 \\
89 & \textbf{0.7629} & 0.7428 & 0.7429 & 0.7428 &  & \textbf{1.0000} & 0.0000 & 0.0000 & 0.0000 &  & \textbf{0.0000} & 0.0147 & 0.0146 & 0.0147 \\
90 & \textbf{0.8093} & 0.7948 & 0.7949 & 0.7949 &  & \textbf{1.0000} & 0.0000 & 0.0000 & 0.0000 &  & \textbf{0.0000} & 0.0102 & 0.0101 & 0.0102 \\ 

\midrule

\textbf{Avg} & \textbf{0.7997} & 0.7837 & 0.7838 & 0.7838 &  & \textbf{0.9999} & 0.0003 & 0.0003 & 0.0003 &  & \textbf{0.0000} & 0.0114 & 0.0114 & 0.0114 \\
\textbf{N. best} & 30 & 0 & 0 & 0 &  & 30 & 0 & 0 & 0 &  & 30 & 0 & 0 & 0 \\
\textit{\textbf{p}} & \multicolumn{4}{c}{1.66E-17} &  & \multicolumn{4}{c}{2.19E-19} &  & \multicolumn{4}{c}{2.39E-15} \\
\textbf{Mean rank} & 4 & 1.5323 & 2.4335 & 2.0323 &  & 4 & 2 & 2 & 2 &  & 1 & 2.95 & 3.05 & 3 \\
\textbf{Rank position} & 1 & 3 & 2 & 2, 3 &  & 1 & 2 & 2 & 2 &  & 1 & 3 & 3 & 3 \\ 

\bottomrule
\end{tabular}
}
\end{table}
\end{landscape}


\begin{landscape}
\begin{table}[!t]
\captionsetup{font=footnotesize}
\centering
\caption{Comparison of the results achieved by SGS-ES, CH-J, NSGA-III and MOEA/D on the MLS instances based on the Spread, Spacing, and CPU time (s) metrics.}
\label{tab:Tab552}
\resizebox{21cm}{!}{
\begin{tabular}{ccccccccccccccc}
\toprule

\textbf{Instance} & \multicolumn{4}{c}{\textbf{Spread}} & \textbf{} & \multicolumn{4}{c}{\textbf{Spacing}} & \textbf{} & \multicolumn{4}{c}{\textbf{CPU}} \\ 

\midrule

\textbf{} & \textbf{SGS-ES} & \textbf{CH-J} & \textbf{NSGA-III} & \textbf{MOEA/D} & \textbf{} & \textbf{SGS-ES} & \textbf{CH-J} & \textbf{NSGA-III} & \textbf{MOEA/D} & \textbf{} & \textbf{SGS-ES} & \textbf{CH-J} & \textbf{NSGA-III} & \textbf{MOEA/D} \\

\midrule

61 & 1.0602 & 1.0473 & \textbf{1.0459} & 1.0471 &  & 0.0077 & 0.0073 & 0.0073 & \textbf{0.0073} &  & \textbf{21.7450} & 50.1929 & 58.8327 & 66.7168 \\
62 & \textbf{1.0119} & 1.0496 & 1.0464 & 1.0496 &  & 0.0055 & \textbf{0.0053} & 0.0054 & 0.0053 &  & \textbf{62.3640} & 472.9695 & 327.7931 & 488.7253 \\
63 & 0.8898 & 0.8261 & 0.8256 & \textbf{0.8255} &  & \textbf{0.0055} & 0.0056 & 0.0056 & 0.0056 &  & \textbf{25.8207} & 38.8928 & 51.8122 & 58.2162 \\
64 & 0.9226 & \textbf{0.8744} & 0.9106 & 0.8880 &  & 0.0051 & \textbf{0.0043} & 0.0059 & 0.0051 &  & \textbf{57.0123} & 457.8403 & 312.8328 & 476.2181 \\
65 & \textbf{0.8744} & 0.9017 & 0.9005 & 0.9017 &  & 0.0059 & \textbf{0.0056} & 0.0056 & \textbf{0.0056} &  & \textbf{30.4711} & 46.5792 & 61.1699 & 67.6016 \\
66 & \textbf{0.9823} & 1.0137 & 1.0132 & 1.0137 &  & 0.0056 & \textbf{0.0053} & 0.0053 & \textbf{0.0053} &  & \textbf{75.7831} & 294.5378 & 215.1681 & 314.6727 \\
67 & 0.8073 & \textbf{0.7758} & 0.7919 & 0.7892 &  & \textbf{0.0053} & 0.0054 & 0.0062 & 0.0062 &  & \textbf{25.9146} & 61.7673 & 73.8247 & 86.6843 \\
68 & 0.8607 & \textbf{0.8158} & 0.8333 & 0.8314 &  & 0.0039 & \textbf{0.0037} & 0.0041 & 0.0043 &  & \textbf{84.8752} & 321.0555 & 244.1829 & 344.4683 \\
69 & 0.8101 & 0.7642 & \textbf{0.7625} & 0.7626 &  & \textbf{0.0054} & 0.0056 & 0.0055 & 0.0055 &  & \textbf{25.9769} & 69.9960 & 81.8046 & 99.3700 \\
70 & 0.8576 & \textbf{0.8543} & 0.8780 & 0.8794 &  & 0.0043 & \textbf{0.0040} & 0.0049 & 0.0051 &  & \textbf{90.3612} & 333.5820 & 254.6311 & 362.0777 \\
71 & 0.9612 & \textbf{0.9179} & 0.9734 & 0.9788 &  & 0.0076 & \textbf{0.0071} & 0.0097 & 0.0101 &  & \textbf{25.2129} & 77.2715 & 77.8492 & 94.0222 \\
72 & 1.0414 & 1.0363 & \textbf{1.0323} & 1.0335 &  & 0.0084 & 0.0084 & 0.0084 & \textbf{0.0084} &  & \textbf{65.3719} & 129.7969 & 108.7024 & 147.3442 \\
73 & 0.9906 & \textbf{0.9181} & 0.9536 & 0.9519 &  & \textbf{0.0068} & 0.0072 & 0.0083 & 0.0084 &  & \textbf{30.7839} & 55.2623 & 64.4688 & 75.9908 \\
74 & 1.0319 & \textbf{1.0087} & 1.0087 & 1.0087 &  & 0.0079 & \textbf{0.0078} & 0.0078 & 0.0078 &  & \textbf{77.8728} & 186.2354 & 145.0289 & 206.7596 \\
75 & 0.9003 & \textbf{0.8709} & 0.8948 & 0.8977 &  & 0.0064 & \textbf{0.0061} & 0.0070 & 0.0072 &  & \textbf{34.7806} & 53.0671 & 65.9283 & 77.1399 \\
76 & \textbf{0.9747} & 0.9768 & 0.9945 & 0.9922 &  & \textbf{0.0074} & 0.0075 & 0.0081 & 0.0081 &  & \textbf{92.2981} & 100.0133 & 94.9817 & 124.1663 \\
77 & 0.8656 & 0.8515 & \textbf{0.8510} & 0.8511 &  & 0.0057 & 0.0055 & 0.0055 & \textbf{0.0055} &  & \textbf{37.2449} & 79.5971 & 86.5105 & 105.8310 \\
78 & 0.9621 & \textbf{0.9338} & 0.9338 & \textbf{0.9338} &  & 0.0052 & \textbf{0.0047} & 0.0047 & \textbf{0.0047} &  & \textbf{107.1587} & 378.6946 & 267.8694 & 405.3375 \\
79 & 0.8274 & \textbf{0.7902} & 0.8018 & 0.7966 &  & 0.0053 & \textbf{0.0051} & 0.0058 & 0.0055 &  & \textbf{45.8857} & 52.1700 & 71.6422 & 83.8278 \\
80 & 0.9324 & \textbf{0.9018} & 0.9218 & 0.9127 &  & \textbf{0.0054} & 0.0055 & 0.0062 & 0.0059 &  & \textbf{101.0368} & 228.2797 & 187.8734 & 261.1875 \\
81 & 0.9468 & \textbf{0.9066} & 0.9523 & 0.9528 &  & 0.0096 & \textbf{0.0084} & 0.0115 & 0.0116 &  & \textbf{37.5365} & 54.1488 & 60.3279 & 75.1196 \\
82 & \textbf{1.1053} & 1.1475 & 1.1439 & 1.1441 &  & 0.0081 & 0.0077 & \textbf{0.0077} & 0.0077 &  & \textbf{92.0232} & 369.0146 & 254.4615 & 389.2578 \\
83 & 1.0663 & 1.0155 & \textbf{1.0145} & 1.0146 &  & \textbf{0.0077} & 0.0078 & 0.0079 & 0.0079 &  & \textbf{42.3040} & 74.4537 & 77.4552 & 98.9680 \\
84 & 0.9831 & \textbf{0.9649} & 0.9666 & 0.9669 &  & 0.0061 & \textbf{0.0058} & 0.0058 & 0.0058 &  & \textbf{113.6161} & 351.1298 & 248.0501 & 375.9355 \\
85 & 1.0316 & \textbf{1.0017} & 1.0243 & 1.0186 &  & 0.0071 & \textbf{0.0069} & 0.0079 & 0.0077 &  & \textbf{49.3493} & 62.4446 & 72.6152 & 89.6216 \\
86 & \textbf{0.9742} & 0.9830 & 0.9828 & 0.9830 &  & 0.0058 & \textbf{0.0055} & 0.0055 & 0.0055 &  & \textbf{129.1932} & 180.4264 & 145.6146 & 207.6757 \\
87 & 0.9481 & \textbf{0.9275} & 0.9276 & \textbf{0.9275} &  & 0.0072 & 0.0068 & \textbf{0.0068} & 0.0068 &  & \textbf{52.3694} & 126.2987 & 119.3945 & 156.4167 \\
88 & 1.0088 & 1.0092 & \textbf{1.0047} & 1.0055 &  & 0.0061 & 0.0059 & 0.0059 & \textbf{0.0059} &  & \textbf{149.7154} & 300.2133 & 229.4651 & 330.7565 \\
89 & 0.9555 & 0.9294 & \textbf{0.9275} & 0.9293 &  & 0.0070 & 0.0066 & 0.0066 & \textbf{0.0066} &  & 72.6986 & \textbf{46.6116} & 68.8165 & 83.1229 \\
90 & 1.0377 & 1.0098 & \textbf{1.0089} & 1.0098 &  & \textbf{0.0052} & 0.0054 & 0.0054 & 0.0054 &  & \textbf{161.0501} & 285.6970 & 220.1420 & 323.1561 \\ 

\midrule

\textbf{Avg} & 0.9541 & \textbf{0.9341} & 0.9442 & 0.9432 &  & 0.0063 & \textbf{0.0061} & 0.0066 & 0.0066 &  & \textbf{67.2609} & 177.9413 & 144.9750 & 202.5463 \\
\textbf{N. best} & 6 & 15 & 8 & 3 &  & 8 & 15 & 2 & 8 &  & 29 & 1 & 0 & 0 \\
\textit{\textbf{p}} & \multicolumn{4}{c}{0.0056} &  & \multicolumn{4}{c}{0.0049} &  & \multicolumn{4}{c}{1.02E-16} \\
\textbf{Mean rank} & 3.1667 & 2.1667 & 2.1667 & 2.5 &  & 2.6333 & 1.8667 & 2.75 & 2.75 &  & 1.0667 & 2.5 & 2.4333 & 4 \\
\textbf{Rank position} & 3 & 2 & 1 & 1 &  & 2 & 1 & 2 & 2 &  & 1 & 2 & 2 & 3\\ 

\bottomrule
\end{tabular}%
}
\end{table}
\end{landscape}

\section{Comparing the mathematical models}
\label{sec:formulations-comparison}

This section compares the two mathematical models introduced in Chapter \ref{chap:problem} from an experimental standpoint, so as to evaluate the impact of the reduction of the solution space introduced by \hyperlink{F2}{Formulation 2} with respect to \hyperlink{F1}{Formulation 1}.
In order to achieve this, the MILP solving step of the exact algorithm was first carried out with \hyperlink{F1}{Formulation 1}, and then with \hyperlink{F2}{Formulation 2}.

Table~\ref{table:mathematical-models-comparisons} reports the results of the comparison for the first 60 instances. The exact algorithm was not able to solve the last 30 instances with \hyperlink{F1}{Formulation 1} due to the RAM constraints of the machine used for the experimental evaluations.
For each instance, the value under the column $\Delta_{F1, F2}$ is a measure of the improvement introduced by \hyperlink{F2}{Formulation 2} in the computational times.
Formally, let us denote the time taken for the exact algorithm with \hyperlink{F1}{Formulation 1} and \hyperlink{F2}{Formulation 2} to solve instance $i$ as $\delta_{\text{F1}, i}$ and $\delta_{\text{F2}, i}$, respectively. 
Then, the value of $\Delta_{F1, F2}$ in row $i$ is equal to $1 - \delta_{\text{F2}, i} / \delta_{\text{F1}, i}$.

\hyperlink{F2}{Formulation 2} outperforms \hyperlink{F1}{Formulation 1} in all the 60 instances, except for instance \addRR{4}. In particular, the average computational time of \hyperlink{F2}{Formulation 2} on instances $31$--$60$ is remarkably two order of magnitudes lower than the average time achieved by \hyperlink{F1}{Formulation 1} on the same instances.
Overall, \hyperlink{F2}{Formulation 2} achieved a $96.68\%$ average improvement on such instances.

\begin{table}[!b]
\captionsetup{font=footnotesize}
\centering
\scalebox{0.85}{
\begin{tabular}{ccccccccc}
\toprule

\textbf{Instance}        & \multicolumn{3}{c}{\textbf{CPU}}                    &  & \textbf{Instance}         & \multicolumn{3}{c}{\textbf{CPU}} \\         


\midrule

& \textbf{F1}            & \textbf{F2}            &  $\Delta_{F1, F2}$      &  &                  & \textbf{F1}      & \textbf{F2}             & $\Delta_{F1, F2}$      \\
                 
\midrule

1                & 0.2853          & \textbf{0.2236} & 21.63  &  & 31               & 7.2461    & \textbf{0.9331}  & 87.12 \\
2                & 0.6087          & \textbf{0.4214} & 30.77  &  & 32               & 21.5001   & \textbf{1.1392}  & 94.70 \\
3                & 0.3469          & \textbf{0.2920} & 15.82  &  & 33               & 67.4728   & \textbf{1.5184}  & 97.75 \\
4                & \textbf{0.3941} & 0.5131          & -30.21  &  & 34               & 175.1797  & \textbf{2.2396}  & 98.72 \\
5                & 0.4617          & \textbf{0.4139} & 10.36  &  & 35               & 264.9398  & \textbf{2.5727}  & 99.03 \\
6                & 0.7080          & \textbf{0.4404} & 37.80  &  & 36               & 10.4937   & \textbf{0.8970}  & 91.45 \\
7                & 0.7839          & \textbf{0.5416} & 30.91  &  & 37               & 29.6982   & \textbf{1.3301}  & 95.52 \\
8                & 0.9900          & \textbf{0.5786} & 41.55  &  & 38               & 130.3063  & \textbf{2.5860}  & 98.02 \\
9                & 0.5386          & \textbf{0.2888} & 46.39  &  & 39               & 219.4452  & \textbf{2.8135}  & 98.72 \\
10               & 0.9039          & \textbf{0.7345} & 18.74  &  & 40               & 417.0358  & \textbf{4.1034}  & 99.02 \\
11               & 0.6921          & \textbf{0.4311} & 37.71  &  & 41               & 16.1518   & \textbf{0.9381}  & 94.19 \\
12               & 1.2225          & \textbf{0.8308} & 32.04  &  & 42               & 37.9442   & \textbf{1.7523}  & 95.38 \\
13               & 0.8033          & \textbf{0.4120} & 48.71  &  & 43               & 124.6872  & \textbf{2.6392}  & 97.88 \\
14               & 1.2348          & \textbf{0.5765} & 53.32  &  & 44               & 334.8600  & \textbf{3.8372}  & 98.85 \\
15               & 0.7733          & \textbf{0.2753} & 64.40  &  & 45               & 652.2999  & \textbf{5.7957}  & 99.11 \\
16               & 1.7666          & \textbf{0.7318} & 58.58  &  & 46               & 45.5896   & \textbf{3.4578}  & 92.42 \\
17               & 1.2358          & \textbf{0.5949} & 51.86  &  & 47               & 98.2345   & \textbf{4.2304}  & 95.69 \\
18               & 2.4041          & \textbf{1.0459} & 56.49  &  & 48               & 296.7794  & \textbf{7.3264}  & 97.53 \\
19               & 1.2549          & \textbf{0.5384} & 57.10  &  & 49               & 1247.3653 & \textbf{17.4488} & 98.60 \\
20               & 1.8719          & \textbf{0.6493} & 65.31  &  & 50               & 1298.3045 & \textbf{11.4346} & 99.12 \\
21               & 2.1630          & \textbf{0.6916} & 68.02  &  & 51               & 134.1745  & \textbf{7.1962}  & 94.64 \\
22               & 2.8599          & \textbf{0.9937} & 65.25  &  & 52               & 102.4868  & \textbf{5.2770}  & 94.85 \\
23               & 1.6491          & \textbf{0.4706} & 71.46  &  & 53               & 512.3011  & \textbf{10.1268} & 98.02 \\
24               & 4.1166          & \textbf{1.2622} & 69.34  &  & 54               & 824.7366  & \textbf{12.4966} & 98.48 \\
25               & 1.2025          & \textbf{0.3022} & 74.87  &  & 55               & 2242.6907 & \textbf{17.9711} & 99.20 \\
26               & 3.1990          & \textbf{0.8453} & 73.58  &  & 56               & 94.1080   & \textbf{4.1502}  & 95.59 \\
27               & 1.5952          & \textbf{0.4340} & 72.79  &  & 57               & 232.9828  & \textbf{11.0508} & 95.26 \\
28               & 4.6481          & \textbf{1.1727} & 74.77 &  & 58               & 547.1832  & \textbf{12.9963} & 97.62 \\
29               & 3.7137          & \textbf{0.9684} & 73.92  &  & 59               & 1729.1312 & \textbf{21.2720} & 98.77 \\
30               & 4.2313          & \textbf{0.9548} & 77.43  &  & 60               & 2359.3696 & \textbf{20.5363} & 99.13 \\

\midrule

\textbf{Avg}     & 1.6220          & \textbf{0.6210} & 49.02  &  & \textbf{Avg}     & 475.8233  & \textbf{6.7356}           & 96.68 \\
\textbf{N. best} & 1               & \textbf{29}     &        &  & \textbf{N. best} & 0         & \textbf{30}               & \\     

\bottomrule
\end{tabular}
}
\caption{Comparison of the first mathematical model with the second one, as a part of the mathematical programming step in the exact algorithm.}
\label{table:mathematical-models-comparisons}
\end{table}

\section{Evaluating the impact of SGH as the initial feasible solution heuristic for the exact algorithm}
\label{sec:warmstart}

The purpose of this section is to evaluate the increase in computational efficiency enabled by the use of SGH as a means to provide an initial feasible solution for the mathematical programming step of the exact algorithm.

Table \ref{table:warmstart-evaluation} shows the performances achieved by the exact algorithm that uses \hyperlink{F2}{Formulation 2}, with (F2-ws) or without (F2) the initial solution provided by SGH.
For each instance, the time limit for computations was set to $7200$ seconds.
The values under the column $\Delta_{F2, F2ws}$ are defined similarly to the values of $\Delta_{F1, F2}$ in Section \ref{sec:formulations-comparison}.
The computational advantage enabled by the initial solution provided by SGH is noteworthy, especially for instances $31$--$60$, where it amounts to a $43.13\%$ average improvement.
Instances \addRR{$20$} and \addRR{$24$} are the only instances where the exact algorithm without the SGH initial solution achieves superior performances. 
For such instances, SGH is revealed as a burden for the already fast mathematical programming step of the exact algorithm.
Finally, it is interesting to observe that the initial solution provided by SGH is still insufficient to enable the completion of the computations on both instance 62 and 83 within the time limit.

\begin{landscape}

\begin{table}[!t]
\centering
\captionsetup{font=footnotesize}
\caption{Comparison of the ad-hoc initial solution provided by SGH with the initial solution heuristics used by CPLEX for the second mathematical model, as part of the mathematical programming step of the exact algorithm.}
\scalebox{0.75}{
\begin{tabular}{cccccccccccccc}

\toprule

\textbf{Instance} & \multicolumn{1}{l}{} & \textbf{CPU}    & \textbf{} & \textbf{} & \textbf{Instance} & \multicolumn{1}{l}{} & \textbf{CPU} & \textbf{} & \textbf{} & \textbf{Instance} & \multicolumn{1}{l}{} & \textbf{CPU}       & \textbf{} \\ 

\midrule

\textbf{}         & \textbf{F2-ws}     & \textbf{F2}     & $\Delta_{F2, F2ws}$ & \textbf{} & \textbf{}         & \textbf{F2-ws}     & \textbf{F2}  & $\Delta_{F2, F2ws}$ & \textbf{} & \textbf{}         & \textbf{F2-ws}     & \textbf{F2}        & $\Delta_{F2, F2ws}$ \\[-0.25ex]

\midrule

1                & \textbf{0.1008}      & 0.2236          & 54.93     &           & 31                & \textbf{0.8512}      & 0.9331       & 8.78      &           & 61                & \textbf{666.9969}    & 824.1859           & 19.07     \\
2                & \textbf{0.1153}      & 0.4214          & 72.63     &           & 32                & \textbf{0.9265}      & 1.1392       & 18.68     &           & 62                & \textbf{7200.0000}   & \textbf{7200.0000} & 0.00      \\
3                & \textbf{0.0727}      & 0.2920          & 75.12     &           & 33                & \textbf{1.1595}      & 1.5184       & 23.64     &           & 63                & \textbf{1108.9581}   & 1255.9919          & 11.71     \\
4                & \textbf{0.0871}      & 0.5131          & 83.03     &           & 34                & \textbf{1.8419}      & 2.2396       & 17.76     &           & 64                & \textbf{1659.7399}   & 1936.1128          & 14.27     \\
5                & \textbf{0.0633}      & 0.4139          & 84.69     &           & 35                & \textbf{1.4991}      & 2.5727       & 41.73     &           & 65                & \textbf{768.7349}    & 905.3403           & 15.09     \\
6                & \textbf{0.0863}      & 0.4404          & 80.39     &           & 36                & \textbf{0.5590}      & 0.8970       & 37.68     &           & 66                & \textbf{1734.9434}   & 2009.4580          & 13.66     \\
7                 & \textbf{0.4766}      & 0.5416          & 12.00     &           & 37                & \textbf{0.9004}      & 1.3301       & 32.31     &           & 67                & \textbf{1087.8168}   & 1354.7480          & 19.70     \\
8                 & \textbf{0.4386}      & 0.5786          & 24.21     &           & 38                & \textbf{2.1188}      & 2.5860       & 18.07     &           & 68                & \textbf{2049.5867}   & 2338.4407          & 12.35     \\
9                 & \textbf{0.2030}      & 0.2888          & 29.71      &           & 39                & \textbf{1.8326}      & 2.8135       & 34.86     &           & 69                & \textbf{1530.0836}   & 1905.9716          & 19.72     \\
10                 & \textbf{0.2011}      & 0.7345          & 72.62     &           & 40                & \textbf{3.1554}      & 4.1034       & 23.10     &           & 70                & \textbf{2169.0729}   & 2469.0323          & 12.15     \\
11                 & \textbf{0.1715}      & 0.4311          & 60.22     &           & 41                & \textbf{0.3282}      & 0.9381       & 65.01     &           & 71                & \textbf{852.8703}    & 983.5572           & 13.29     \\
12                 & \textbf{0.3166}      & 0.8308          & 61.89     &           & 42                & \textbf{0.6280}      & 1.7523       & 64.16     &           & 72                & \textbf{722.9097}    & 884.6691           & 18.28     \\
13                 & \textbf{0.2909}      & 0.4120          & 29.40     &           & 43                & \textbf{1.9229}      & 2.6392       & 27.14     &           & 73                & \textbf{989.8646}    & 1147.7326          & 13.75     \\
14                 & \textbf{0.4330}      & 0.5765          & 24.88     &           & 44                & \textbf{2.8586}      & 3.8372       & 25.50     &           & 74                & \textbf{1049.9244}   & 1268.2838          & 17.22     \\
15                 & \textbf{0.2621}      & 0.2753          & 4.78      &           & 45                & \textbf{4.0976}      & 5.7957       & 29.30     &           & 75                & \textbf{1026.3015}   & 1202.6681          & 14.66     \\
16                & \textbf{0.3315}      & 0.7318          & 54.70      &           & 46                & \textbf{0.9616}      & 3.4578       & 72.19     &           & 76                & \textbf{1079.7493}   & 1294.5702          & 16.59     \\
17                & \textbf{0.3683}      & 0.5949          & 38.09     &           & 47                & \textbf{2.8655}      & 4.2304       & 32.26     &           & 77                & \textbf{1360.9421}   & 1509.7140          & 9.85      \\
18                & \textbf{0.6615}      & 1.0459          & 36.76     &           & 48                & \textbf{5.2100}      & 7.3264       & 28.89     &           & 78                & \textbf{2367.6941}   & 2675.9165          & 11.52     \\
19                & \textbf{0.2933}      & 0.5384          & 45.53     &           & 49                & \textbf{13.6404}     & 17.4488      & 21.83     &           & 79                & \textbf{1569.5841}   & 1787.6726          & 12.20     \\
20                & 0.6525               & \textbf{0.6493} & -0.50     &           & 50                & \textbf{8.5745}      & 11.4346      & 25.01     &           & 80                & \textbf{1988.7290}   & 2338.7394          & 14.97     \\
21                & \textbf{0.6331}      & 0.6916          & 8.46     &           & 51                & \textbf{0.7239}      & 7.1962       & 89.94     &           & 81                & \textbf{655.7005}    & 776.1643           & 15.52     \\
22                & \textbf{0.9480}      & 0.9937          & 4.60     &           & 52                & \textbf{1.4469}      & 5.2770       & 72.58     &           & 82                & \textbf{1407.1212}   & 2595.9413          & 45.80     \\
23                & \textbf{0.2800}      & 0.4706          & 40.50     &           & 53                & \textbf{3.9724}      & 10.1268      & 60.77     &           & 83                & \textbf{7200.0000}   & \textbf{7200.0000} & 0.00      \\
24                & 1.2983               & \textbf{1.2622} & -2.85     &           & 54                & \textbf{4.7098}      & 12.4966      & 62.31     &           & 84                & \textbf{1730.3946}   & 2075.4408          & 16.63     \\
25                & \textbf{0.2478}      & 0.3022          & 17.99     &           & 55                & \textbf{11.6508}     & 17.9711      & 35.17     &           & 85                & \textbf{1121.9333}   & 1344.9506          & 16.58     \\
26                & \textbf{0.4783}      & 0.8453          & 43.42     &           & 56                & \textbf{1.4056}      & 4.1502       & 66.13     &           & 86                & \textbf{1635.3323}   & 1953.5618          & 16.29     \\
27                & \textbf{0.2995}      & 0.4340          & 30.99     &           & 57                & \textbf{1.6399}      & 11.0508      & 85.16     &           & 87                & \textbf{1100.0952}   & 1294.0546          & 14.99     \\
28                & \textbf{0.8081}      & 1.1727          & 31.09     &           & 58                & \textbf{2.7634}      & 12.9963      & 78.74     &           & 88                & \textbf{1804.7383}   & 2141.9116          & 15.74     \\
29                & \textbf{0.6944}      & 0.9684          & 28.29     &           & 59                & \textbf{14.8052}     & 21.2720      & 30.40     &           & 89                & \textbf{1325.5667}   & 1553.8070          & 14.69     \\
30                & \textbf{0.8524}      & 0.9548          & 10.72     &           & 60                & \textbf{7.2250}      & 20.5363      & 64.82     &           & 90                & \textbf{2634.9724}   & 3066.1234          & 14.06     \\

\midrule

\textbf{Avg}      & 0.4055               & 0.6210          & 38.61     &           & \textbf{Avg}      & 3.5425               & 6.7356       & 43.13     &           & \textbf{Avg}      & \textbf{1786.6786}   & 2043.1587          & 15.01     \\
\textbf{N. best}  & 28                   & 2               &           &           & \textbf{N. best}  & 30                   & 0            &           &           & \textbf{N. best}  & 30                   & 2                  & \\         
\bottomrule
\end{tabular}
}
\end{table}
\label{table:warmstart-evaluation}
\end{landscape}

\section{Comparing the exact algorithm with SGS-ES}
\label{sec:SGS-ES-vs-exact-algorithm}

This section concludes the presentation of the numerical results.
The purpose of this section is to compare the most computationally efficient exact solution approach with the whole heuristic schema presented in this thesis to evaluate the trade-off between solutions quality and required computational times.
This section achieves so by comparing the performances of the exact algorithm that exploits \hyperlink{F2}{Formulation 2} for the mathematical programming step with SGH as a heuristic for the initial feasible solution, and SGS-ES.
The optimality of the solutions obtained by the exact algorithm allows \addRR{us }to disregard the distribution metrics.
Indeed, the purpose of the comparison is to evaluate the quality of the Pareto front computed by SGS-ES with respect to the reference optimal Pareto front computed by the exact algorithm, regardless of its structure, i.e., the uniformity of the distribution of the Pareto-optimal solutions.

Table \ref{table:final-comparison} shows the results of the comparisons.
For each instance, the time limit for computations was again set to 7200 seconds.
The values under the column $\Delta_{F2ws, SE}$ are defined similarly to the values of $\Delta_{F1, F2}$ in Section \ref{sec:formulations-comparison}.
Formally, let $\delta_{SE, i}$ and $\delta_{F2ws, i}$ be the time taken for SGS-ES and the exact algorithm, with \hyperlink{F2}{Formulation 2} for the MILP step and SGH as the initial feasible solution heuristic, to solve instance $i$, respectively.
Then, the value of $\Delta_{F2ws, SE}$ for instance $i$ is $1 - \delta_{SE, i} / \delta_{F2ws, i}$.
SGS-ES achieves significantly better computational times on the whole benchmark. In particular, the values under the column $\Delta_{F2ws, SE}$ suggest an average improvement of $73.56\%$ for the MLS instances, which is as high as $96.08\%$ and $96.55\%$ for the VLS and the small-scale instances, respectively.

Hypervolume is the only quality metric reported in the table, as it is sufficient to highlight the difference in solutions quality attained by the two algorithms.
Although the exact algorithm considerably outperforms SGS-ES on the MLS and the VLS instances, except for instances 62 and 83, SGS-ES succeeds in computing the optimal solutions for ten of the small-scale instances (1, 11, 13, 20, 25, 26, 27, 28, 29, and 30).
The performances of SGS-ES may be improved by considering a larger set of EPS moves, at the expense of a higher computational burden.

\begin{landscape}
\begin{table}[b]
\centering
\captionsetup{font=footnotesize}
\caption{Comparison of the results achieved by the (warmstarted) exact algorithm with Formulation 2 and SGS-ES on each benchmark instance based on the Hypervolume and the CPU time (s) metrics.}
\scalebox{0.6}{
\begin{tabular}{cccccccccccccccccccc}
\toprule

\textbf{Instance} & \multicolumn{2}{c}{\textbf{Hypervolume}} & \multicolumn{3}{c}{\textbf{CPU}}                     &  & \textbf{Instance} & \multicolumn{2}{c}{\textbf{Hypervolume}} & \multicolumn{3}{c}{\textbf{CPU}}                     &  & \textbf{Instance} & \multicolumn{2}{c}{\textbf{Hypervolume}} & \multicolumn{3}{c}{\textbf{CPU}}                       \\

\midrule

\textbf{}         & \textbf{F2-ws}      & \textbf{SGS-ES}    & \textbf{F2-ws} & \textbf{SGS-ES} & $\Delta_{F2ws, SE}$ &  & \textbf{}         & \textbf{F2-ws}      & \textbf{SGS-ES}    & \textbf{F2-ws} & \textbf{SGS-ES} & $\Delta_{F2ws, SE}$ &  & \textbf{}         & \textbf{F2-ws}      & \textbf{SGS-ES}    & \textbf{F2-ws} & \textbf{SGS-ES}   & $\Delta_{F2ws, SE}$ \\

\midrule

1                 & \textbf{0.7460}     & \textbf{0.7460}    & 0.4766         & \textbf{0.0059} & 98.77             &  & 31                & \textbf{0.8829}     & 0.8813             & 0.8512         & \textbf{0.0384} & 95.48             &  & 61                & \textbf{0.8223}     & 0.8162             & 666.9969       & \textbf{18.6766}  & 97.20             \\
2                 & \textbf{0.7756}     & 0.7745             & 0.4386         & \textbf{0.0098} & 97.76             &  & 32                & \textbf{0.8122}     & 0.8109             & 0.9265         & \textbf{0.0551} & 94.06             &  & 62                & 0.2722              & \textbf{0.8262}    & \addRR{7200.0000}      & \textbf{53.8304}  & 99.25             \\
3                 & \textbf{0.7603}     & 0.7408             & 0.2030         & \textbf{0.0046} & 97.72             &  & 33                & \textbf{0.7783}     & 0.7769             & 1.1595         & \textbf{0.0767} & 93.38             &  & 63                & \textbf{0.7704}     & 0.7645             & 1108.9581      & \textbf{22.1941}  & 98.00             \\
4                 & \textbf{0.7566}     & 0.7563             & 0.2011         & \textbf{0.0137} & 93.18             &  & 34                & \textbf{0.7021}     & 0.6916             & 1.8419         & \textbf{0.0946} & 94.86             &  & 64                & \textbf{0.7886}     & 0.7827             & 1659.7399      & \textbf{49.0354}  & 97.05             \\
5                 & \textbf{0.8019}     & 0.8017             & 0.1715         & \textbf{0.0077} & 95.52             &  & 35                & \textbf{0.6012}     & 0.6011             & 1.4991         & \textbf{0.0483} & 96.78             &  & 65                & \textbf{0.7298}     & 0.7205             & 768.7349       & \textbf{25.6876}  & 96.66             \\
6                 & \textbf{0.8247}     & 0.8234             & 0.3166         & \textbf{0.0148} & 95.34             &  & 36                & \textbf{0.8563}     & 0.8541             & 0.5590         & \textbf{0.0641} & 88.54             &  & 66                & \textbf{0.8064}     & 0.7984             & 1734.9434      & \textbf{65.3891}  & 96.23             \\
7                 & \textbf{0.7259}     & 0.7256             & 0.2909         & \textbf{0.0031} & 98.92             &  & 37                & \textbf{0.8731}     & 0.8710             & 0.9004         & \textbf{0.0914} & 89.85             &  & 67                & \textbf{0.7590}     & 0.7549             & 1087.8168      & \textbf{22.0816}  & 97.97             \\
8                 & \textbf{0.8153}     & 0.8137             & 0.4330         & \textbf{0.0104} & 97.59             &  & 38                & \textbf{0.8473}     & 0.8407             & 2.1188         & \textbf{0.1351} & 93.63             &  & 68                & \textbf{0.7446}     & 0.7372             & 2049.5867      & \textbf{72.8456}  & 96.45             \\
9                 & \textbf{0.7687}     & 0.7559             & 0.2621         & \textbf{0.0054} & 97.94             &  & 39                & \textbf{0.7804}     & 0.7771             & 1.8326         & \textbf{0.2147} & 88.29             &  & 69                & \textbf{0.7418}     & 0.7356             & 1530.0836      & \textbf{22.0728}  & 98.56             \\
10                & \textbf{0.7732}     & 0.7722             & 0.3315         & \textbf{0.0153} & 95.39             &  & 40                & \textbf{0.7823}     & 0.7730             & 3.1554         & \textbf{0.2913} & 90.77             &  & 70                & \textbf{0.7563}     & 0.7450             & 2169.0729      & \textbf{77.6561}  & 96.42             \\
11                & \textbf{0.8077}     & \textbf{0.8077}    & 0.3683         & \textbf{0.0079} & 97.85             &  & 41                & \textbf{0.9294}     & 0.9282             & 0.3282         & \textbf{0.0651} & 80.17             &  & 71                & \textbf{0.8055}     & 0.7983             & 852.8703       & \textbf{21.7973}  & 97.44             \\
12                & \textbf{0.8506}     & 0.8484             & 0.6615         & \textbf{0.0193} & 97.09             &  & 42                & \textbf{0.8535}     & 0.8529             & 0.6280         & \textbf{0.1035} & 83.52             &  & 72                & \textbf{0.8763}     & 0.8704             & 722.9097       & \textbf{56.3769}  & 92.20             \\
13                & \textbf{0.6900}     & \textbf{0.6900}    & 0.2933         & \textbf{0.0038} & 98.72             &  & 43                & \textbf{0.8630}     & 0.8614             & 1.9229         & \textbf{0.1566} & 91.86             &  & 73                & \textbf{0.7987}     & 0.7918             & 989.8646       & \textbf{26.4831}  & 97.32             \\
14                & \textbf{0.7548}     & 0.7533             & 0.6525         & \textbf{0.0104} & 98.40             &  & 44                & \textbf{0.8279}     & 0.8223             & 2.8586         & \textbf{0.2399} & 91.61             &  & 74                & \textbf{0.8522}     & 0.8471             & 1049.9244      & \textbf{67.1380}  & 93.61             \\
15                & \textbf{0.7520}     & 0.7509             & 0.6331         & \textbf{0.0060} & 99.06             &  & 45                & \textbf{0.7936}     & 0.7899             & 4.0976         & \textbf{0.4059} & 90.09             &  & 75                & \textbf{0.7890}     & 0.7835             & 1026.3015      & \textbf{30.1399}  & 97.06             \\
16                & \textbf{0.7993}     & 0.7987             & 0.9480         & \textbf{0.0196} & 97.94             &  & 46                & \textbf{0.8266}     & 0.8162             & 0.9616         & \textbf{0.3468} & 63.94             &  & 76                & \textbf{0.8550}     & 0.8488             & 1079.7493      & \textbf{79.8189}  & 92.61             \\
17                & \textbf{0.7086}     & 0.7082             & 0.2800         & \textbf{0.0082} & 97.07             &  & 47                & \textbf{0.8642}     & 0.8612             & 2.8655         & \textbf{0.6161} & 78.50             &  & 77                & \textbf{0.7514}     & 0.7442             & 1360.9421      & \textbf{31.9174}  & 97.65             \\
18                & \textbf{0.8156}     & 0.8106             & 1.2983         & \textbf{0.0185} & 98.58             &  & 48                & \textbf{0.8760}     & 0.8725             & 5.2100         & \textbf{0.8131} & 84.39             &  & 78                & \textbf{0.7935}     & 0.7866             & 2367.6941      & \textbf{92.9936}  & 96.07             \\
19                & \textbf{0.5794}     & 0.5785             & 0.2478         & \textbf{0.0042} & 98.29             &  & 49                & \textbf{0.8064}     & 0.8020             & 13.6404        & \textbf{1.1943} & 91.24             &  & 79                & \textbf{0.7519}     & 0.7451             & 1569.5841      & \textbf{40.2177}  & 97.44             \\
20                & \textbf{0.7294}     & \textbf{0.7294}    & 0.4783         & \textbf{0.0084} & 98.24             &  & 50                & \textbf{0.8324}     & 0.8162             & 8.5745         & \textbf{1.5257} & 82.21             &  & 80                & \textbf{0.8171}     & 0.8118             & 1988.7290      & \textbf{86.7287}  & 95.64             \\
21                & \textbf{0.7458}     & 0.7455             & 0.2995         & \textbf{0.0070} & 97.66             &  & 51                & \textbf{0.8557}     & 0.8451             & 0.7239         & \textbf{0.6305} & 12.90             &  & 81                & \textbf{0.8230}     & 0.8152             & 655.7005       & \textbf{32.5687}  & 95.03             \\
22                & \textbf{0.7970}     & 0.7968             & 0.8081         & \textbf{0.0143} & 98.23             &  & 52                & \textbf{0.8681}     & 0.8656             & 1.4469         & \textbf{0.9671} & 33.16             &  & 82                & \textbf{0.8684}     & 0.8616             & 1407.1212      & \textbf{79.0559}  & 94.38             \\
23                & \textbf{0.7743}     & 0.7703             & 0.6944         & \textbf{0.0075} & 98.93             &  & 53                & \textbf{0.8903}     & 0.8875             & 3.9724         & \textbf{1.3249} & 66.65             &  & 83                & 0.2324              & \textbf{0.8159}    & \addRR{7200.0000}      & \textbf{36.6585}  & 99.49             \\
24                & \textbf{0.8413}     & 0.8403             & 0.8524         & \textbf{0.0217} & 97.45             &  & 54                & \textbf{0.8848}     & 0.8836             & 4.7098         & \textbf{1.8315} & 61.11             &  & 84                & \textbf{0.8314}     & 0.8261             & 1730.3946      & \textbf{97.6430}  & 94.36             \\
25                & \textbf{0.6976}     & \textbf{0.6976}    & 0.1008         & \textbf{0.0024} & 97.59             &  & 55                & \textbf{0.8454}     & 0.8445             & 11.6508        & \textbf{2.8734} & 75.34             &  & 85                & \textbf{0.8251}     & 0.8201             & 1121.9333      & \textbf{42.2191}  & 96.24             \\
26                & \textbf{0.7683}     & \textbf{0.7683}    & 0.1153         & \textbf{0.0059} & 94.86             &  & 56                & \textbf{0.8899}     & 0.8868             & 1.4056         & \textbf{0.9294} & 33.88             &  & 86                & \textbf{0.8337}     & 0.8276             & 1635.3323      & \textbf{111.6060} & 93.18             \\
27                & \textbf{0.7425}     & \textbf{0.7425}    & 0.0727         & \textbf{0.0037} & 94.88             &  & 57                & \textbf{0.7517}     & 0.7432             & 1.6399         & \textbf{1.4716} & 10.26             &  & 87                & \textbf{0.8107}     & 0.8041             & 1100.0952      & \textbf{45.2810}  & 95.88             \\
28                & \textbf{0.7438}     & \textbf{0.7438}    & 0.0871         & \textbf{0.0091} & 89.56             &  & 58                & \textbf{0.9060}     & 0.9057             & 2.7634         & \textbf{2.0238} & 26.76             &  & 88                & \textbf{0.8402}     & 0.8351             & 1804.7383      & \textbf{128.1692} & 92.90             \\
29                & \textbf{0.4531}     & \textbf{0.4531}    & 0.0633         & \textbf{0.0054} & 91.42             &  & 59                & \textbf{0.8994}     & 0.8950             & 14.8052        & \textbf{3.8813} & 73.78             &  & 89                & \textbf{0.7685}     & 0.7590             & 1325.5667      & \textbf{63.2728}  & 95.23             \\
30                & \textbf{0.7532}     & \textbf{0.7532}    & 0.0863         & \textbf{0.0116} & 86.58             &  & 60                & \textbf{0.8198}     & 0.8169             & 7.2250         & \textbf{3.6342} & 49.70             &  & 90                & \textbf{0.8131}     & 0.8068             & 2634.9724      & \textbf{136.8648} & 94.81             \\

\midrule

\textbf{Avg}      & 0.7517              & 0.7499             & 0.4055         & 0.0095          & 96.55             &  & \textbf{Avg}      & 0.8333              & 0.8292             & 3.5425         & 0.8715          & 73.56             &  & \textbf{Avg}      & 0.7643              & 0.7960             & 1786.7791      & 57.8807           & 96.08             \\
\textbf{N. best}  & 30                  & 10                 & 0              & 90              &                   &  & \textbf{N. best}  & 30                  & 0                  & 0              & 90              &                   &  & \textbf{N. best}  & 28                  & 2                  & 0              & 90                &                  \\
\bottomrule
\end{tabular}
}
\label{table:final-comparison}
\end{table}
\end{landscape}

%% file: chap5.tex
\chapter{Conclusions}
\label{chap:conclusions}

The achievement of energy efficiency in manufacturing production has become a compelling matter in the latest years, due to the pressing environmental issues and the consequent desire to shift towards a sustainable industry model.
The pricing scheme induced by the TOU-based tariffs policy indeed allows to drive customers demand so as to relieve peak energy generation, and enabling financial benefits for the energy providers as well.


This thesis considered a representative, hard TOU scheduling problem, the BPMSTP.
It provided two distinct mathematical models, as well as the two heuristics SGH and ES, based on several insights on the problem itself.
The exact algorithm based on the models, and the whole heuristic scheme based on SGH and ES, achieved high performances on the wide test benchmark.
The combinatorial properties of the BPMSTP, and the novel intuitions underlying the algorithms presented in this thesis, may also enable further research progress in problems that display a similar structure.

Among the immediate future developments of this thesis, it is important to discuss the scope for improvement of the exact algorithm, SGS, and SGS-ES.
Currently, such algorithms tackle the BPMSTP by solving a sequence of distinct single-objective optimization problems that constrain the maximum makespan of the solution.
Such problems are considered independently.
In fact, the three algorithms do not exploit the information provided by the job assignments performed at the previous iterations, despite the makespan varies by little between them.
Such a small variation intuitively suggests that neighboring solutions may share several job assignments.
In the case of SGS and SGS-ES, such an intuition may lead to increase the computational efficiency.
Instead, its application to the case of the exact algorithm may lead to a matheuristic where the previous assignments are treated as additional constraints in the mathematical program.

Furthermore, as specifically regards ES, the considered subset of EPS moves may be reduced by disregarding part of the EPS-I's only consisting of idle slots.
As a matter of fact, many such EPS-I's are already implicitly disregarded by SGH while performing the assignments of the jobs.
Such information could be exploited by ES so as to avoid considering EPS-I's that cannot lead to improvement.
Another improvement of the heuristic scheme may rely on using the exact algorithm to improve the initial solution provided by SGH as much as possible within a strict time limit, before applying ES.

\addRR{Finally, expanding the experimental benchmark by including instances with particular properties, such as non-increasing slot costs or distinct processing times, may be beneficial to gain further insight into the combinatorics of the BPMSTP.
Furthermore, the BPMSTP itself could be extended to take into account different machines environments, such as uniform or unrelated machines. The jobs could be characterized by machine-dependent setup times, while the relationships between distinct jobs could be captured by introducing sequence-dependent setup times.
Classical objectives in scheduling, such as the total weighted tardiness or the maximum lateness, could be considered along with the makespan or the TEC in order to increase the capability of the problem in modeling real manufacturing systems.
It would be interesting to investigate how the existing mathematical formulations and heuristics for the BPMSTP may translate to new problems with the aforementioned characteristics.}

\bigskip

In conclusion, the difficulty of the TOU scheduling problems, along with their practical nature, naturally drives research towards operational solutions for practitioners. 
Yet, several research efforts showed how observations on the solutions structure and properties often led to more efficient heuristics, in terms of both computational efficiency and solutions quality. 
Future research should address the difficulty of the problems by developing exact or fast heuristic algorithms resulting from a thorough investigation of the structure of the problem.
Specifically, at the time of writing, while the literature in TOU scheduling offers several mathematical modeling efforts,
it is lacking in matheuristics, and most of all exact algorithms that exploits mathematical programming.

%% file: appa.tex
\chapter{Additional notation}
\label{sec:notation}

This appendix presents a reference for some additional notation used in the thesis.
In order to describe the considered scheduling problems, the thesis extensively uses the triplet notation $\alpha \, | \, \beta \, | \, \gamma$ \citep{Graham1977OptimizationAA}, and adopts the naming conventions used by \citet{Pinedo2016}. 
For clarity, Table \ref{table:alfa-beta-gamma} reports it to the extent of the scope of the thesis.
Moreover, Table \ref{table:algorithms-acronyms} shows the acronyms used by the thesis for mathematical programming and algorithms.

\begin{table}[!b]
\captionsetup{font=footnotesize}
\centering
\scalebox{0.75}{
\begin{tabular}{cccc}
\toprule
\textbf{Field} & \textbf{Characteristic} & \textbf{Meaning} \\

\midrule

\multirow{8}{10em}{\makecell{$\alpha$ field}}
& $1$ & single machine\\
& $Pm$ & $m$ parallel identical machines\\
& $Qm$ & $m$ parallel machines with different speeds\\
& $Rm$ & $m$ unrelated parallel machines)\\
& $Fm$ & flow shop on $m$ machines\\
& $Jm$ & job shop on $m$ machines\\
& $FFc$ & flexible flow shop on $c$ machines\\
& $FJc$ & flexible job shop on $c$ machines\\

\midrule

\multirow{8}{10em}{\makecell{$\beta$ field}} 
& $r_j$ & release dates\\ 
& $d_j$ & due dates\\
& $w_j$ & weights\\ 
& $prmp$ & preemption\\ 
& $batch(b)$ & batch processing\\ 
& $prmu$ & permutation\\ 
& $nwt$ & no-wait)\\
& $rcrc$ & recirculation\\

\midrule

\multirow{6}{10em}{\makecell{$\gamma$ field}} 
& $C_{max}$ & makespan\\
& $L_{max}$ & maximum lateness\\
& $T_{max}$ & maximum tardiness\\
& $\sum w_j C_j$ & total weighted completion time\\
& $\sum w_j T_j$ & total weighted tardiness\\
& $\sum w_j U_j$ & weighted number of tardy jobs\\

\bottomrule
\end{tabular}
}
\caption{Three-field notation for classical scheduling problems.}
\label{table:alfa-beta-gamma}
\end{table}

\begin{landscape}
\centering
\begin{table}[p]
\captionsetup{font=footnotesize}
\centering
\scalebox{0.9}{
\begin{tabular}{cccc}
\toprule
\textbf{Class} & \textbf{Algorithm} & \textbf{Acronym} \\

\midrule

\multirow{6}{18em}{\makecell{Mathematical programming\\and approximation algorithms \citep{KorteVygen2018}}}
& Linear Programming & LP\\
& Integer Programming & IP\\
& Mixed-Integer Programming & MIP\\
& Mixed-Integer Linear Programming & MILP\\
& Mixed-Integer Non-Linear Programming & MINLP\\
& Polynomial-Time Approximation Scheme & PTAS\\

\midrule

\multirow{13}{18em}{\makecell{Heuristics and metaheuristics \citep{Luke2013Metaheuristics}}}
& Local Search & LS\\
& Iterated Local Search & ILS\\
& Genetic Algorithm & GA\\
& Evolution Algorithm & EA\\
& Multi-objective Evolutionary Algorithm based on Decomposition & MOEA/D\\
& Non-dominated Sorted Genetic Algorithm & NSGA\\
& Particle Swarm Optimization & PSO\\
& Tabu Search & TS\\
& Variable Neighborhood Search & VNS\\
& Ant Colony Optimization & ACO\\
& Single-Population Genetic Algorithm & SPGA\\
& Multi-Population Genetic Algorithm & MPGA\\
& Strength Pareto-archived Evolutionary Algorithm 2 & SPEA \\  

\bottomrule
\end{tabular}
}
\caption{Acronyms of some algorithms for discrete optimization.}
\label{table:algorithms-acronyms}
\end{table}
\end{landscape}

%% file: appb.tex
\chapter{Symbols}
\label{chap:symbol-table}

This appendix reports the main mathematical symbols used in this thesis as a reference.
Table \ref{table:symbol-table} reports the symbols related to the problem statement, and introduced in Chapter \ref{chap:problem}.
Table \ref{table:symbol-table-2} reports the symbols related to the algorithms described in the thesis, and introduced in Chapter \ref{chap:algorithms}.

\begin{table}[!b]
\captionsetup{font=footnotesize}
\centering
\scalebox{0.95}{
\begin{tabular}{p{0.45\linewidth}  p{0.5\linewidth}}
\toprule
\textbf{Symbol} & \textbf{Meaning} \\

\midrule

$\mathcal{J} = \{1, 2, \ldots, N\}$                   & Set of jobs\\

$p_j$, $j \in \mathcal{J}$                           & Processing time of job $j$\\

$\mathcal{H} = \{1, 2, \ldots, M\}$                   & Set of machines\\

$u_h$, $h \in \mathcal{H}$                           & Energy consumption rate of machine $h$ \\

$\mathcal{T} = \{1, 2, \ldots, K\}$                   & Set of time slots\\

$c_t$, $t \in \mathcal{T}$                           & Cost of slot $t$\\

$\mathcal{S}$                   & Schedule\\

$C_j(\mathcal{S})$, $j \in \mathcal{J}$              & Completion time of job $j$ in $\mathcal{S}$\\

$C^\text{max}(\mathcal{S})$                  & Makespan of $\mathcal{S}$\\

$E(\mathcal{S})$                             & Total energy cost of $\mathcal{S}$\\

$\mathcal{I}$                               & Instance $(\mathcal{J}, \{p_j, j \in \mathcal{J}\}, \mathcal{H}, \{u_h, h \in \mathcal{H}\}, \mathcal{T}, \{c_t, t \in \mathcal{T}\})$ of the BPMSTP\\

$\mathcal{P}_\mathcal{J'}$                  & Distinct processing times of the jobs in $\mathcal{J'} \subseteq \mathcal{J}$\\

$\mathcal{J}_d$, $d \in \mathcal{P}_\mathcal{J}$    & Set of jobs with processing time $d$\\

$b_{d, t}$, $d \in \mathcal{P}_\mathcal{J}$, $t = 1, 2, \ldots, K - d + 1$  & Sum of the costs of the slots in $\{t, t + 1, \ldots, t + d - 1\}$\\

\bottomrule
\end{tabular}
}
\caption{Table of the main mathematical symbols introduced in Chapter \ref{chap:problem}, and related to the problem statement.}
\label{table:symbol-table}
\end{table}

\begin{table}[!ht]
\captionsetup{font=footnotesize}
\centering
\scalebox{0.95}{
\begin{tabular}{p{0.49\linewidth}  p{0.51\linewidth}}
\toprule
\textbf{Symbol} & \textbf{Meaning} \\

\midrule

$(h, \mathcal{A})$              & Location in slots $\mathcal{A} \subseteq \mathcal{T}$ on machine $h \in \mathcal{H}$ (for a job with processing time $|\mathcal{A}|$)\\

$\mathcal{B}$           & Schedule block\\

$\mathcal{S}_h$         & Schedule on machine $h \in \mathcal{H}$\\

$\mathcal{D}(\hat{K})$  & Instance $(\mathcal{J}, \{p_j, j \in \mathcal{J}\}, \mathcal{H}, \{u_h, h \in \mathcal{H}\}, \hat{\mathcal{T}} = \{1, 2, \ldots, \hat{K}\} \subseteq \mathcal{T}, \{c_t, t \in \hat{\mathcal{T}}\})$ of the BPMSTP\\

$\Omega$        & Minimum between $N$ and $\hat{K}$, $1 \le \hat{K} \le K$\\

&\\

$\mathcal{S}_\mathcal{E}$ ($\mathcal{S}_{\mathcal{E}, h}$) & EPS subschedule of $\mathcal{E}$ (on $h \in \mathcal{H}$ in $\mathcal{S}$)\\

$\mathcal{J}(\mathcal{S}_\mathcal{E})$ & Jobs scheduled in the subschedule $\mathcal{S}_\mathcal{E}$\\

$p_\text{max}$ & Maximum processing time in $\mathcal{P}_\mathcal{J}$\\

$\mu_{t, h}$, $t \in \mathcal{T}$, $h \in \mathcal{H}$ & Sum of the costs of the slots in $\{1, 2, \ldots, t\}$, multiplied by $u_h$\\

$\mathcal{L}^J_p(\mathcal{S})$, $\mathcal{L}^I_p(\mathcal{S})$ & Sets of subschedules of EPS-J's and EPS-I's in $\mathcal{S}$ of cardinality $p \in \mathcal{P}_\mathcal{J}$, respectively\\

$\mathcal{M}^J_p(\mathcal{S}) : \mathcal{H} \times \mathcal{T} \rightarrow \mathcal{L}^J_p(\mathcal{S}) \cup \emptyset, \quad p \in \mathcal{P}_\mathcal{J}$ & Function that associates a time slot $t$ on a machine $h$ with a subschedule $\mathcal{S}_\mathcal{E}$ associated to the EPS-J $\mathcal{E}$ on $h$, and such that the smallest slot of $\mathcal{E}$ is $t$.\\

$\mathcal{M}^I_p(\mathcal{S}) : \mathcal{H} \times \mathcal{T} \rightarrow \mathcal{L}^I_p(\mathcal{S}) \cup \emptyset, \quad p \in \mathcal{P}_\mathcal{J}$ & Function that associates a time slot $t$ on a machine $h$ with a subschedule $\mathcal{S}_\mathcal{E}$ associated to the EPS-I $\mathcal{E}$ on $h$, and such that the smallest slot of $\mathcal{E}$ is $t$.\\

$Q_\mathcal{E}$ & List of the slots in the EPS $\mathcal{E}$ in non-decreasing order of their costs\\

$\eta_{h, n}^\text{lb}(Q_\mathcal{E})$ & Sum of the costs of the first $n$ slots in $Q_\mathcal{E}$, multiplied by $u_h$\\

$\eta_{h}^\text{ub}(\mathcal{E})$ & Sum of the costs of the slots in $\mathcal{E}$\\

$\alpha(\mathcal{S}_\mathcal{E})$ & Number of assigned slots in the subschedule $\mathcal{S}_\mathcal{E}$\\

$\underline{K}(\mathcal{I})$ & Lower bound $\max\left\{ \lfloor \: \sum_{j \in \mathcal{J}} p_j / M \rfloor, \max_{j \in \mathcal{J}}\{p_j\}\right\}$ for the makespan in SGS and the exact algorithm\\

\bottomrule
\end{tabular}
}
\caption{Table of the main mathematical symbols introduced in Chapter \ref{chap:algorithms}, and used in the description of SGH, ES, SGS, and the exact algorithm.}
\label{table:symbol-table-2}
\end{table}

%% file: appc.tex
\chapter{Reference}
\label{chap:reference}

This appendix provides a summary of the main definitions and notions introduced in the thesis, by means of Table \ref{table:reference-1} and Table \ref{table:reference-2}.

\begin{table}[!b]
\captionsetup{font=footnotesize}
\centering
\scalebox{0.875}{
\begin{tabular}{p{0.425\linewidth}  p{0.625\linewidth}}
\toprule
\textbf{Notion} & \textbf{Meaning} \\
\midrule

Adjacent slots & Two slots $t$ and $t + 1$, with $t \in \mathcal{T}$, $t \neq K$\\

Assigned location & Location that only contains the assigned slots of one job\\

Assigned-consecutive slots & Two slots $t$ and $t + k$, $1 \le t \le t + 2 \le t + k \le K$, on machine $h \in \mathcal{H}$, that are assigned slots of some job $j \in \mathcal{J}$, and such that $t + 1, t + 2, \ldots, t + k - 1$ are assigned slots of a subset of jobs of $\mathcal{J} \setminus \{j\}$.\\


Exchange Search (ES) & The local search for the BPMSTP presented in this thesis\\

Exchangeable Period Sequence (EPS) & A subset $\mathcal{E} \subseteq \mathcal{T}$ of adjacent time slots on a machine $h \in \mathcal{H}$ such that if $h \in \mathcal{H}$ processes some job $j \in \mathcal{J}$
in a time slot $t \in \mathcal{E}$, then all the assigned slots of $j$ are in $\mathcal{E}$\\

EPS move & An EPS swap of two EPS's $\mathcal{E}$ and $\mathcal{E}'$, followed by an EPS rearrangement of $\mathcal{E}$, and an EPS rearrangement of $\mathcal{E}'$\\

EPS rearrangement & A procedure that reassigns each job scheduled in an EPS $\mathcal{E}$ on machine $h \in \mathcal{H}$ to a subset of adjacent slots in $\mathcal{E}$ on $h$.\\

EPS subschedule & The single-machine schedule associated with an EPS\\

EPS swap & For two given EPS's $\mathcal{E}$ and $\mathcal{E}'$ on machines $h$ and $h'$, respectively, an algorithm that schedules, in $\mathcal{E}'$ on machine $h'$, each job originally scheduled in $\mathcal{E}$ on machine $h$, and vice-versa, without changing the relative assignments of the jobs.\\

\bottomrule
\end{tabular}
}
\caption{A summary of the main definitions and notions introduced in the thesis.}
\label{table:reference-1}
\end{table}

\begin{table}[!]
\captionsetup{font=footnotesize}
\centering
\scalebox{0.875}{
\begin{tabular}{p{0.35\linewidth}  p{0.65\linewidth}}
\toprule
\textbf{Notion} & \textbf{Meaning} \\
\midrule

Free-consecutive slots & Two free slots $t$ and $t + k$ in $\mathcal{T}$ on a machine $h \in \mathcal{T}$, such that $t + 1, t + 2, \ldots, t + k - 1$ are assigned\\

Free location & Location that contains only free slots\\

Free (or idle) slot & A slot $t \in \mathcal{T}$ on a machine $h \in \mathcal{H}$, such that no job is assigned to $t$ on $h$\\

Location & An ordered pair $(h, \mathcal{A})$, $h \in \mathcal{H}$, $\mathcal{A} \subseteq \mathcal{T}$\\

Schedule block & Single-machine schedule that involves only subsets of non-free slots delimited by two free slots\\

Split-Greedy Heuristic (SGH) & The constructive heuristic for the BPMSTP presented in this thesis\\

Split-Greedy Scheduler (SGS) & The algorithm presented in this thesis that iteratively exploits SGH to find a heuristic Pareto front for the BPMSTP\\

Split-Greedy Scheduler with Exchange Search (SGS-ES) & The algorithm presented in this thesis that iteratively exploits SGH and ES to find a heuristic Pareto front for the BPMSTP\\

Split-location & Location including at least two free-consecutive slots\\

Split-schedule & A schedule with at least a split-scheduled job\\

Split-scheduled job & A job that is assigned to a split-location\\

\bottomrule
\end{tabular}
}
\caption{A summary of the main definitions and notions introduced in the thesis (continued).}
\label{table:reference-2}
\end{table}

%% file: biblio.tex
\begin{singlespace}
\bibliography{main.bib}
\bibliographystyle{plainnat}
\end{singlespace}